\documentclass[aps,pra,twocolumn,superscriptaddress,letterpaper]{revtex4-1}
\usepackage{graphicx}
\usepackage{amsmath}
\usepackage{amssymb}
\usepackage{esint}
\usepackage{verbatim}
\usepackage{xcolor}
\usepackage{soul}
\usepackage{siunitx}
\usepackage{hyperref}

\newcommand{\be}{\begin{equation}}
\newcommand{\ee}{\end{equation}}
\newcommand{\ba}{\begin{array}}
\newcommand{\ea}{\end{array}}
\newcommand{\bqa}{\begin{eqnarray}}
\newcommand{\eqa}{\end{eqnarray}}

\begin{document}

\title{Tunneling in the Brillouin Zone:\\ Theory of Backscattering in Valley Hall Edge Channels}

\author{Tirth Shah}
\affiliation{Max Planck Institute for the Science of Light, Staudtstrasse 2, 91058 Erlangen, Germany}
\affiliation{Department of Physics, Friedrich-Alexander Universit\"at Erlangen-N\"urnberg, Staudtstrasse 7, 91058 Erlangen, Germany}
\author{Florian Marquardt}
\affiliation{Max Planck Institute for the Science of Light, Staudtstrasse 2, 91058 Erlangen, Germany}
\affiliation{Department of Physics, Friedrich-Alexander Universit\"at Erlangen-N\"urnberg, Staudtstrasse 7, 91058 Erlangen, Germany}
\author{Vittorio Peano}
\affiliation{Max Planck Institute for the Science of Light, Staudtstrasse 2, 91058 Erlangen, Germany}
\email{vittorio.peano@mpl.mpg.de}

\date{\today}

\begin{abstract} 
A large set of recent experiments has been exploring topological transport in bosonic systems, e.g. of photons or phonons. In the vast majority, time-reversal symmetry is preserved, and band structures are engineered by a suitable choice of geometry, to produce topologically nontrivial bandgaps in the vicinity of high-symmetry points. However, this leaves open the possibility of large-quasimomentum backscattering, destroying the topological protection. Up to now, it has been unclear what precisely are the conditions where this effect can be sufficiently suppressed. In the present work, we introduce a comprehensive semiclassical theory of tunneling transitions in momentum space, describing backscattering for one of the most important system classes, based on the valley Hall effect. We predict that even for a smooth domain wall effective scattering centres develop at locations determined by both the local slope of the wall and the energy. Moreover, our theory provides a quantitative analysis of the exponential suppression of the overall reflection amplitude with increasing domain wall smoothness.
\end{abstract}

\maketitle

\section{Introduction}  

The quest for low-imprint high-frequency devices for the robust transport of classical waves such as light and vibrations has pushed research towards devices where the wavelength of the relevant excitations is of the order of  the lattice scale, which itself is limited by the fabrication precision. In time-symmetry broken topological phononic and photonic systems, backscattering from defects and scatterers is completely suppressed, however it is challenging to break time-reversal symmetry at the nanoscale \cite{peano_topological_2015,nash_topological_2015, mathew_synthetic_2020,wang_observation_2009,bahari_nonreciprocal_2017}.

Time-symmetric topological insulators support helical edge states that are protected by  Kramers degeneracy \cite{kane_quantum_2005, bernevig_quantum_2006, hasan_colloquium_2010}. Kramers degeneracy prevents any coupling between these counter-propagating  states and, thus, any backscattering. It is automatically realized in any time-symmetric fermionic system because   ${\cal T}^2=-1\!\!1$ for the time-reversal operator ${\cal T}$ of fermionic particles. On the other hand, ${\cal T}^2=1\!\!1$ for bosons and, thus, time-reversal-symmetric bosonic systems do not naturally have Kramers degeneracy. Nevertheless, they can mimic the physics of topological time-symmetric fermions in the presence of an  engineered  anti-unitary symmetry  ${\cal T}_{\rm en}$ with ${\cal T}_{\rm en}^2=-1\!\!1$. In practice, this is achieved by designing a Hamiltonian that is identical to the Hamiltonian of a fermionic topological insulator across the Brillouin zone (BZ) \cite{ningyuan_time-_2015,susstrunk_observation_2015}. The topological transport will then be protected against any perturbation that commutes with ${\cal T}_{\rm en}$ or, equivalently,  the engineered unitary symmetry $U={\cal T}{\cal T}_{\rm en}$. This approach allows to implement edge states that are able to turn any arbitrary sharp corners but it requires a high degree of control of the Hamiltonian engineering. For this reason, it is not easily transferable to  miniaturized devices.


An alternative approach for  implementations  of topological transport in classical bosonic systems at the micro- and nanoscale consists in  reproducing the Hamiltonian of a topological fermionic counterpart only in the vicinity of one or more  high-symmetry points in the Brillouin zone \cite{martin_topological_2008,ju_topological_2015,ma_all-si_2016,lu_observation_2017,dong_valley_2017,vila_observation_2017,wu_direct_2017,gao_valley_2017,kang_pseudo-spinvalley_2018,noh_observation_2018,shalaev_robust_2019,zeng_electrically_2020,ren_topological_2020,arora_direct_2021,wu_scheme_2015,he_acoustic_2016,brendel_snowflake_2018,yang_visualization_2016,cha_experimental_2018,barik_topological_2018,parappurath_direct_2020,shao_high-performance_2020}. 
In these approaches,  the smooth envelope of each helical edge state is described by a different Dirac Hamiltonian. The two  Dirac  Hamiltonians  are mapped onto each other via the time-reversal symmetry ${\cal T}$, but are otherwise decoupled. 
This approach is more suitable to the small scale because it  is based on robust symmetry-based principles (more on this below). On the other hand, the topological protection is only guaranteed within a smooth-envelope approximation. This approximation does not capture backscattering induced by large quasi-momentum transfer.  Heuristically, one should expect that these backscattering processes should be suppressed as long as  the envelope is smooth on the lattice scale.  Empirically, many experiments and numerical studies of  smooth-envelope topological systems have convincingly demonstrated good protection. However, most works did not attempt to quantify the residual backscattering, see  \cite{lu_observation_2017,brendel_snowflake_2018} for two notable exceptions. Even these two pioneering works did not pursue an analytical approach and, thus,  their findings are difficult to transfer to future investigations. Thus,  the nature and extent of the topological protection for smooth-envelope topological insulators remains unclear.

In this paper, we  present a theory of  backscattering for smooth-envelope topological insulators.  We  show that, in this setting, backscattering can be interpreted as tunneling on the surface of a torus, the quasi-momentum space. This insight allows us to employ advanced WKB techniques to quantify this phenomenon.  This in turn provides  guidance in improving future devices. Our results are most relevant for the widely investigated so-called Valley Hall effect, where the topological edge states are localized in two different quasi-momentum valleys \cite{martin_topological_2008,ju_topological_2015,ma_all-si_2016,lu_observation_2017,dong_valley_2017,vila_observation_2017,wu_direct_2017,gao_valley_2017,kang_pseudo-spinvalley_2018,noh_observation_2018,shalaev_robust_2019,zeng_electrically_2020,ren_topological_2020,arora_direct_2021}. 
However, the physical insight that we provide, as well as some of our analytical  results, can also be transferred to other smooth-envelope topological insulators where both helical edge states are localized around the $\Gamma$ point \cite{wu_scheme_2015,he_acoustic_2016,brendel_snowflake_2018,yang_visualization_2016,cha_experimental_2018,barik_topological_2018,parappurath_direct_2020,shao_high-performance_2020}. Our work ties to other investigations that have adopted the WKB approximation to investigate the electronic band structure or density of states in  graphene and other materials in the presence of smooth electromagnetic fields \cite{vogl_semiclassics_2017,gosselin_berry_2009,fuchs_topological_2010,delplace_wkb_2010,reijnders_electronic_2018,doost_foldy-wouthuysen_2021}. 

\section{Review of the smooth-envelope approach}

 Each of the two  edge states of a smooth-envelope topological insulator is described by a Dirac equation in the form 
\begin{equation}
i\dot{\boldsymbol{\Psi}}(t)=\hat{H}_{D}\boldsymbol{\Psi}
,\quad \hat{H}_{D}= m(\hat{\mathbf{x}})\hat{\sigma}_{z}+v \hat{\mathbf{p}}\cdot\hat{\boldsymbol{\sigma}}.  \label{eq:H_Dirac}  
\end{equation}
Here,  $\mathbf{x}=(x,y)$ is the position, $\mathbf{p}=-i\boldsymbol{\nabla}$,
$\hat{\sigma}_{z}$ is the $z$-Pauli matrix, and the 2D vector $\hat{\boldsymbol{\sigma}}$ groups the $x$- and $y$-Pauli matrices. Moreover,  the components $\Psi_{1}(\mathbf{x})$ and $\Psi_{2}(\mathbf{x})$ of the vector field $\boldsymbol{\Psi}(\mathbf{x})$ are the smooth envelopes modulating two rotationally symmetric Bloch waves. In other words, $\mathbf{p}$ is the quasi-momentum counted off from a rotationally symmetric high-symmetry point.
More specifically, this Hamiltonian  with mass parameter $m(\mathbf{x})=0$ is relevant for any periodic structure with an underlying hexagonal Bravais lattice that supports a pair of Dirac cones  at the rotationally-invariant high-symmetry points  $\boldsymbol{\Gamma}$ (two-fold degenerate), or   $\mathbf{K}$  and $\mathbf{K}'$.
The gap-opening perturbation $m$  is engineered by changing the geometrical parameters to move away from an accidental degeneracy \cite{mousavi_topologically_2015,he_acoustic_2016,miniaci_experimental_2018} or by breaking a symmetry to split an essential degeneracy. Examples of the latter include enlarging the unit cell \cite{wu_scheme_2015,brendel_snowflake_2018,cha_experimental_2018,parappurath_direct_2020}, breaking the two-fold symmetry  in a structure with $C_{6}$ symmetry \cite{martin_topological_2008,ju_topological_2015,ma_all-si_2016,dong_valley_2017,vila_observation_2017,wu_direct_2017,noh_observation_2018,shalaev_robust_2019,zeng_electrically_2020,arora_direct_2021}, and  breaking the mirror symmetry in a structure with $C_{3v}$  symmetry \cite{lu_observation_2017,kang_pseudo-spinvalley_2018,gao_valley_2017}.  This allows to tune the mass parameter $m(\mathbf{x})$.
We assume that the mass $m(\mathbf{x})$ defines  two adjacent bulk regions separated by a domain wall where $m(\mathbf{x})=0$. The mass can abruptly change across the domain walls or smoothly vary to  reach the asymptotic values $m(\mathbf{x})\approx \pm m_{\rm bk}$ in the two adjacent bulk regions.
The resulting Dirac cones bulk band structure $E_n(\mathbf{p})=(-1)^{n} \sqrt{m_{\rm bk}^2+v^2|\mathbf{p}|^2}$ ($n=1,2$) is identical  in the two domains and has band gap $2m_{\rm bk}$. The two domains are, however, topologically distinct because they have half-integer Chern numbers (here, defined as the integral of the Berry connection over the 2D plane) with opposite sign, ${\cal C}_n=(-1)^n{\rm sign}(mv)/2$.

An exact solution of Eq.(\ref{eq:H_Dirac}),  originally derived by Jackiw and Rebbi \cite{jackiw_solitons_1976}, shows that a translationally invariant domain wall supports  a chiral gapless edge state. If we choose a Cartesian coordinate system with unit vectors  $\mathbf{e}_s=\cos\varphi \mathbf{e}_x+\sin\varphi \mathbf{e}_y$ along the domain wall and $\mathbf{e}_r=\mathbf{e}_z\wedge \mathbf{e}_s$ normal to it, and fix the origin and direction of $\mathbf{e}_s$ such that with $m>0$ ($m<0$) for $r<0$ ($r>0$),  the edge state solution reads
\begin{equation}
E_{p_s}=vp_s,\quad\boldsymbol{\Psi}_{p_s}= C\begin{pmatrix}1\\
e^{i\varphi}
\end{pmatrix}e^{ip_s s}  e^{\int_0^r dr' m(r')/v },\label{eq:JackiwandRebbi}
\end{equation}
where $C$ is a normalization constant.
This is in agreement with the bulk-boundary correspondence because ${\cal C}_0(m_{\rm bk})-{\cal C}_0(-m_{\rm bk})=1$. 

In this work, we will be interested eventually in situations where waves traveling along an edge state are backscattered because the domain wall is curved or possibly even has sharp corners. This is obviously a practically very relevant scenario for real applications of topological transport. One way to characterize backscattering in such situations is to consider a closed domain wall, which produces a 'topological cavity', i.e. the energy eigenstates become quantized according to the total circumference of the domain wall loop. In that case, backscattering reveals itself in terms of a level splitting emerging from ideally degenerate counterpropagating solutions \cite{zhang_directional_2018,ren_topological_2020}.

More specifically, in a sufficiently smooth, closed  domain wall Eq.~(\ref{eq:H_Dirac}) will still apply, but now  with $s$ being the arc length along the domain wall (from a reference point on the domain wall),  $r$ the local coordinate transverse to the domain wall, and with the angle $\varphi$ being  $s$-dependent. The periodic boundary conditions will then lead to the  quantization condition 
\begin{equation}\label{eq:Bohr_quantization}
 p_n=\frac{2\pi}{L} n.   
\end{equation}
where $L$ is the arc length of the domain wall.
As we discussed above, each of these  running wave approximate solutions will have a time-reversed partner solution with the same energy within the smooth-envelope approximation.   Unlike for Kramers doublets in fermionic systems, here, the degeneracy
 is not protected by an exact symmetry. Thus, one should expect that large quasi-momentum transfer beyond the smooth-envelope approximation will induce a small coupling $\Delta(E)$ between the two partner states.  This will give rise   to a spectrum formed by equidistant quasi-degenerate pairs of standing-wave  solutions  with splitting $\Delta(E)$. In this setting,  the  backscattering probability $|r|^2$ over one roundtrip for a Gaussian wave-packet with average energy $E$ is connected to the splitting, $|r|^2\approx|\Delta(E)L/v|^2$. 

Most experiments so far have used sharp domain walls, where the mass has opposite sign in the two domains and the domain wall has a polygonal shape. In this setting  Eq.~(\ref{eq:JackiwandRebbi}) is valid only away from the polygon corners. In this case, one can still expect weak backscattering and, thus, a spectrum formed by equidistant quasidegenerate pairs if the Jackiw-Rebbi solutions for neighboring sides can be smoothly connected in the region around  the corners. 


The simplest  way to roughly estimate whether one should expect weak backscattering is to require that the Jackiw-Rebbi solution  for a straight domain wall is consistent with the smooth envelope assumption, i.e. it is smooth on the lattice scale.  In other words, the transverse localization length $\xi$ should be much larger than the lattice constant $a$, $\xi\ll 1/a$ and the longitudinal quasi-momentum $p_s$ much smaller than the inverse lattice constant $p_s\ll a^{-1}$. From Eq.~(\ref{eq:JackiewandRebbi}) one can calculate that for sharp domain walls $\boldsymbol{\Psi}_{p_s}\propto \exp [-m_{\rm bk}|r|/v]$ and, thus, $\xi=v/m_{\rm bk}$. This leads to the condition 
\begin{equation}\label{eq:cond_hard_bound}
m_{\rm bk}\ll v/a,  
\end{equation}
which also ensures that $p_s$ remains small for energies $E$ inside the bulk band gap, $-m_{\rm bk}<E<m_{\rm bk}$. Since the bulk band gap defines the bandwidth available for topological transport, the smooth-envelope condition Eq.~(\ref{eq:cond_hard_bound}) can be interpreted as imposing a fundamental limit on the bandwidth.
We note that the condition $\xi\gg a$
 ensures that the momentum spread $1/\xi$ of  the Fourier transform $\tilde{\boldsymbol{\Psi}}
(\mathbf{p})$ of the Jackiw-Rebbi solution Eq.~(\ref{eq:JackiewandRebbi}) is small. Even in this case, some residual backscattering will be observed, because  the tails of $\tilde{\boldsymbol{\Psi}}(\mathbf{p})$ penetrate the large quasi-momentum regions, inducing a coupling of the counter-propagating edge states. For sharp boundaries, the tails decay slowly, $\tilde{\boldsymbol{\Psi}}(\mathbf{p})\propto m/(vp_x)$.  This implies that to strongly  suppress the residual backscattering, very small values of $m_{\rm bk}$ will be required.

\begin{figure}
\begin{center}
\includegraphics[width=1\columnwidth]{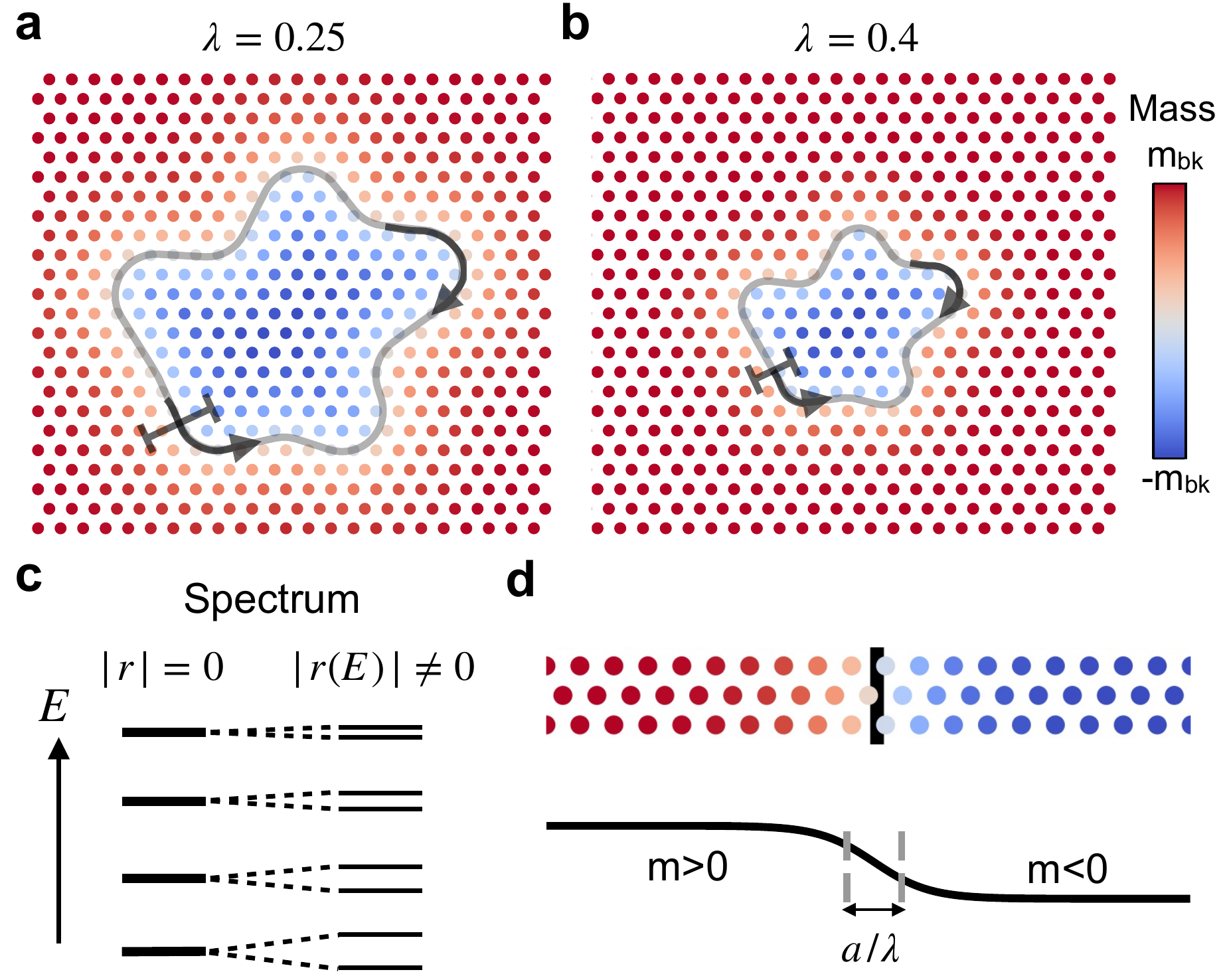}
\caption{\textbf{a, b}, Schematic of an arbitrarily shaped smooth closed domain wall on a triangular lattice, for two different values of the scale parameter $\lambda$. The topological edge states (indicated with grey arrows) travel along the domain wall in counter-propagating directions. The domain wall length scales as $\lambda^{-1}$.
\textbf{c}, Schematic of the standing wave spectrum in the presence (absence) of backscattering $\vert r \vert =0$ $(\vert r(E) \vert \neq 0)$. The degenerate doublets split for non-zero backscattering. 
\textbf{d}, Smooth domain wall transition (for $\lambda = 0.25$), extending over $a/\lambda$ sites.
}
\label{fig1}
\end{center}
\end{figure}

\section{Smooth domain walls and effective Planck Constant}

It has  been suggested and  demonstrated with numerical experiments that an effective strategy to reduce the backscattering without reducing the bulk mass $m_{\rm bk}$ (and, thus, the topological bandwidth) consists in implementing smooth domain walls \cite{brendel_snowflake_2018}.  Here, we formalize this intuition by introducing a WKB theory of backscattering for smooth domain walls. The first step is to introduce a quantity that will formally play the role of the Planck constant in quantum mechanics. This can be achieved by introducing a  rescaling of the position dependence of the mass term,  replacing $m(\mathbf{x})$ in Eq.~(\ref{eq:H_Dirac}) with $m(\lambda\mathbf{x})$. In this way, the domain wall defined by $m(\lambda\mathbf{x})=0$ maintains the original shape but its length is rescaled by a factor of $\lambda^{-1}$, cf Fig.1(a-b). We emphasize that this is not just a trivial rescaling because in the underlying microscopic model the lattice constant remains fixed. Thus, for decreasing $\lambda$ the domain wall becomes smoother and we expect reduced backscattering.
It is convenient to introduce the rescaled coordinate $\mathbf{Q}=\lambda\mathbf{x}$ and time  $\tau=\lambda t$. In terms of the rescaled variables,  the Dirac equation takes the form $i\lambda\dot{\boldsymbol{\Psi}}(\mathbf{Q},\tau)=\hat{H}_{D}\boldsymbol{\Psi}(\mathbf{Q},\tau)$ with $\hat{H}_D$ as in Eq.(\ref{eq:H_Dirac}) but now with the mass term $m(\mathbf{Q})\hat{\sigma}_z$ and $[\hat{Q}_l,\hat{p}_j]=i\lambda\delta_{lj}$. Thus, we can interpret $\lambda$ as an effective Planck constant. We note that the speed $v$ is not rescaled and that the rescaled domain wall length $L_{\rm rs}=\lambda L$ becomes independent of $\lambda$.    While our theory is general, for concreteness we will consider a scenario where the mass varies as a smooth step function in the direction $\mathbf{e}_r$ perpendicular to the domain wall tangent,
\begin{equation}
m(\mathbf{Q})=- m_{{\rm bk}}\tanh\left(\frac{ Q_r}{a}\right),\label{eq:sigmoid}
\end{equation}
where $Q_r$ is the rescaled  local coordinate, $Q_r=\lambda r$. We note that  the lattice constant has rescaled length $\lambda a$. Thus, $\lambda^{-1}$ is the typical number of unit cells  over which the mass is varied before reaching the asymptotic value $m_{\rm bk}$, cf Fig.~1(b). In this way, the  sharp domain walls scenario is included as the 'deep quantum' limit  $\lambda^{-1}\ll 1$ of our theory. 


\section{The Valley Hall effect on the honeycomb lattice}

Next, we move to the central focus of this work, i.e. to develope a description of the edge states that goes beyond the smooth-envelope approximation and allows to incorporate backscattering.  For this purpose we use as a case study the simplest and most well-known implementation of Valley Hall Physics, which is based on  the graphene tight-binding Hamiltonian.  In this model, the gap-opening interaction consists in a staggered onsite potential, assuming the values $m$ and $-m$ on the sublattices $A$ and $B$, respectively, cf Fig.~2(a). For simplicity, we consider only nearest-neighbor hopping transitions with rate $J$.

Our ultimate goal is to describe the tunneling between counter-propagating Jackiw-Rebbi solutions localized at different valleys.
Since these semi-classical solutions are localized in quasi-momentum space, it is convenient to adopt the quasi-momentun  representation
\begin{equation}
\tilde{\boldsymbol{\psi}}(\mathbf{k})=A^{-1/2}_{\rm BZ}\sum_{\{\mathbf{Q}\}}e^{-i\mathbf{k}\cdot\mathbf{Q}/\lambda}\boldsymbol{\psi}(\mathbf{Q}),
\end{equation}
where  $\boldsymbol{\psi}(\mathbf{Q})=(\psi_A(\mathbf{Q}),\psi_B(\mathbf{Q}))$ is the wavefunction in position space and $\{\mathbf{Q}\}$ indicates that the sum runs over all rescaled lattice vectors $\mathbf{Q}$. 
As ususal, the quasi-momentum  is defined modulus a reciprocal lattice vector and, thus, $\tilde{\boldsymbol{\psi}}(\mathbf{k})$  can be viewed as being defined on a torus of surface  area $A_{\rm BZ}=8\pi^2 3^{-3/2}/a^2$. 
Thus, the wave functions $\tilde{\boldsymbol{\psi}}(\mathbf{k})$ are periodic solutions of the Schr\"odinger Equation 
\begin{equation}
\hat{H}\tilde{\boldsymbol{\psi}}(\mathbf{k})=E\tilde{\boldsymbol{\psi}}(\mathbf{k}),\quad\hat{H}=m(\hat{\mathbf{Q}})\hat{\sigma}_{z}+\mathbf{h}(\hat{\mathbf{k}})\cdot\hat{\boldsymbol{\sigma}}
\label{eq:H_WKB}  
\end{equation}
with $\mathbf{h}=(h_x,h_y)$
\begin{eqnarray}
&&h_x=-J-2J\cos\left(\frac{\sqrt{3}k_xa}{2}\right)\cos\left(\frac{3k_ya}{2}\right),\nonumber\\
&&h_y=-2J\cos\left(\frac{\sqrt{3}k_xa}{2}\right)\sin\left(\frac{3k_ya}{2}\right).  \label{eq:h_of_k}
\end{eqnarray}
As usual, in the quasi-momentum representation the position operator $\hat{\mathbf{Q}}$ can be expressed in terms of the derivative of the quasi-momentum, $\hat{\boldsymbol{Q}}=i\lambda\nabla_\mathbf{k}$. We note that the Dirac Hamiltonian Eq.~(\ref{eq:H_Dirac}) for the smooth envelopes $\boldsymbol{\Psi}(\mathbf{Q})=\exp[-i\mathbf{K}\cdot\mathbf{Q}/\lambda]\boldsymbol{\psi}(\mathbf{Q})$ is obtained by expanding the tight-binding Hamiltonian Eq.~(\ref{eq:H_WKB}) about the high-symmetry point $\mathbf{K}=2\pi(-3^{-1/2},1)/(3a)$. In this setting, $v=3Ja/2$ and $\mathbf{p}=\mathbf{k}-\mathbf{K}$. 

Before considering an arbitrary domain wall  shape, we go back to the conceptually simpler special case of a straight  domain wall. Below,  we refer to such a translationally invariant configuration as a strip.

\begin{figure*}
\begin{center}
\includegraphics[width=2\columnwidth]{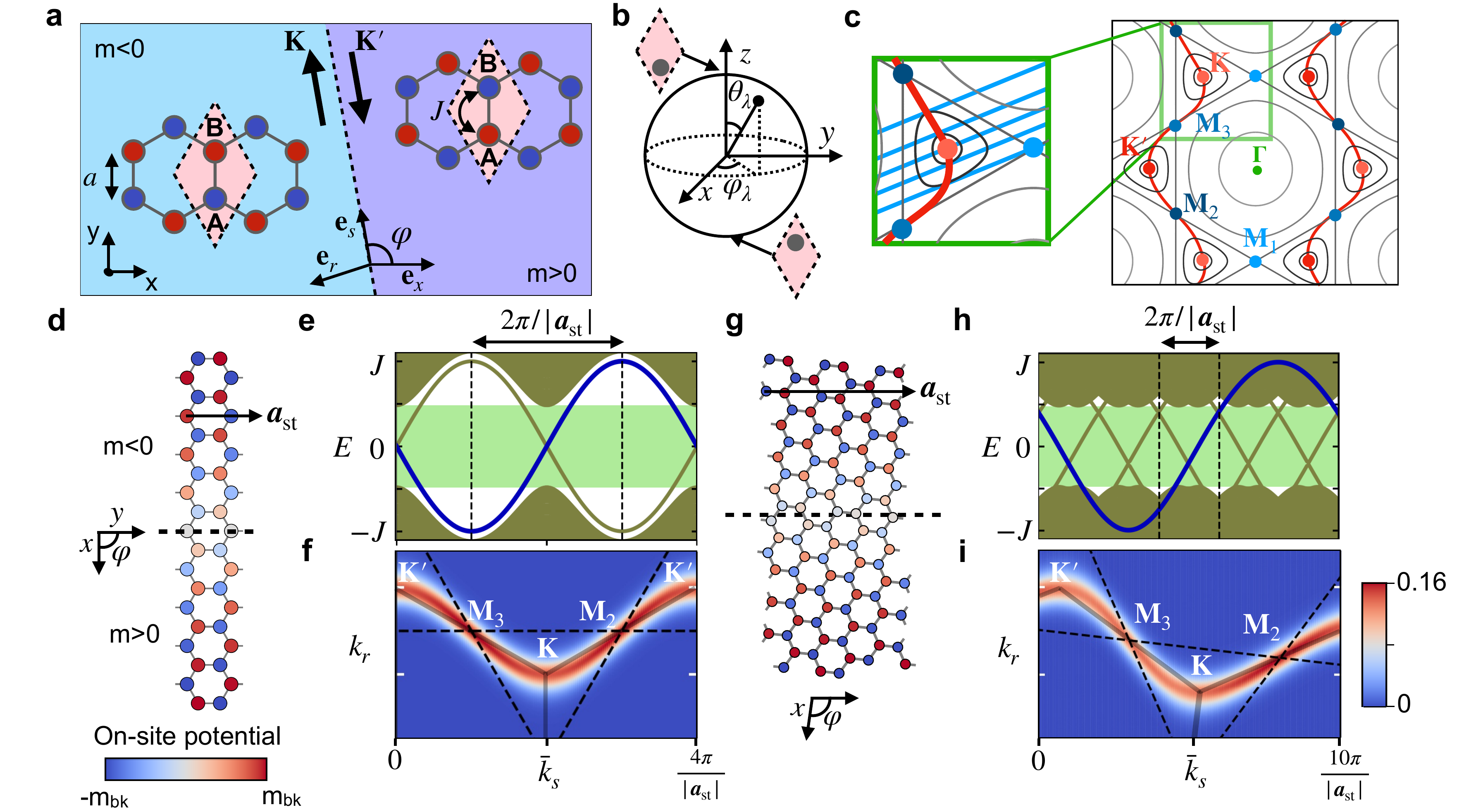}
\caption{\textbf{}
\textbf{a}, Sketch of the implementation of the Valley Hall effect on a honeycomb lattice. The color of the lattice sites represents the onsite potential, cf. colour bar in \textbf{d}. Each thick arrow shows the edge state propagation direction in  a valley, $\mathbf{K}$ or $\mathbf{K}'$.
\textbf{b}, Sketch of the Bloch sphere for the sublattice pseudospin. In the WKB approximation the angles $\theta_\lambda$ and $\varphi_\lambda$ are expanded in powers of the effective Planck constant $\lambda$. \textbf{c}, Sketch showing the closed path $\bar{k}(\bar{k}_s)$ formed by the 'classical' quasi-momenta for the topological edge state band (red line).  The path is formed by the quasi-momenta where the lines $\cos\varphi k_x+\sin\varphi k_y=\bar{k}_s$ (in blue in the zoom-in) are tangent to the contour  lines of  the graphene upper band bulk energy $|\mathbf{h}(\mathbf{k})|$ (in grey). The contour line $|\mathbf{h}(\mathbf{k})|=J$ is formed by two equilateral triangles. Each triangle can be viewed as defining a valley rim. Equivalent high-symmetry points are marked by dots of the same color. \textbf{d} and \textbf{g}, Onsite potentials for two straight-domain-wall configurations with the same effective Planck constant $\lambda=1/4$ but different domain wall orientations $\varphi$. For \textbf{d}, $\varphi=\pi/2$, corresponding to an armchair strip. \textbf{e} and \textbf{h} show the corresponding band structures for  $m_{\rm bk}=0.5J=v/(3a)$. The edge dispersion (blue line) is well approximated by the semi-classical solutions. Its period $T(\varphi)$   is $N$ times the width  $2\pi/|\boldsymbol{a}_{\rm st}|$ of the strip BZ, with $N=2$ and $N=5$, in (\textbf{e}) and (\textbf{h}), respectively. \textbf{f} and \textbf{i} show the underlying wavefunctions $|{\boldsymbol{\tilde\phi}}_{\bar{k}_s}(k_r)|^2$. The probability density is approximately a Gaussian, peaked around the classical quasi-momentum $\bar{k}$. We note that $\textbf{i}$ represents a zoom-in because  $\tilde{\boldsymbol{\phi}}_{\bar{k}_s}(k_r)$ is defined on a quasi-momentum loop that traverses multiple times both valleys, cf Fig.~3(d).
}
\label{fig2}
\end{center}
\end{figure*}

\section{semi-classical edge band}
\label{Sec:edge_band}

As a first step towards a full WKB calculation of the edge state spectrum  in the presence of a straight domain wall, we find a semi-classical solution that is no longer restricted to quasi-momenta in the vicinity of the high-symmetry points $\mathbf{K}$ and $\mathbf{K}'$, but  does not yet include tunneling. In other words, our solution extends  the Jackiw-Rebbi solution to the full Brillouin zone.

 We consider an arbitrary domain wall with mass $m(Q_r)$ depending  on the coordinate $Q_r=Q_y\cos\varphi -Q_x\sin\varphi$. For concreteness, we restrict our discussion to $\varphi$ in the interval $\pi/3 \leq \varphi < 2\pi/3$ throughout this Section. This does not imply any real loss of generality because the honeycomb lattice has six-fold rotational symmetry. We emphasize that   while the Hamiltonian does not depend on the coordinate $Q_s$, it is only translationally invariant if the domain wall orientation is  aligned to a lattice vector, for rational values of $\alpha\equiv\sqrt{3}\cot\varphi$, see Appendix \ref{APP:period}. Thus, for irrational $\alpha$ the edge states cannot be expressed as Bloch waves with a conserved quasi-momentum. This intricated angle-dependence of the discrete translational symmetry is well known in the  framework  of carbon nanotubes \cite{charlier_electronic_2007} and graphene nanoribbons \cite{akhmerov_boundary_2008}.  Delplace et al.~ \cite{delplace_zak_2011} have investigated the topological  states at the physical boundary of the latter graphene-based structures for arbitrary rational angles. Here, instead we investigate our  topological domain-wall states for arbitrary angles, including irrational angles. This simpler description is introduced  by viewing the edge-state energy as a function of its 'classical' quasi-momentum (defined below) instead of a conserved quasi-momentum.   

 We can reduce the problem of calculating the strip eigenstates  to a  1D problem using the ansatz
\begin{equation}
 \tilde{\boldsymbol{\psi}}_{\bar{k}_s}(\mathbf{k})=\delta(k_s-\bar{k}_s)\tilde{\boldsymbol{\phi}}_{\bar{k}_s}(k_r).\label{eq:Blochwave_gen}
\end{equation}
For an edge state solution, the transverse wave function $\tilde{\boldsymbol{\phi}}_{\bar{k}_s}(k_r)$ is  peaked about a radial quasi-momentum $\bar{k}_r$ but has a non-vanishing width. The 2D quasi-momentum $\bar{\mathbf{k}}_{\bar{k}_s}\equiv(\bar{k}_s,\bar{k}_r)$ can be viewed as the 'classical' quasi-momentum of the edge state solution.
We note that strictly speaking Eq.~(\ref{eq:Blochwave_gen}) is not yet a valid solution because it is not a periodic function of the quasi-momentum $\mathbf{k}$. However, one can  use it to build such  a periodic solution, see Appendix \ref{APP:Bloch_wave_ansatz} for a formal definition. Intuitively, it is enough to  view the wave function $\tilde{\boldsymbol{\psi}}_{\bar{k}_s}(\mathbf{k})$ as having support on the path $\mathbf{k}_{\bar {k}_s}(k_r)=(\bar{k}_s,k_r)$ defined on the torus and parametrized by the  radial quasi-momentum $k_r$.  When the quasi-momentum $\mathbf{k}$ is taken within the first Brillouin zone instead of on a single line, such a path traverses the BZ multiple times, defining several parallel lines (for irrational $\alpha$ infinitely many of them) --  more on this below.

The Jackiw-Rebbi solution Eq.~(\ref{eq:JackiwandRebbi}) is   localized  about a high-symmetry point, $\mathbf{K}$ or $\mathbf{K}'$.  These points are the two  global minima of $|\mathbf{h}(\mathbf{k})|$, the energy of the upper bulk band for the massless case. In our  generalized solution, each edge state wavefunction $\tilde{\boldsymbol{\psi}}_{\bar{k}_s}(\mathbf{k})$  is localized about a  quasi-momentum $\bar{\mathbf{k}}_{\bar{k}_s}=(\bar{k}_{s},\bar{k}_r)$  whose radial component   $\bar{k}_{r}$  is a local minimum of $|\mathbf{h}(\mathbf{k})|$   for fixed $k_s$, $k_s=\bar{k}_s$. In general, $|\mathbf{h}(\mathbf{k})|$ has more than one local minimum for fixed $k_s$ (the number depends on $\varphi$). However, one can  follow the same local minimum as a function of $\bar{k}_s$ to define a  continuous path
 $\bar{\mathbf{k}}_{\bar{k}_s}=(\bar{k}_s,\bar{k}_r)$ in the BZ. By inspecting  the contour plot  of $|\mathbf{h}(\mathbf{k})|$ one can easily verify that the path $\bar{\mathbf{k}}_{\bar{k}_s}$ is  unique (apart from a trivial reparametrization). For $\pi/3<\varphi<2\pi/3$ it is also closed, as it connects the high-symmetry points $\mathbf{K}$ and $\mathbf{K}'$ via the $\mathbf{M}_2$ and the $\mathbf{M}_3$ points, cf Fig.~2(c). For the critical angle $\varphi=\pi/3$ (corresponding to  a so-called 'zig-zag' orientation), $\bar{\mathbf{k}}_{\bar{k}_s}$ reaches asymptotically the two mid-points between the $\mathbf{M}_1$ and $\mathbf{M}_2$ points. (For even smaller angles it passes through $\mathbf{M}_1$ instead of $\mathbf{M}_2$, see Appendix \ref{APP:armchair-zigzag-anal} and Appendix \ref{Appendix:fourier_transform_different_orientations}).
 
 The wave functions $\tilde{\boldsymbol{\psi}}_{\bar{k}_s}(\mathbf{k})$   are obtained by plugging the ansatz Eq.~(\ref{eq:Blochwave_gen}) into Eq.~(\ref{eq:H_WKB}) while also expanding $\mathbf{h}(\mathbf{k})$ up to linear order about $\bar{\mathbf{k}}_{\bar{k}_s}$, see Appendix \ref{app:edge_state_calculation}. We find that the wave function $\tilde{\boldsymbol{\phi}}_{\bar{k}_s}(k_r)$ has a Gaussian profile, but otherwise has  the same pseudospin and energy as one of the two  bulk solutions for mass $m=0$ and with quasi-momentum $\mathbf{k}$ equal to the classical quasi-momentum $\bar{\mathbf{k}}_{\bar{k}_s}$,
\begin{equation}
 E_{\bar{k}_s}=g(\bar{\mathbf{k}}_{\bar{k}_s})|h(\bar{\mathbf{k}}_{\bar{k}_s})|,\quad g(\mathbf{k})={\rm sign}[(\partial_{k_r}\mathbf{h}\wedge \mathbf{e}_z)\cdot\mathbf{h}].\label{eq:En_WKB}
\end{equation}
The edge spectrum $E_{\bar{k}_s}$ defines a periodic band,   $E_{\bar{k}_s+T}=E_{\bar{k}_s}$ with period $T=4\pi/(3a)\sin\varphi$, cf Fig.~2(a).  For the special case $\varphi=\pi/2$, corresponding to  an armchair strip, the energy dispersion has a simple closed form  $E_{\bar{k}_s}=J\sin\left(3\bar{k}_s a/2\right)$ ($\mathbf{e}_s=\mathbf{e}_y$ in this case), see Fig.~2(d-e) and Appendix \ref{APP:armchair-zigzag-anal}. For a generic angle $\varphi$, the precise energy dispersion $E_{\bar{k}_s}$   has to be evaluated numerically, solving an algebraic equation for ${\bar{k}_r}$, see Appendix \ref{APP:num_calc_classical_quasimom}. 
However,  its qualitative shape is robust. It is positive (negative) in the segment that connects the $\mathbf{K}$ and $\mathbf{K}'$ points via the $\mathbf{M}_2$ ($\mathbf{M}_3$) point. In addition, the speed only vanishes at the $\mathbf{M}_3$ and $\mathbf{M}_2$ points where the energy $E_{\bar{k}_s}$ assumes its  maximum and minimum values  $E=\pm J$, respectively (for details see Appendix \ref{app:edge_state_calculation}). Thus, these points divide $E_{\bar{k}_s}$ into  two counter-propagating branches that are localized in different valleys, Fig.~2(e-f,h-i).   In other words,  for any energy $E$ in the interval $-J<E<J$ there are exactly two counter-propagating edge state solutions (one in each valley) mapped onto each other via the time-reversal symmetry. We remark  that, in contrast to the Jackiw-Rebbi solution, here, the pseudospin  depends on the classical quasi-momentum $\bar{\mathbf{k}}_{\bar{k}_s}$ and completes a full revolution of the Bloch sphere equator over the period $T(\varphi)$. 

Next we focus on domain-wall orientations for which  the domain wall is aligned to a lattice vector, giving rise to a translationally  invariant Hamiltonian.  This scenario  is realized for rational values of $\alpha=\sqrt{3}\cot\varphi$.
In this case, the Hamiltonian is diagonalized by Bloch waves whose quasi-momentum $k$ can be chosen in the the interval $-\pi/|\boldsymbol{a}_{\rm st}|<k\leq \pi/|\boldsymbol{a}_{\rm st}|$ where  $\boldsymbol{a}_{\rm st}$ is the strip unit vector.  The strip unit vector is a discontinuous function of $\varphi$, $|\boldsymbol{a}_{\rm st}(\varphi)|=3qa/\{[1+(pq\!\!\mod 2)]\sin\varphi\}$ where  $p$ and $q$ are relatively prime integers defined by $\alpha\equiv p/q$, see Appendix \ref{APP:period}. This  angle dependence is also relevant for the  intricate band structure of  carbon nanotubes, formed by rolling up graphene \cite{charlier_electronic_2007}.  

Our semi-classical edge state solutions $\tilde{\boldsymbol{\psi}}_{\bar{k}_s}(\mathbf{k})$ are Bloch waves with strip quasi-momentum $k=\bar{k}_s\!\!\mod (2\pi/|\boldsymbol{a}_{\rm st}|)$. 
This enables us to compare our semi-classical calculations with exact numerical results (Appendix \ref{Appendix:tight_binding_setup_strip}). In Fig.2 only the numerical results are shown because the corresponding analytical results would not be distinguishable with the bare eyes. This indicates that  tunneling is  strongly suppressed for the parameters  considered ($\lambda=1/4$ and $m_{\rm bk}=0.5J$).
We note  that the period $T(\varphi)=4\pi/(3a)\sin\varphi$ of the semi-classical edge band is an integer multiple of the width  $2\pi/|\boldsymbol{a}_{\rm st}|$ of the strip BZ, $T(\varphi)|\boldsymbol{a}_{\rm st}|/2\pi=2q/[1+(pq\!\!\mod 2)]\equiv N(\varphi)$. 
Thus, when plotted inside the strip BZ, the semi-classical edge band $E_{\bar{k}_s}$ is folded into $N$ bands $E_{n,k}$, $n=0,\ldots,N-1$ (cf Fig.~2(e) and (h) where $N=2$ and $5$, respectively). Just like the strip unit vector, also the number of edge bands $N$ is a discontinuous function of $\varphi$.

\begin{figure}
\begin{center}
\includegraphics[width=1\columnwidth]{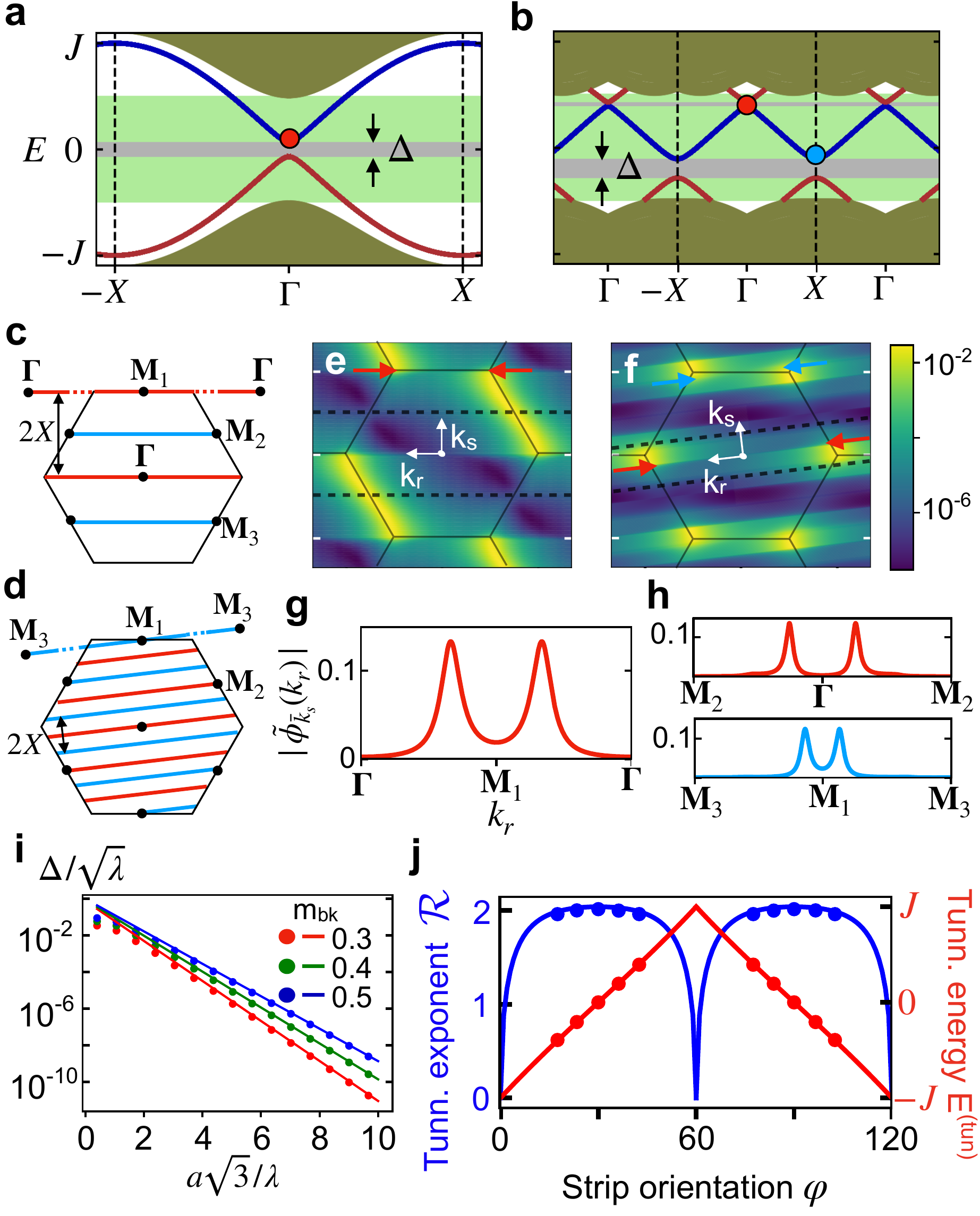}
\caption{\textbf{}
\textbf{a} and \textbf{b}, Band structures for  the same domain wall orientations  as in Fig.~2(d) and (g), respectively. The mass  parameter $m_{\rm bk}$  is also the same as in Fig.~2(e-h) but, here,  $\lambda=2.5$ (ten times large), corresponding to a sharper interface. We note that the edge band structure is now visibly gapped. Similar band gaps are present also in Fig.~2(d) and (g) but are not visible with the bare eye.  \textbf{c} and \textbf{d},
Quasi-momentum loops $\mathbf{k}_{\bar{k}_s}(\cdot)$ for the time-symmetric strip quasi-momenta $\Gamma$ and $X$ ($k=\bar{k}_s\!\!\mod(2\pi/\boldsymbol{a}_{\rm st})=0,\pi/\boldsymbol{a}_{\rm st}$) red and blue lines, respectively.
\textbf{e} and \textbf{f}, Bloch waves of the band highlighted in blue in (a) and (b), respectively. The color that can be inspected by following a quasi-momentum loop encodes the probability density of the corresponding Bloch wave. \textbf{g} and \textbf{h}, Resonant tunneling Bloch waves. The corresponding energies are marked by the dots of the same color in (a) and (b), respectively. \textbf{i}, The edge band gap decays exponentially for increasing smoothness of the domain wall (comparing numerics vs analytical result for an armchair strip). \textbf{j}, Tunneling exponent ${\cal R}$ and tunneling energy $E^{(\rm tun)}$ for the dominant tunneling pathway.
}
\label{fig3}
\end{center}
\end{figure}

\section{tunneling-induced gaps in the  edge band structure}

The folded semi-classical edge band can be viewed as a gapless band structure. Similar to the edge band structure of a time-symmetric topological insulator, subsequent edge bands cross at a time-reversal symmetric strip quasi-momentum $\Gamma$ or $X$, corresponding to $k=0$ and $k=\pi/|\boldsymbol{a}_{\rm st}|$. Once tunneling is taken into account, such crossings turn into avoided crossing.   Interestingly, the number of edge-band gaps  is a discontinuous function of the domain wall orientation  $\varphi$. In other words,  a tiny variation of the domain wall orientation  can change substantially the number of band gaps. This physics is reminiscent of the (bulk) Hofstadter butterfly spectrum of lattice electrons in a magnetic field \cite{hofstadter_energy_1976}, with $\varphi$  playing the role of the magnetic field flux. We will show that the edge-band gaps are induced by tunneling transitions in quasi-momentum space and, thus, decay exponentially with the inverse effective Planck constant $\lambda^{-1}$. Importantly, the different band gaps are of very different magnitudes.  Identifying the underlying tunneling pathways in the quasi-momentum space allows us to calculate the edge-band gaps (up to logarithmic precision) and to identify a dominant tunneling pathway, that will also play an important role for the backscattering in  closed domain walls.

In the absence of tunneling, subsequent edge bands $E_{n,k}$ and $E_{n+1,k}$ touch whenever two edge states  with the same energy $E_{\bar{k}_s}$  also have the same strip quasi-momentum $k=\bar{k}_s\!\!\mod{2\pi/|\boldsymbol{a}_{\rm st}|}$. Since the semi-classical edge states with equal energy are also time-reversed partners, the band crossings occur only at the time-reversal-invariant  quasi-momenta $\Gamma$ and  $X$.  At a crossing,  two counter-propagating solutions are resonantly coupled via tunneling.  Hence, once tunneling is taken into account, the exact crossings turn into avoided crossings, leading  to  the opening of  small edge-band gaps, cf Fig.3(a-b). We note that (when neglecting tunneling)  there are $N-1$ crossings. However,  a crossing does not necessarily give rise to a band gap because of the spectral overlap of the edge- and the bulk bands for $J>m_{\rm bk}$, cf   Fig.2(e) where $N=5$ but only two edge-band gaps are present.

At the $\Gamma$ and $X$ points the Bloch waves can always be chosen to be time-reversal invariant (because ${\cal T}^2=1$). In the special case of an avoided crossing (between the $n$-th and $n+1$-th band), this implies that the Bloch waves are equal superpositions of two time-reversal-partner semiclassical solutions, cf Fig~3(g,h). This leads to the tunneling band structure (see Appendix \ref{eq:APP_avoided_crossings})
\begin{equation}
E_{n+1/2\pm 1/2,k+\delta k}=E_{\bar{k}_s}\pm\sqrt{(\Delta/2)^2+(v_{\bar{k}_s}\delta k)^2}   
\end{equation}
where $v_{\bar{k}_s}$ is the group velocity $v_{\bar{k}_s}=dE_{\bar{k}_s}/d\bar{k}_s$, $k=0$ or $\pi/|\boldsymbol{a}_{\rm st}|$, $\delta k$ is the distance from the relevant time-symmetric quasi-momentum and $\Delta$ is the tunneling rate that we set out to calculate.

\subsection{WKB formula for the Tunneling exponent}

Two semi-classical edge state solutions localized near distant classical quasi-momenta are coupled via their tails. The solutions calculated so far  by expanding the Hamiltonian Eq.~(\ref{eq:H_WKB}) about the relevant classical quasi-momenta $\mathbf{k}_{\bar{k}_s}$ are accurate only near these quasi-momenta. Thus, an important preliminary step towards calculating the tunneling rate $\Delta$ consists in generalizing our solution to correctly evaluate the tails. Such a solution can be found using the WKB ansatz
\begin{equation}\label{eq:WKB_ansatz_k}
\tilde{\boldsymbol{\phi}}_{\bar{k}_s}(k_r)=C'\begin{pmatrix}e^{-i\varphi_{\lambda,\bar{k}_s}(k_r)/2}\\
e^{i\varphi_{\lambda,\bar{k}_s}(k_r)/2}
\end{pmatrix}\exp[-\frac{i}{\lambda}S_{\lambda,\bar{k}_s}(k_r)].
\end{equation}
Here, $\varphi_{\lambda(k_r),\bar{k}_s}$ is the azimuthal Bloch sphere angle, see Fig.~2(b), and $S_{\lambda,\bar{k}_s}$ is the action. We expand  these functions in powers of $\lambda$, 
\begin{equation}
 \varphi_{\lambda,\bar{k}_s}=\sum_{n=0}^\infty \varphi_{n,\bar{k}_s}\lambda^n,\quad S_{\lambda,\bar{k}_s}=\sum_{n=0}^\infty S_{n,\bar{k}_s}\lambda^n.
\end{equation}
This expansion effectively  divides the Schrödinger equation into an infinite series of equations obtained by grouping the terms  with the same power-law dependence on $\lambda$. This allows to calculate $S_\lambda$, and $\varphi_\lambda$  recursively, starting from the leading order which in the standard setting describes the classical limit.

For the purpose  of estimating the tunneling rate, it is  sufficient to calculate the leading order $S_{0,\bar{k}_s}$ of the action. This allows to calculate the exponent  ${\cal R}=-\lim_{\lambda\to0}\lambda\ln\Delta$ of the tunneling rate $\Delta$ which in turn determines the order of magnitude of the tunneling rate, see below. We note in passing that to calculate also the prefactor is considerably more elaborate and usually  requires to take advantage of additional symmetries \cite{landau_quantum_1981,marthaler_quantum_2007,zhang_time-translation-symmetry_2017}. In Appendix \ref{App:tunneling_armchair}, we perform such a calculation  for an armchair strip, adapting to our problem a  trick invented by Landau.
For arbitrary domain wall angles,  we can show (see Appendix \ref{Appendix:details_WKB_wavefunction}) that the leading order of the action $S_{0,\bar{k}_s}(k_r)$ is equal to the action for a classical 1D problem with the effective  Hamiltonian
\begin{equation}\label{eq:effective_class_Ham}
{\cal H}_{\bar{k}_s}(Q_r,k_r)= g(\bar{\mathbf{k}}_{\bar{k}_s})\sqrt{m(Q_r)^2+|\mathbf{h}(\bar{k}_s,k_r)|^2}.
\end{equation}
In practice, we find
\begin{equation}\label{eq:action_k_space}
 S_{0,\bar{k}_s}(k_r)=\int_{\bar{k}_r}^{k_r}Q_{\bar{k}_s}(k'_r)dk'_r
\end{equation}
where the (imaginary) position $Q_{\bar{k}_s}(k_r)$ is calculated by solving
\begin{equation}\label{eq:Hamilton-Jacobi}
 {\cal H}(Q_{\bar{k}_s},\bar{k}_s,k_r)=E_{\bar{k}_s}.  
\end{equation}
For the mass-dependence $m(Q_r)$ in Eq.~(\ref{eq:sigmoid}) we find
\begin{equation}\label{eq:Q_of_k_tail}
Q_{\bar{k}_s} =-ia\arctan\sqrt{\left[
    |\mathbf{h}(\bar{k}_s,k_r)|^2-E_{\bar{k}_s}^2\right]/m^2_{\rm bk}}.
\end{equation}
Taking into account that Eq.~(\ref{eq:action_k_space}) is equivalent to $\partial_{k_r}S_{0,\bar{k}_s}(k_r)=Q_{\bar{k}_s}$, we are calculating the action solving an equation analogous to the  Hamilton-Jacobi equation but, here, exchanging the role of position and (quasi-)momentum. From Eqs.~(\ref{eq:WKB_ansatz_k}),(\ref{eq:action_k_space}), and (\ref{eq:Q_of_k_tail}) we gain the powerful insight that the massless bulk band structure  $|\mathbf{h}(\bar{k}_s,k_r)|$ can be interpreted as a (dimensionless) potential barrier seen by the  edge excitation while tunneling in quasi-momentum space (with the bulk mass parameter $m_{\rm bk}$ playing the role of a rescaling of such barrier). This intuition can be transferred also to the more complex scenario in which the domain wall is not straight and the tunneling induces backscattering between two counter-propagating edge states. More on this below.

Taking into account that in the WKB approximation  the tunneling rate has the same exponent ${\cal R}$ as the  overlap of the two tunneling wavefunctions on the tunneling pathway, we arrive at the formula
\begin{equation}\label{eq:tunneling_exponent}
{\cal R}= -i \int_{\gamma^{(\rm tun)}} Q_{\bar{k}^{(\rm tun)}_s}(k_r)dk_r,
\end{equation}
where $Q_{\bar{k}^{(\rm tun)}_s}(k_r)$ is evaluated using Eq.~(\ref{eq:Q_of_k_tail}) along the tunneling path ${\gamma^{(\rm tun)}}$, connecting the semi-classical quasi-momenta $\pm\bar{\mathbf{k}}_{\bar{k}_s}$  of two semi-classical time-reversal-partner solutions.  Formula (\ref{eq:tunneling_exponent}) reduces the problem of   finding the  edge-band gaps and estimating their magnitude to the problem of identifying the corresponding  tunneling pathways ${\gamma^{(\rm tun)}}$, discussed in the next section.

\subsection{Tunneling pathways}

For rational values of $\alpha$, corresponding to translationally invariant domain wall  configurations,  the number  of  band gaps  is finite, but it is a discontinuous function of  the domain wall angle $\varphi$, see discussion above.  Because of this  intricate behavior, the task of systematically investigating  the band gaps for arbitrary $\varphi$ looks daunting. Below we show that this endeavour turns out to be surprisingly  simple if one switches the focus to the available  pathways for resonant tunneling and considers a generic irrational $\alpha$.

The first step towards classifying the available tunneling paths is to gain insight about the 2D quasi-momentum  paths $\mathbf{k}(\cdot)_{\bar{k}_s}$ on which a 1D semi-classical solution $\tilde{\boldsymbol{\psi}}_{\bar{k}_s}(\mathbf{k})$  obtained using the ansatz Eq.~(\ref{eq:Blochwave_gen}) has non-zero probability density.
The path $\mathbf{k}(\cdot)_{\bar{k}_s}$ with classical longitudinal quasi-momentum $\bar{k}_s$  is equivalent to the straight line $k_x\cos\varphi+k_y\sin\varphi=\bar{k}_s$.
 Since the quasi-momentum    $\mathbf{k}$ is defined up to a reciprocal lattice vector, we can view   $\mathbf{k}(\cdot)_{\bar{k}_s}$  as continuing as a parallel line  inside the first BZ after crossing the BZ hexagonal perimeter, cf Fig.3(c-d). For rational $\alpha$,  the path $\mathbf{k}(\cdot)_{\bar{k}_s}$  is a closed loop, crossing $N$ times ($N=2q/[1+(pq\!\!\mod 2)]$) the BZ perimeter before returning to the initial quasi-momentum, cf Fig.3(c-d). Its length $T_r(\varphi)$ is set by the period of $\mathbf{h}(\bar{k}_s,k_r)$ (as a function of $k_r$), $T_r(\varphi)=A_{\rm BZ}|\boldsymbol{a}_{\rm st}|/2\pi=4\pi q/\{\sqrt{3}a[1+(pq\!\!\mod 2)]\sin\varphi\}$. Importantly, all semi-classical quasi-momenta $\bar{k}_s$ corresponding to the same strip quasi-momentum $k=\bar{k}_s\mod 2\pi/|\boldsymbol{a}_{\rm st}|$ give rise to the same path (up to a reparametrization).  Since  the paths for different $k$ do not overlap, it is possible  to represent the Bloch waves for a full band as a single density plot, cf Fig.~3(e,f). For irrational $\alpha$, $\mathbf{h}(k_s,k_r)$ is not a periodic function of $k_r$. In this scenario, the path $\mathbf{k}(\cdot)_{\bar{k}_s}$  is infinitely long, crossing the perimeter of the BZ infinitely many times.  We expect that along the way it will come arbitrarily close to any point in the BZ. Moreover, the classical quasi-momenta $\bar{k}_s$ giving rise to the same path form a countably infinite set [with one element for  each local minimum of $\mathbf{h}(\bar{k}_s,k_r)$].

As discussed above for rational $\alpha$, a precondition  for resonant tunneling is that the strip quasi-momentum $k$ is time-reversal invariant, $k=\bar{k}_s\mod 2\pi/|\boldsymbol{a}_{\rm st}|=0$ or $\pi/|\boldsymbol{a}_{\rm st}|$. This precondition can be generalized to irrational $\alpha$ as a precondition for   the corresponding  path $\mathbf{k}(\cdot)_{\bar{k}_s}$: This path should be time-reversal invariant (up to a reparametrization). Only in this case, it can and will pass  through both classical quasi-momenta $\pm\bar{\mathbf{k}}_{\bar{k}_s}$ of two time-reversal-partner semi-classical solutions. It turns out that this condition is fulfilled if and only if  $\mathbf{k}(\cdot)_{\bar{k}_s}$ passes through a time-symmetric high-symmetry point $\mathbf{k}^{(\rm T)}$,
$\mathbf{k}^{(\rm T)}=\boldsymbol{\Gamma}$, $\mathbf{M}_1$, $\mathbf{M}_2$, or $\mathbf{M}_3$, see Appendix \ref{APP:details_tunn_path}. Thus,   this precondition identifies four  distinct paths.
Each of these four paths  traverses infinitely many times each of the two triangle-shaped valleys (cf caption of Fig.~2) and,  at each passage, passes through a different local minimum of $\mathbf{h}(\bar{k}_s,k_r)$ which in turn corresponds to a valid semi-classical solution, see discussion in Section \ref{Sec:edge_band}. For each such semi-classical solution there will be also a corresponding tunneling pathway  connecting it to the time-reversed quasi-momentum via $\mathbf{k}^{(\rm T)}$. Thus, we can classify all possible tunneling pathways based on the time-reversal-symmetric high-symmetry point $\mathbf{k}^{(\rm T)}$ they go through and the number of times $j$ they traverse each valley, $\gamma^{(\rm tun) }_{\mathbf{k}^{(\rm T)},j}$ with $j\in\mathbb{N}$. For rational $\alpha$, the path $\mathbf{k}(\cdot)_{\bar{k}_s}$ traverses only a finite number of times each valley, setting a limit on the maximum $j$. In addition, since  $\mathbf{k}(\cdot)_{\bar{k}_s}$ is a closed loop  passing through two time-symmetric high-symmetry points, two tunneling pathways connect the same pair of time-reversal-partner solutions, cf Fig.3~(c,d,g,h). In this case the tunneling occurs via the pathway with smaller tunneling exponent, cf Eq.~(\ref{eq:tunneling_exponent}). 

Our classification of the tunneling pathways allows us to easily calculate the corresponding tunneling exponents ${\cal R}_{\mathbf{k}^{({\rm T})},j}$. This only requires to solve a simple algebraic equation to calculate the classical quasi-momentum $\bar{\mathbf{k}}^{(\rm tun)}_{\mathbf{k}^{({\rm T})},j}$ as a function of $\varphi$, plug it in Eq.(\ref{eq:En_WKB}) to obtain the tunneling energy $E^{(\rm tun)}_{\mathbf{k}^{({\rm T})},j}$, and evaluate the integral in Eq.~(\ref{eq:tunneling_exponent}), see Appendix \ref{APP:details_tunn_path} for more details. We note that the exponent ${\cal R}_{\mathbf{k}^{({\rm T})},j}$ will be smaller for smaller $j$, corresponding to shorter tunneling paths $\gamma^{(\rm tun) }_{\mathbf{k}^{(\rm T)},j}$. Out of the four shortest tunneling paths (with $j=1$) only  $\gamma^{(\rm tun) }_{\mathbf{M}_1,1}$  directly connects the two  valleys  without entering the region outside the triangular valley rims (where the tunneling barrier  is larger, $|\mathbf{h}(\mathbf{k})|>J$).  Thus,  one should expect ${\cal R}_{\mathbf{M}_1,1}$  to be the smaller exponent, which is confirmed  by numerical calculations. We emphasize that different exponents lead to tunneling rates $\Delta_{\mathbf{k}^{({\rm T})},j}$ that can differ by orders of magnitudes in the semi-classical regime $\lambda\ll1$. Even for $\lambda=2.5$, our exact numerical simulations show  that $\Delta_{\mathbf{M}_1,1}\gg \Delta_{\boldsymbol{\Gamma},1}$, see Fig.~3(b) where $\Delta_{\mathbf{M}_1,1}$ and $\Delta_{\boldsymbol{\Gamma},1}$ correspond, respectively, to the lower and upper edge-band gaps (marked in grey).  The angle-dependence of the exponent ${\cal R}_{\mathbf{M}_1,1}$ and the tunneling energy $E^{(\rm tun)}_{\mathbf{M}_1,1}$ for $m_{\rm bk}=0.5J$ are shown in Fig.~3(j).  In the next Section, we show that the dominant exponent is able to capture the magnitude of the backscattering in a setup featuring an arbitrary smooth closed domain wall.


\section{Transport in closed domain walls}

\begin{figure*}
\begin{center}
\includegraphics[width=2\columnwidth]{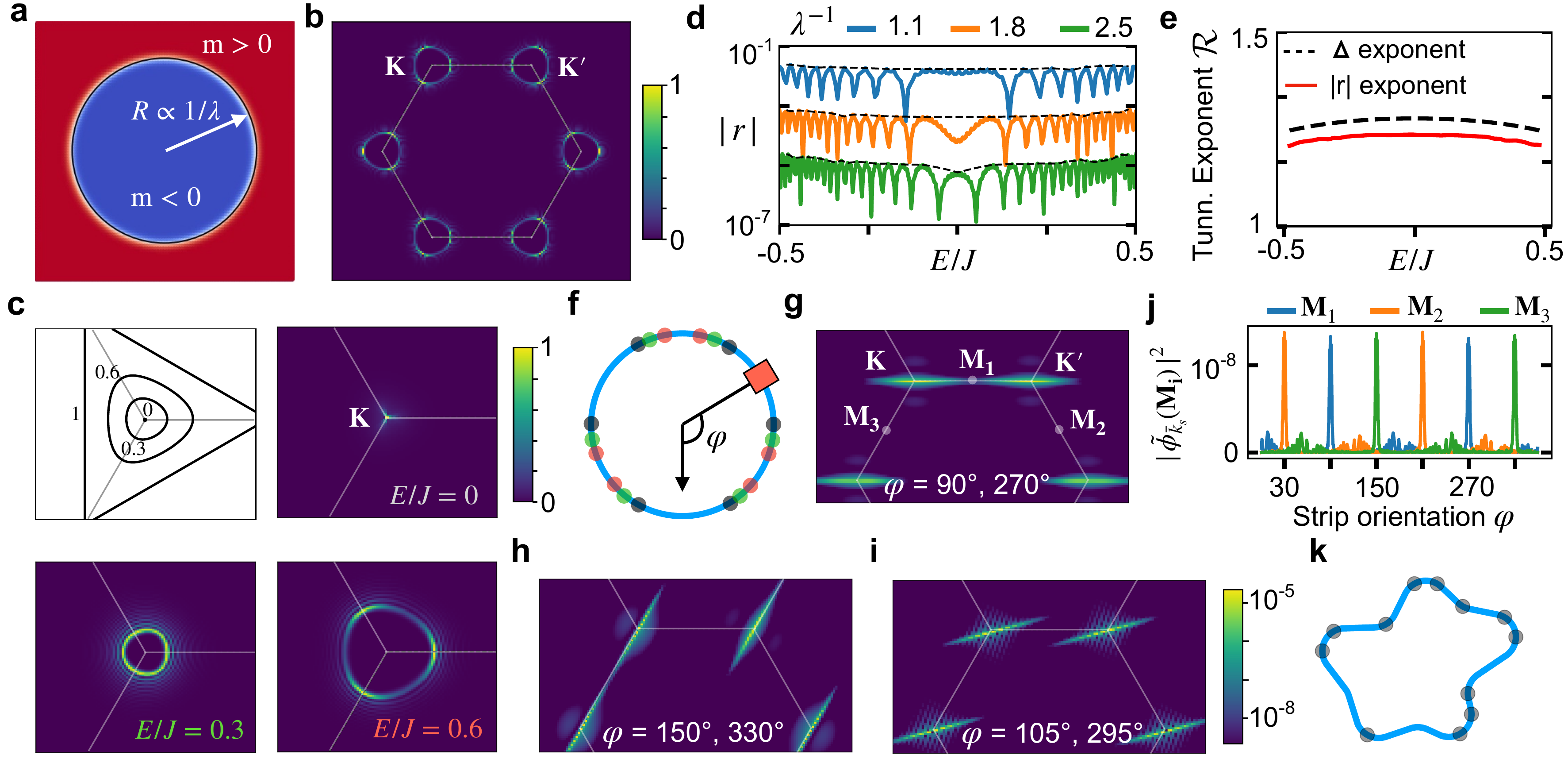}
\caption{Backscattering in arbitrarily shaped smooth domain walls.
\textbf{a}, Setup of the circular domain wall interface. Mass parameter $m_{bk}=J$, Radius $R=R_{\rm rs}/\lambda$.
\textbf{b}, Fourier transform $\vert \tilde{\phi}(E)\vert ^2$ (maximum value normalized to 1) of the standing wave eigenmode at energy $E=0.6J$. The standing wave is formed by a superoposition of two counter-propagating edge states that are localized near to the $\mathbf{K}$ and $\mathbf{K'}$ valleys.
\textbf{c}, Fourier transform of the eigenmodes for three different energies in the vicinity of the $\mathbf{K}$ point. The densities are localised on the contour lines of the graphene upper band bulk energy $\vert \mathbf{h}(\mathbf{k}) \vert $, shown in the first panel. \textbf{d}, Reflection coefficient $\vert r(E) \vert $ for the three equidistant values of $\lambda^{-1}$. The envelope of the maxima are equidistant on the log scale of the y-axis, demonstrating the exponential decay of $\vert r \vert$ with $\lambda ^{-1}$. The lineshape features a complicated interference pattern, that arises because of the changing scatterer locations with energy.
\textbf{e}, Comparison of the backscattering and tunneling exponent as a function of energy. The two exponents should converge to be identical for larger rescaled Radius  ($ R_{\rm rs}/a \rightarrow \infty$), see Appendix \ref{app:WKB_closed}. 
\textbf{f}, Position of effective energy-dependent scatterers on the circular domain wall, for the three energies in \textbf{c}. A local Fourier transform of the eigenmode at the different positions on the topological cavity, indicated by an orange box at angle $\varphi$, is taken to investigate the backscattering in the reciprocal space. Note that $\varphi$ is also the domain wall orientation. 
\textbf{g,h,i}, Local Fourier transforms of wave function at the different positions along the domain wall, for the zero energy eigenmode. At the scatterer locations i.e. at $\varphi=90^\circ (150^\circ)$, the tunneling between the two counter-propagating edge states is through the time-symmetric $\mathbf{M_1} (\mathbf{M_3})$ point. On the other hand, tunneling is negligible at $\varphi=105^\circ$. 
\textbf{j}, Fourier transform of the eigenmode at the three $\mathbf{M}$ points as a function of position along the domain wall. As expected, the probability density shoots up at the scatterer locations. 
\textbf{k}, Position of effective scatterers for an arbitrary shape of the smooth domain wall (here: at zero energy; locations are energy-dependent). [The Fourier transform plots in \textbf{b,c,g--j}  corresponds to $\lambda^{-1}=1.1$ and $R_{\rm rs}/a=1500$, while in panel \textbf{d,e} $R_{\rm rs}/a=50$.]
}
\label{fig4}
\end{center}
\end{figure*}

\begin{figure}
\begin{center}
\includegraphics[width=1\columnwidth]{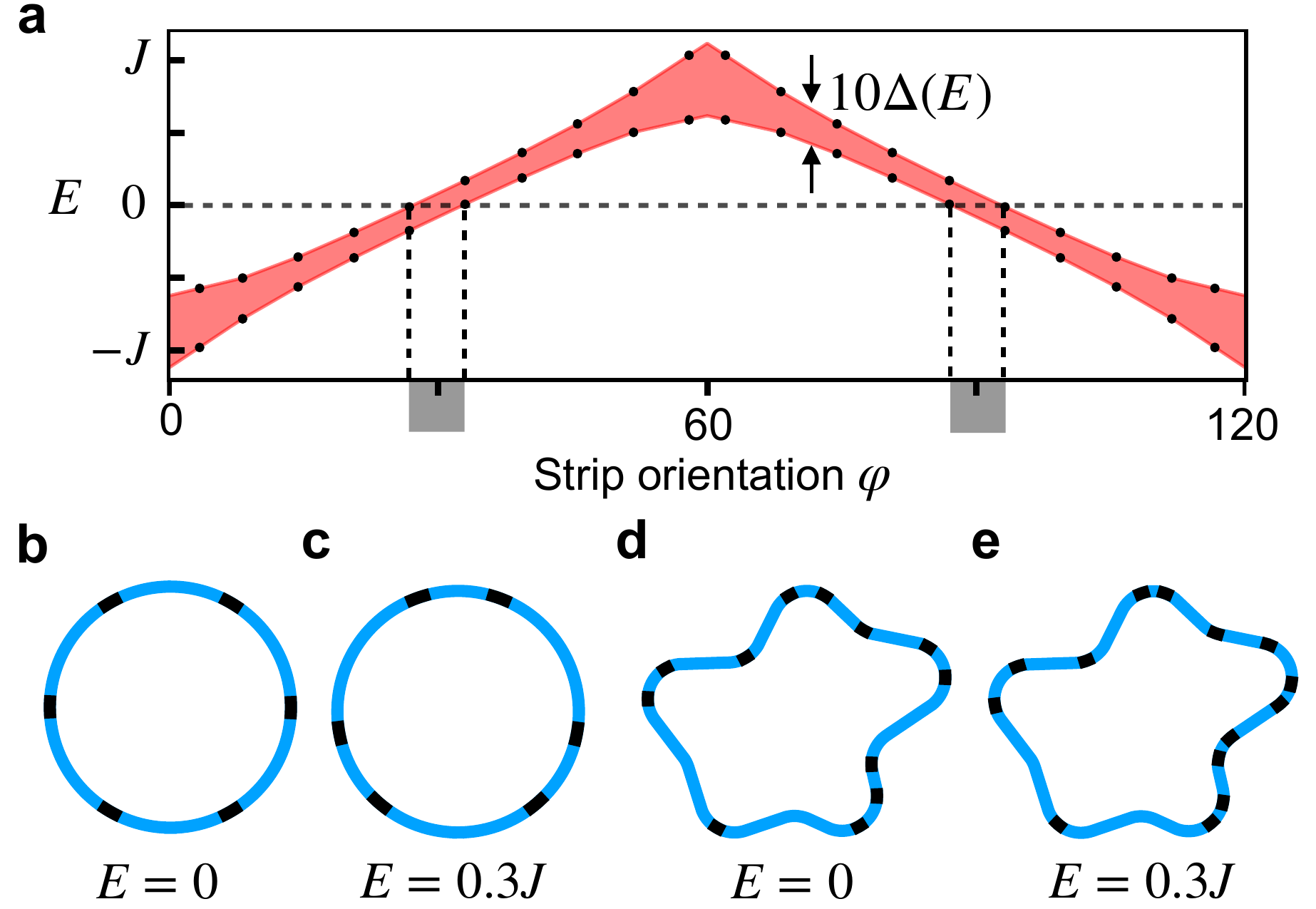}
\caption{Backscattering in edge states propagating along arbitrarily shaped smooth domain walls. \textbf{a}, Dominant edge-band gap (obtained numerically) as a function of the domain wall orientation $\varphi$ ($m_{bk}=1.5J, \lambda = 0.25$). Here, the band gap is shown 10 times larger for visualisation. The grey regions on the x-axis indicate the interval of $\varphi$ where there is a band gap at zero energy. Note that the true width of the grey region will be $\approx 10$ times smaller.
\textbf{b, c, d, e}, Finite scattering regions (depicted in black) on a circular \textbf {b,c} and an arbitrary \textbf {d,e} shaped domain wall for two different values of energy. Note that the true size of the scattering regions will be $\approx 10$ times smaller.
}
\label{fig5}
\end{center}
\end{figure}

In this section, we show how our understanding of the strip band structure for different orientations, developed above, can be utilized to interpret numerical results for the transport of edge states along curved domain walls similar to that of Fig.~\ref{fig1}. We consider scenarios where the domain wall has a radius of curvature that is larger than the typical transverse transition length $a/\lambda$ between the two domains. We will show that, even then, some backscattering exists. This scattering is localized at effective scattering centres whose position along the domain wall is determined by the energy and the local slope of the wall. 

The central idea can be explained easily by revisiting Fig.~\ref{fig3}j. There, we see that the band gap that is induced by tunneling between counterpropagating edge states moves up and down in energy, depending on the orientation of the strip. Translating this to an arbitrary smooth domain wall, this means the following: When we inject a wave packet at some fixed energy, there will be certain orientation angles $\varphi$ at which backscattering takes place. As the orientation of the domain wall changes smoothly along the wall, this defines a condition where certain locations (where the local angle $\varphi$ is just right) become effective scattering centres.

As the curved domain wall is interrupted not only by one but by several scattering centres in this manner, we will moreover obtain the typical behaviour to be expected in such a scenario: Interference between backscattered waves.

We will now employ direct numerical simulations to confirm this picture, i.e. the existence of effective scattering centres that can be predicted from the shape of the domain wall and interference effects arising on this basis.

The theory we develop here will give insights into completely arbitrarily shaped smooth domain walls. However, we will start by describing the backscattering of edge states in the simplest case of a circular domain wall (Fig.~\ref{fig4}a). This allows us to visualize and discuss the results more easily.

Numerics in this context is not entirely trivial, since we want to go to relatively large system sizes, in order to be able to investigate smooth and long domain walls (extending over many lattice sites) and get rid of finite-size effects, see Appendix \ref{Appendix:tight_binding_setup_circle_domain_wall}. Using exact numerical diagonalization of the Schr\"odinger equation (Lanczos diagonalization on sparse matrices) on a tight-binding lattice (of approximately $2500/\lambda^2$ sites, leading to a maximum size of $6 \times 10^4$ lattice sites), we obtain the energy eigenstates in a certain energy interval. Among these, we are able to select the edge state eigenfunctions inside the bulk band gap. Due to the finite amount of backscattering, these are automatically superpositions of the two counterpropagating waves, with the formerly degenerate solutions being split into doublets (as indicated already in Fig.~\ref{fig1}). A closer inspection of these wave functions in momentum space (Fig.~\ref{fig4}b,c) confirms the soundness of the semi-classical picture which we have employed in our analysis up to now.

We are interested in backscattering, and in how this effect depends on the smoothness of the domain wall (as controlled by the scale parameter $\lambda$). In the numerics, it is most convenient to work with eigenstates (and not wave packets or scattering solutions). Still, we are able to extract the reflection coefficient $|r(E)|$ by using its connection to the  splitting of ideally degenerate counterpropagating solutions. The results are shown in Fig.~\ref{fig4}d. We witness two important features: (i) an exponential suppression of reflection with smoothness and (ii) an intricate interference pattern in energy space.

The exponential suppression of the reflection coefficient actually follows the suppression of the dominant bandgap $\Delta_{\mathbf{M}_1,1}$, as can be seen from the numerical results in Fig.~\ref{fig4}e. In that figure, we show the energy dependence for the two exponents, governing the decay of $|r|$ and of $\Delta_{\mathbf{M}_1,1}$, respectively. This valuable link allows us to refer back to our detailed analysis of the band gap that we have provided in previous sections of this work.  In a more formal setting,  we have calculated a WKB edge-state solution for an arbitrary curved domain wall and proved that the solution for a straight domain wall (with a locally varying angular coordinate $\varphi$)  can be viewed as the leading order approximation of our more general solution  in the small parameter $a/R_{\rm rs}$ (with $R_{\rm rs}$  the rescaled radius of curvature), see  Appendix \ref{app:WKB_closed}. 

The main features of the interference pattern observed in Fig.~\ref{fig4}d can be explained even quantitatively by taking into account two effects. The first, more standard effect is the change of the phases accumulated in the different segments between effective scattering centres along the domain wall, as the energy and the wavenumber are varied. The second effect is due to the displacement of the effective scattering centres with energy (Fig.~\ref{fig4}f), related to the shift of band gap with orientation (as explained above; see Fig.~\ref{fig3}j). 
While the location of the scattering centres can be obtained by referencing Fig.~\ref{fig3}j and tracking the slope of the domain wall, we can also use our numerics to give a more detailed insight into what exactly sets these locations apart. We can take a {\em local} Fourier transform of the energy eigenstate, in a certain finite region at any selected point along the domain wall. This enables us to discuss the momentum space behaviour at any point, connecting back to our semi-classical arguments about tunneling between different valleys. As can be observed in Figs.~\ref{fig4}g,h,i, the effective scattering centres are exactly those locations where the momentum space wave function has a peculiar property: The tails of the two parts of the wave function centred around $\mathbf{K}$ and $\mathbf{K'}$ overlap, at an $\mathbf{M}$ point (halfway in-between). This opens an efficient tunneling pathway, giving rise to backscattering. In Fig.~\ref{fig4}j, we show the momentum space wave function of the zero energy mode at the $\mathbf{M}$ point, as a function of orientation angle, visualizing the locations of the effective scattering centres (see Appendix \ref{Appendix:scat_points_non_0_E} for a similar demonstration for non-zero energy modes).

Our choice of a circular domain wall was only for ease of visualization. 
The general situation is shown in Fig.~\ref{fig4}k: At a given fixed energy, the effective scattering centres are located at certain spots along the domain wall, which can be determined easily by applying the reasoning presented here. In a more refined picture, we observe that due to the finite size $\Delta$ of the edge-band gap, these scattering locations actually turn into domain wall regions of finite length, each of them encompassing an interval where the strip orientation leads to a band gap that includes the given wave packet energy, cf Fig.~\ref{fig5}.

One straightforward but helpful consequence of this analysis is an understanding of what happens in a typical scenario encountered in many topological transport experiments: In such experiments one often has straight segments connected by corners. If we think of a smooth domain wall and correspondingly smooth corners (to suppress backscattering), then the remaining backscattering is typically located at the corners. In our picture, this is simply due to the fact that the corner represents a segment where a whole interval of orientation angles is assumed, such that it is likely we encounter an effective backscattering centre there.

\section{Conclusion}

In conclusion, we have introduced a novel analysis of the backscattering of edge states in smooth-envelope topological insulators, based on the insight that they can be understood as tunneling in reciprocal space. We have exploited this insight to derive a detailed semiclassical calculation of the tunneling rate. We find that by increasing the domain wall smoothness even slightly, one can suppress the backscattering rate by a huge amount due to its exponential scaling. In doing so, it also allows to increase the available  bandwidth eliminating  a trade off between backscattering  and bandwidth that affects devices with sharp domain walls. Moreover, we have shown that in an edge channel propagating along a smooth domain wall, the backscattering actually occurs at specific scattering locations which we can predict based on our analysis.

The theory of backscattering, developed in this work, can be used as a basis to engineer backscattering-reduced wavelength-scale topological bosonic waveguides. For instance, the design parameters $m_{bk}$ and $\lambda$ can be carefully chosen in a given experimental situation constrained by the maximum allowed device footprint. Within the smooth-envelope approximation, the footprint scales as $A \sim \lambda ^{-2}$, while the backscattering rate scales as $\vert r \vert \sim \exp[-\mathcal{R}(m_{bk})/\lambda]$. Thus, $m_{bk}$ and $\lambda$ can be optimised to minimise $\vert r \vert$ subject to the constraint of maximum allowed footprint. Furthermore, the theory can be utilized to engineer the domain wall shape in order to avoid as far as possible the appearance of effective scatterers. These strategies can be implemented to build increasingly robust future topological devices.





\vspace{2mm}
\noindent\textbf{Acknowledgements}\\ 
We acknowledge Hermann Schulz-Baldes for discussion. T.S. and F.M. acknowledge support from the European Union’s Horizon 2020 research and innovation programme under the Marie Sklodowska-Curie grant agreement No. 722923 (OMT).  F.M. acknowledges support from the European Union’s Horizon 2020 Research and Innovation program under Grant No. 732894, Future and Emerging Technologies (FET)-Proactive Hybrid Optomechanical Technologies (HOT).





\bibliographystyle{apsrev4-1}
\bibliography{ChiralSound}

\begin{thebibliography}{49}%
\makeatletter
\providecommand \@ifxundefined [1]{%
 \@ifx{#1\undefined}
}%
\providecommand \@ifnum [1]{%
 \ifnum #1\expandafter \@firstoftwo
 \else \expandafter \@secondoftwo
 \fi
}%
\providecommand \@ifx [1]{%
 \ifx #1\expandafter \@firstoftwo
 \else \expandafter \@secondoftwo
 \fi
}%
\providecommand \natexlab [1]{#1}%
\providecommand \enquote  [1]{``#1''}%
\providecommand \bibnamefont  [1]{#1}%
\providecommand \bibfnamefont [1]{#1}%
\providecommand \citenamefont [1]{#1}%
\providecommand \href@noop [0]{\@secondoftwo}%
\providecommand \href [0]{\begingroup \@sanitize@url \@href}%
\providecommand \@href[1]{\@@startlink{#1}\@@href}%
\providecommand \@@href[1]{\endgroup#1\@@endlink}%
\providecommand \@sanitize@url [0]{\catcode `\\12\catcode `\$12\catcode
  `\&12\catcode `\#12\catcode `\^12\catcode `\_12\catcode `\%12\relax}%
\providecommand \@@startlink[1]{}%
\providecommand \@@endlink[0]{}%
\providecommand \url  [0]{\begingroup\@sanitize@url \@url }%
\providecommand \@url [1]{\endgroup\@href {#1}{\urlprefix }}%
\providecommand \urlprefix  [0]{URL }%
\providecommand \Eprint [0]{\href }%
\providecommand \doibase [0]{http://dx.doi.org/}%
\providecommand \selectlanguage [0]{\@gobble}%
\providecommand \bibinfo  [0]{\@secondoftwo}%
\providecommand \bibfield  [0]{\@secondoftwo}%
\providecommand \translation [1]{[#1]}%
\providecommand \BibitemOpen [0]{}%
\providecommand \bibitemStop [0]{}%
\providecommand \bibitemNoStop [0]{.\EOS\space}%
\providecommand \EOS [0]{\spacefactor3000\relax}%
\providecommand \BibitemShut  [1]{\csname bibitem#1\endcsname}%
\let\auto@bib@innerbib\@empty
\bibitem [{\citenamefont {Peano}\ \emph {et~al.}(2015)\citenamefont {Peano},
  \citenamefont {Brendel}, \citenamefont {Schmidt},\ and\ \citenamefont
  {Marquardt}}]{peano_topological_2015}%
  \BibitemOpen
  \bibfield  {author} {\bibinfo {author} {\bibfnamefont {V.}~\bibnamefont
  {Peano}}, \bibinfo {author} {\bibfnamefont {C.}~\bibnamefont {Brendel}},
  \bibinfo {author} {\bibfnamefont {M.}~\bibnamefont {Schmidt}}, \ and\
  \bibinfo {author} {\bibfnamefont {F.}~\bibnamefont {Marquardt}},\ }\href
  {\doibase 10.1103/PhysRevX.5.031011} {\bibfield  {journal} {\bibinfo
  {journal} {Physical Review X}\ }\textbf {\bibinfo {volume} {5}},\ \bibinfo
  {pages} {031011} (\bibinfo {year} {2015})}\BibitemShut {NoStop}%
\bibitem [{\citenamefont {Nash}\ \emph {et~al.}(2015)\citenamefont {Nash},
  \citenamefont {Kleckner}, \citenamefont {Read}, \citenamefont {Vitelli},
  \citenamefont {Turner},\ and\ \citenamefont
  {Irvine}}]{nash_topological_2015}%
  \BibitemOpen
  \bibfield  {author} {\bibinfo {author} {\bibfnamefont {L.~M.}\ \bibnamefont
  {Nash}}, \bibinfo {author} {\bibfnamefont {D.}~\bibnamefont {Kleckner}},
  \bibinfo {author} {\bibfnamefont {A.}~\bibnamefont {Read}}, \bibinfo {author}
  {\bibfnamefont {V.}~\bibnamefont {Vitelli}}, \bibinfo {author} {\bibfnamefont
  {A.~M.}\ \bibnamefont {Turner}}, \ and\ \bibinfo {author} {\bibfnamefont
  {W.~T.~M.}\ \bibnamefont {Irvine}},\ }\href {\doibase
  10.1073/pnas.1507413112} {\bibfield  {journal} {\bibinfo  {journal}
  {Proceedings of the National Academy of Sciences}\ }\textbf {\bibinfo
  {volume} {112}},\ \bibinfo {pages} {14495} (\bibinfo {year}
  {2015})}\BibitemShut {NoStop}%
\bibitem [{\citenamefont {Mathew}\ \emph {et~al.}(2020)\citenamefont {Mathew},
  \citenamefont {Pino},\ and\ \citenamefont
  {Verhagen}}]{mathew_synthetic_2020}%
  \BibitemOpen
  \bibfield  {author} {\bibinfo {author} {\bibfnamefont {J.~P.}\ \bibnamefont
  {Mathew}}, \bibinfo {author} {\bibfnamefont {J.~d.}\ \bibnamefont {Pino}}, \
  and\ \bibinfo {author} {\bibfnamefont {E.}~\bibnamefont {Verhagen}},\ }\href
  {\doibase 10.1038/s41565-019-0630-8} {\bibfield  {journal} {\bibinfo
  {journal} {Nature Nanotechnology}\ }\textbf {\bibinfo {volume} {15}},\
  \bibinfo {pages} {198} (\bibinfo {year} {2020})}\BibitemShut {NoStop}%
\bibitem [{\citenamefont {Wang}\ \emph {et~al.}(2009)\citenamefont {Wang},
  \citenamefont {Chong}, \citenamefont {Joannopoulos},\ and\ \citenamefont
  {Solja{\v c}i{\'c}}}]{wang_observation_2009}%
  \BibitemOpen
  \bibfield  {author} {\bibinfo {author} {\bibfnamefont {Z.}~\bibnamefont
  {Wang}}, \bibinfo {author} {\bibfnamefont {Y.}~\bibnamefont {Chong}},
  \bibinfo {author} {\bibfnamefont {J.~D.}\ \bibnamefont {Joannopoulos}}, \
  and\ \bibinfo {author} {\bibfnamefont {M.}~\bibnamefont {Solja{\v
  c}i{\'c}}},\ }\href@noop {} {\bibfield  {journal} {\bibinfo  {journal}
  {Nature}\ }\textbf {\bibinfo {volume} {461}},\ \bibinfo {pages} {772}
  (\bibinfo {year} {2009})}\BibitemShut {NoStop}%
\bibitem [{\citenamefont {Bahari}\ \emph {et~al.}(2017)\citenamefont {Bahari},
  \citenamefont {Ndao}, \citenamefont {Vallini}, \citenamefont {El~Amili},
  \citenamefont {Fainman},\ and\ \citenamefont
  {Kant{\'e}}}]{bahari_nonreciprocal_2017}%
  \BibitemOpen
  \bibfield  {author} {\bibinfo {author} {\bibfnamefont {B.}~\bibnamefont
  {Bahari}}, \bibinfo {author} {\bibfnamefont {A.}~\bibnamefont {Ndao}},
  \bibinfo {author} {\bibfnamefont {F.}~\bibnamefont {Vallini}}, \bibinfo
  {author} {\bibfnamefont {A.}~\bibnamefont {El~Amili}}, \bibinfo {author}
  {\bibfnamefont {Y.}~\bibnamefont {Fainman}}, \ and\ \bibinfo {author}
  {\bibfnamefont {B.}~\bibnamefont {Kant{\'e}}},\ }\href {\doibase
  10.1126/science.aao4551} {\bibfield  {journal} {\bibinfo  {journal}
  {Science}\ }\textbf {\bibinfo {volume} {358}},\ \bibinfo {pages} {636}
  (\bibinfo {year} {2017})}\BibitemShut {NoStop}%
\bibitem [{\citenamefont {Kane}\ and\ \citenamefont
  {Mele}(2005)}]{kane_quantum_2005}%
  \BibitemOpen
  \bibfield  {author} {\bibinfo {author} {\bibfnamefont {C.~L.}\ \bibnamefont
  {Kane}}\ and\ \bibinfo {author} {\bibfnamefont {E.~J.}\ \bibnamefont
  {Mele}},\ }\href@noop {} {\bibfield  {journal} {\bibinfo  {journal} {Physical
  review letters}\ }\textbf {\bibinfo {volume} {95}},\ \bibinfo {pages}
  {226801} (\bibinfo {year} {2005})}\BibitemShut {NoStop}%
\bibitem [{\citenamefont {Bernevig}\ \emph {et~al.}(2006)\citenamefont
  {Bernevig}, \citenamefont {Hughes},\ and\ \citenamefont
  {Zhang}}]{bernevig_quantum_2006}%
  \BibitemOpen
  \bibfield  {author} {\bibinfo {author} {\bibfnamefont {B.~A.}\ \bibnamefont
  {Bernevig}}, \bibinfo {author} {\bibfnamefont {T.~L.}\ \bibnamefont
  {Hughes}}, \ and\ \bibinfo {author} {\bibfnamefont {S.-C.}\ \bibnamefont
  {Zhang}},\ }\href {\doibase 10.1126/science.1133734} {\bibfield  {journal}
  {\bibinfo  {journal} {Science}\ }\textbf {\bibinfo {volume} {314}},\ \bibinfo
  {pages} {1757} (\bibinfo {year} {2006})}\BibitemShut {NoStop}%
\bibitem [{\citenamefont {Hasan}\ and\ \citenamefont
  {Kane}(2010)}]{hasan_colloquium_2010}%
  \BibitemOpen
  \bibfield  {author} {\bibinfo {author} {\bibfnamefont {M.~Z.}\ \bibnamefont
  {Hasan}}\ and\ \bibinfo {author} {\bibfnamefont {C.~L.}\ \bibnamefont
  {Kane}},\ }\href@noop {} {\bibfield  {journal} {\bibinfo  {journal} {Reviews
  of modern physics}\ }\textbf {\bibinfo {volume} {82}},\ \bibinfo {pages}
  {3045} (\bibinfo {year} {2010})}\BibitemShut {NoStop}%
\bibitem [{\citenamefont {Ningyuan}\ \emph {et~al.}(2015)\citenamefont
  {Ningyuan}, \citenamefont {Owens}, \citenamefont {Sommer}, \citenamefont
  {Schuster},\ and\ \citenamefont {Simon}}]{ningyuan_time-_2015}%
  \BibitemOpen
  \bibfield  {author} {\bibinfo {author} {\bibfnamefont {J.}~\bibnamefont
  {Ningyuan}}, \bibinfo {author} {\bibfnamefont {C.}~\bibnamefont {Owens}},
  \bibinfo {author} {\bibfnamefont {A.}~\bibnamefont {Sommer}}, \bibinfo
  {author} {\bibfnamefont {D.}~\bibnamefont {Schuster}}, \ and\ \bibinfo
  {author} {\bibfnamefont {J.}~\bibnamefont {Simon}},\ }\href {\doibase
  10.1103/PhysRevX.5.021031} {\bibfield  {journal} {\bibinfo  {journal}
  {Physical Review X}\ }\textbf {\bibinfo {volume} {5}},\ \bibinfo {pages}
  {021031} (\bibinfo {year} {2015})}\BibitemShut {NoStop}%
\bibitem [{\citenamefont {Susstrunk}\ and\ \citenamefont
  {Huber}(2015)}]{susstrunk_observation_2015}%
  \BibitemOpen
  \bibfield  {author} {\bibinfo {author} {\bibfnamefont {R.}~\bibnamefont
  {Susstrunk}}\ and\ \bibinfo {author} {\bibfnamefont {S.~D.}\ \bibnamefont
  {Huber}},\ }\href {\doibase 10.1126/science.aab0239} {\bibfield  {journal}
  {\bibinfo  {journal} {Science}\ }\textbf {\bibinfo {volume} {349}},\ \bibinfo
  {pages} {47} (\bibinfo {year} {2015})}\BibitemShut {NoStop}%
\bibitem [{\citenamefont {Martin}\ \emph {et~al.}(2008)\citenamefont {Martin},
  \citenamefont {Blanter},\ and\ \citenamefont
  {Morpurgo}}]{martin_topological_2008}%
  \BibitemOpen
  \bibfield  {author} {\bibinfo {author} {\bibfnamefont {I.}~\bibnamefont
  {Martin}}, \bibinfo {author} {\bibfnamefont {Y.~M.}\ \bibnamefont {Blanter}},
  \ and\ \bibinfo {author} {\bibfnamefont {A.}~\bibnamefont {Morpurgo}},\
  }\href@noop {} {\bibfield  {journal} {\bibinfo  {journal} {Physical review
  letters}\ }\textbf {\bibinfo {volume} {100}},\ \bibinfo {pages} {036804}
  (\bibinfo {year} {2008})}\BibitemShut {NoStop}%
\bibitem [{\citenamefont {Ju}\ \emph {et~al.}(2015)\citenamefont {Ju},
  \citenamefont {Shi}, \citenamefont {Nair}, \citenamefont {Lv}, \citenamefont
  {Jin}, \citenamefont {Velasco~Jr}, \citenamefont {Ojeda-Aristizabal},
  \citenamefont {Bechtel}, \citenamefont {Martin}, \citenamefont {Zettl},\ and\
  \citenamefont {{others}}}]{ju_topological_2015}%
  \BibitemOpen
  \bibfield  {author} {\bibinfo {author} {\bibfnamefont {L.}~\bibnamefont
  {Ju}}, \bibinfo {author} {\bibfnamefont {Z.}~\bibnamefont {Shi}}, \bibinfo
  {author} {\bibfnamefont {N.}~\bibnamefont {Nair}}, \bibinfo {author}
  {\bibfnamefont {Y.}~\bibnamefont {Lv}}, \bibinfo {author} {\bibfnamefont
  {C.}~\bibnamefont {Jin}}, \bibinfo {author} {\bibfnamefont {J.}~\bibnamefont
  {Velasco~Jr}}, \bibinfo {author} {\bibfnamefont {C.}~\bibnamefont
  {Ojeda-Aristizabal}}, \bibinfo {author} {\bibfnamefont {H.~A.}\ \bibnamefont
  {Bechtel}}, \bibinfo {author} {\bibfnamefont {M.~C.}\ \bibnamefont {Martin}},
  \bibinfo {author} {\bibfnamefont {A.}~\bibnamefont {Zettl}}, \ and\ \bibinfo
  {author} {\bibnamefont {{others}}},\ }\href@noop {} {\bibfield  {journal}
  {\bibinfo  {journal} {Nature}\ }\textbf {\bibinfo {volume} {520}},\ \bibinfo
  {pages} {650} (\bibinfo {year} {2015})}\BibitemShut {NoStop}%
\bibitem [{\citenamefont {Ma}\ and\ \citenamefont
  {Shvets}(2016)}]{ma_all-si_2016}%
  \BibitemOpen
  \bibfield  {author} {\bibinfo {author} {\bibfnamefont {T.}~\bibnamefont
  {Ma}}\ and\ \bibinfo {author} {\bibfnamefont {G.}~\bibnamefont {Shvets}},\
  }\href {\doibase 10.1088/1367-2630/18/2/025012} {\bibfield  {journal}
  {\bibinfo  {journal} {New Journal of Physics}\ }\textbf {\bibinfo {volume}
  {18}},\ \bibinfo {pages} {025012} (\bibinfo {year} {2016})}\BibitemShut
  {NoStop}%
\bibitem [{\citenamefont {Lu}\ \emph {et~al.}(2017)\citenamefont {Lu},
  \citenamefont {Qiu}, \citenamefont {Ye}, \citenamefont {Fan}, \citenamefont
  {Ke}, \citenamefont {Zhang},\ and\ \citenamefont
  {Liu}}]{lu_observation_2017}%
  \BibitemOpen
  \bibfield  {author} {\bibinfo {author} {\bibfnamefont {J.}~\bibnamefont
  {Lu}}, \bibinfo {author} {\bibfnamefont {C.}~\bibnamefont {Qiu}}, \bibinfo
  {author} {\bibfnamefont {L.}~\bibnamefont {Ye}}, \bibinfo {author}
  {\bibfnamefont {X.}~\bibnamefont {Fan}}, \bibinfo {author} {\bibfnamefont
  {M.}~\bibnamefont {Ke}}, \bibinfo {author} {\bibfnamefont {F.}~\bibnamefont
  {Zhang}}, \ and\ \bibinfo {author} {\bibfnamefont {Z.}~\bibnamefont {Liu}},\
  }\href@noop {} {\bibfield  {journal} {\bibinfo  {journal} {Nature Physics}\
  }\textbf {\bibinfo {volume} {13}},\ \bibinfo {pages} {369} (\bibinfo {year}
  {2017})}\BibitemShut {NoStop}%
\bibitem [{\citenamefont {Dong}\ \emph {et~al.}(2017)\citenamefont {Dong},
  \citenamefont {Chen}, \citenamefont {Zhu}, \citenamefont {Wang},\ and\
  \citenamefont {Zhang}}]{dong_valley_2017}%
  \BibitemOpen
  \bibfield  {author} {\bibinfo {author} {\bibfnamefont {J.-W.}\ \bibnamefont
  {Dong}}, \bibinfo {author} {\bibfnamefont {X.-D.}\ \bibnamefont {Chen}},
  \bibinfo {author} {\bibfnamefont {H.}~\bibnamefont {Zhu}}, \bibinfo {author}
  {\bibfnamefont {Y.}~\bibnamefont {Wang}}, \ and\ \bibinfo {author}
  {\bibfnamefont {X.}~\bibnamefont {Zhang}},\ }\href@noop {} {\bibfield
  {journal} {\bibinfo  {journal} {Nature materials}\ }\textbf {\bibinfo
  {volume} {16}},\ \bibinfo {pages} {298} (\bibinfo {year} {2017})}\BibitemShut
  {NoStop}%
\bibitem [{\citenamefont {Vila}\ \emph {et~al.}(2017)\citenamefont {Vila},
  \citenamefont {Pal},\ and\ \citenamefont {Ruzzene}}]{vila_observation_2017}%
  \BibitemOpen
  \bibfield  {author} {\bibinfo {author} {\bibfnamefont {J.}~\bibnamefont
  {Vila}}, \bibinfo {author} {\bibfnamefont {R.~K.}\ \bibnamefont {Pal}}, \
  and\ \bibinfo {author} {\bibfnamefont {M.}~\bibnamefont {Ruzzene}},\
  }\href@noop {} {\bibfield  {journal} {\bibinfo  {journal} {Physical Review
  B}\ }\textbf {\bibinfo {volume} {96}},\ \bibinfo {pages} {134307} (\bibinfo
  {year} {2017})}\BibitemShut {NoStop}%
\bibitem [{\citenamefont {Wu}\ \emph {et~al.}(2017)\citenamefont {Wu},
  \citenamefont {Meng}, \citenamefont {Tian}, \citenamefont {Huang},
  \citenamefont {Xiang}, \citenamefont {Han},\ and\ \citenamefont
  {Wen}}]{wu_direct_2017}%
  \BibitemOpen
  \bibfield  {author} {\bibinfo {author} {\bibfnamefont {X.}~\bibnamefont
  {Wu}}, \bibinfo {author} {\bibfnamefont {Y.}~\bibnamefont {Meng}}, \bibinfo
  {author} {\bibfnamefont {J.}~\bibnamefont {Tian}}, \bibinfo {author}
  {\bibfnamefont {Y.}~\bibnamefont {Huang}}, \bibinfo {author} {\bibfnamefont
  {H.}~\bibnamefont {Xiang}}, \bibinfo {author} {\bibfnamefont
  {D.}~\bibnamefont {Han}}, \ and\ \bibinfo {author} {\bibfnamefont
  {W.}~\bibnamefont {Wen}},\ }\href@noop {} {\bibfield  {journal} {\bibinfo
  {journal} {Nature communications}\ }\textbf {\bibinfo {volume} {8}},\
  \bibinfo {pages} {1} (\bibinfo {year} {2017})}\BibitemShut {NoStop}%
\bibitem [{\citenamefont {Gao}\ \emph {et~al.}(2017)\citenamefont {Gao},
  \citenamefont {Yang}, \citenamefont {Gao}, \citenamefont {Xue}, \citenamefont
  {Yang}, \citenamefont {Dong},\ and\ \citenamefont {Zhang}}]{gao_valley_2017}%
  \BibitemOpen
  \bibfield  {author} {\bibinfo {author} {\bibfnamefont {Z.}~\bibnamefont
  {Gao}}, \bibinfo {author} {\bibfnamefont {Z.}~\bibnamefont {Yang}}, \bibinfo
  {author} {\bibfnamefont {F.}~\bibnamefont {Gao}}, \bibinfo {author}
  {\bibfnamefont {H.}~\bibnamefont {Xue}}, \bibinfo {author} {\bibfnamefont
  {Y.}~\bibnamefont {Yang}}, \bibinfo {author} {\bibfnamefont {J.}~\bibnamefont
  {Dong}}, \ and\ \bibinfo {author} {\bibfnamefont {B.}~\bibnamefont {Zhang}},\
  }\href@noop {} {\bibfield  {journal} {\bibinfo  {journal} {Physical Review
  B}\ }\textbf {\bibinfo {volume} {96}},\ \bibinfo {pages} {201402} (\bibinfo
  {year} {2017})}\BibitemShut {NoStop}%
\bibitem [{\citenamefont {Kang}\ \emph {et~al.}(2018)\citenamefont {Kang},
  \citenamefont {Ni}, \citenamefont {Cheng}, \citenamefont {Khanikaev},\ and\
  \citenamefont {Genack}}]{kang_pseudo-spinvalley_2018}%
  \BibitemOpen
  \bibfield  {author} {\bibinfo {author} {\bibfnamefont {Y.}~\bibnamefont
  {Kang}}, \bibinfo {author} {\bibfnamefont {X.}~\bibnamefont {Ni}}, \bibinfo
  {author} {\bibfnamefont {X.}~\bibnamefont {Cheng}}, \bibinfo {author}
  {\bibfnamefont {A.~B.}\ \bibnamefont {Khanikaev}}, \ and\ \bibinfo {author}
  {\bibfnamefont {A.~Z.}\ \bibnamefont {Genack}},\ }\href@noop {} {\bibfield
  {journal} {\bibinfo  {journal} {Nature communications}\ }\textbf {\bibinfo
  {volume} {9}},\ \bibinfo {pages} {1} (\bibinfo {year} {2018})}\BibitemShut
  {NoStop}%
\bibitem [{\citenamefont {Noh}\ \emph {et~al.}(2018)\citenamefont {Noh},
  \citenamefont {Huang}, \citenamefont {Chen},\ and\ \citenamefont
  {Rechtsman}}]{noh_observation_2018}%
  \BibitemOpen
  \bibfield  {author} {\bibinfo {author} {\bibfnamefont {J.}~\bibnamefont
  {Noh}}, \bibinfo {author} {\bibfnamefont {S.}~\bibnamefont {Huang}}, \bibinfo
  {author} {\bibfnamefont {K.~P.}\ \bibnamefont {Chen}}, \ and\ \bibinfo
  {author} {\bibfnamefont {M.~C.}\ \bibnamefont {Rechtsman}},\ }\href@noop {}
  {\bibfield  {journal} {\bibinfo  {journal} {Physical review letters}\
  }\textbf {\bibinfo {volume} {120}},\ \bibinfo {pages} {063902} (\bibinfo
  {year} {2018})}\BibitemShut {NoStop}%
\bibitem [{\citenamefont {Shalaev}\ \emph {et~al.}(2019)\citenamefont
  {Shalaev}, \citenamefont {Walasik}, \citenamefont {Tsukernik}, \citenamefont
  {Xu},\ and\ \citenamefont {Litchinitser}}]{shalaev_robust_2019}%
  \BibitemOpen
  \bibfield  {author} {\bibinfo {author} {\bibfnamefont {M.~I.}\ \bibnamefont
  {Shalaev}}, \bibinfo {author} {\bibfnamefont {W.}~\bibnamefont {Walasik}},
  \bibinfo {author} {\bibfnamefont {A.}~\bibnamefont {Tsukernik}}, \bibinfo
  {author} {\bibfnamefont {Y.}~\bibnamefont {Xu}}, \ and\ \bibinfo {author}
  {\bibfnamefont {N.~M.}\ \bibnamefont {Litchinitser}},\ }\href {\doibase
  10.1038/s41565-018-0297-6} {\bibfield  {journal} {\bibinfo  {journal} {Nature
  Nanotechnology}\ }\textbf {\bibinfo {volume} {14}},\ \bibinfo {pages} {31}
  (\bibinfo {year} {2019})}\BibitemShut {NoStop}%
\bibitem [{\citenamefont {Zeng}\ \emph {et~al.}(2020)\citenamefont {Zeng},
  \citenamefont {Chattopadhyay}, \citenamefont {Zhu}, \citenamefont {Qiang},
  \citenamefont {Li}, \citenamefont {Jin}, \citenamefont {Li}, \citenamefont
  {Davies}, \citenamefont {Linfield}, \citenamefont {Zhang}, \citenamefont
  {Chong},\ and\ \citenamefont {Wang}}]{zeng_electrically_2020}%
  \BibitemOpen
  \bibfield  {author} {\bibinfo {author} {\bibfnamefont {Y.}~\bibnamefont
  {Zeng}}, \bibinfo {author} {\bibfnamefont {U.}~\bibnamefont {Chattopadhyay}},
  \bibinfo {author} {\bibfnamefont {B.}~\bibnamefont {Zhu}}, \bibinfo {author}
  {\bibfnamefont {B.}~\bibnamefont {Qiang}}, \bibinfo {author} {\bibfnamefont
  {J.}~\bibnamefont {Li}}, \bibinfo {author} {\bibfnamefont {Y.}~\bibnamefont
  {Jin}}, \bibinfo {author} {\bibfnamefont {L.}~\bibnamefont {Li}}, \bibinfo
  {author} {\bibfnamefont {A.~G.}\ \bibnamefont {Davies}}, \bibinfo {author}
  {\bibfnamefont {E.~H.}\ \bibnamefont {Linfield}}, \bibinfo {author}
  {\bibfnamefont {B.}~\bibnamefont {Zhang}}, \bibinfo {author} {\bibfnamefont
  {Y.}~\bibnamefont {Chong}}, \ and\ \bibinfo {author} {\bibfnamefont {Q.~J.}\
  \bibnamefont {Wang}},\ }\href {\doibase 10.1038/s41586-020-1981-x} {\bibfield
   {journal} {\bibinfo  {journal} {Nature}\ }\textbf {\bibinfo {volume}
  {578}},\ \bibinfo {pages} {246} (\bibinfo {year} {2020})}\BibitemShut
  {NoStop}%
\bibitem [{\citenamefont {Ren}\ \emph {et~al.}(2020)\citenamefont {Ren},
  \citenamefont {Shah}, \citenamefont {Pfeifer}, \citenamefont {Brendel},
  \citenamefont {Peano}, \citenamefont {Marquardt},\ and\ \citenamefont
  {Painter}}]{ren_topological_2020}%
  \BibitemOpen
  \bibfield  {author} {\bibinfo {author} {\bibfnamefont {H.}~\bibnamefont
  {Ren}}, \bibinfo {author} {\bibfnamefont {T.}~\bibnamefont {Shah}}, \bibinfo
  {author} {\bibfnamefont {H.}~\bibnamefont {Pfeifer}}, \bibinfo {author}
  {\bibfnamefont {C.}~\bibnamefont {Brendel}}, \bibinfo {author} {\bibfnamefont
  {V.}~\bibnamefont {Peano}}, \bibinfo {author} {\bibfnamefont
  {F.}~\bibnamefont {Marquardt}}, \ and\ \bibinfo {author} {\bibfnamefont
  {O.}~\bibnamefont {Painter}},\ }\href {http://arxiv.org/abs/2009.06174}
  {\bibfield  {journal} {\bibinfo  {journal} {arXiv:2009.06174 [cond-mat,
  physics:physics]}\ } (\bibinfo {year} {2020})}\BibitemShut {NoStop}%
\bibitem [{\citenamefont {Arora}\ \emph {et~al.}(2021)\citenamefont {Arora},
  \citenamefont {Bauer}, \citenamefont {Barczyk}, \citenamefont {Verhagen},\
  and\ \citenamefont {Kuipers}}]{arora_direct_2021}%
  \BibitemOpen
  \bibfield  {author} {\bibinfo {author} {\bibfnamefont {S.}~\bibnamefont
  {Arora}}, \bibinfo {author} {\bibfnamefont {T.}~\bibnamefont {Bauer}},
  \bibinfo {author} {\bibfnamefont {R.}~\bibnamefont {Barczyk}}, \bibinfo
  {author} {\bibfnamefont {E.}~\bibnamefont {Verhagen}}, \ and\ \bibinfo
  {author} {\bibfnamefont {L.}~\bibnamefont {Kuipers}},\ }\href {\doibase
  10.1038/s41377-020-00458-6} {\bibfield  {journal} {\bibinfo  {journal}
  {Light: Science \& Applications}\ }\textbf {\bibinfo {volume} {10}},\
  \bibinfo {pages} {9} (\bibinfo {year} {2021})}\BibitemShut {NoStop}%
\bibitem [{\citenamefont {Wu}\ and\ \citenamefont {Hu}(2015)}]{wu_scheme_2015}%
  \BibitemOpen
  \bibfield  {author} {\bibinfo {author} {\bibfnamefont {L.-H.}\ \bibnamefont
  {Wu}}\ and\ \bibinfo {author} {\bibfnamefont {X.}~\bibnamefont {Hu}},\ }\href
  {\doibase 10.1103/PhysRevLett.114.223901} {\bibfield  {journal} {\bibinfo
  {journal} {Physical Review Letters}\ }\textbf {\bibinfo {volume} {114}},\
  \bibinfo {pages} {223901} (\bibinfo {year} {2015})}\BibitemShut {NoStop}%
\bibitem [{\citenamefont {He}\ \emph {et~al.}(2016)\citenamefont {He},
  \citenamefont {Ni}, \citenamefont {Ge}, \citenamefont {Sun}, \citenamefont
  {Chen}, \citenamefont {Lu}, \citenamefont {Liu},\ and\ \citenamefont
  {Chen}}]{he_acoustic_2016}%
  \BibitemOpen
  \bibfield  {author} {\bibinfo {author} {\bibfnamefont {C.}~\bibnamefont
  {He}}, \bibinfo {author} {\bibfnamefont {X.}~\bibnamefont {Ni}}, \bibinfo
  {author} {\bibfnamefont {H.}~\bibnamefont {Ge}}, \bibinfo {author}
  {\bibfnamefont {X.-C.}\ \bibnamefont {Sun}}, \bibinfo {author} {\bibfnamefont
  {Y.-B.}\ \bibnamefont {Chen}}, \bibinfo {author} {\bibfnamefont {M.-H.}\
  \bibnamefont {Lu}}, \bibinfo {author} {\bibfnamefont {X.-P.}\ \bibnamefont
  {Liu}}, \ and\ \bibinfo {author} {\bibfnamefont {Y.-F.}\ \bibnamefont
  {Chen}},\ }\href {\doibase 10.1038/nphys3867} {\bibfield  {journal} {\bibinfo
   {journal} {Nature Physics}\ }\textbf {\bibinfo {volume} {12}},\ \bibinfo
  {pages} {1124} (\bibinfo {year} {2016})}\BibitemShut {NoStop}%
\bibitem [{\citenamefont {Brendel}\ \emph {et~al.}(2018)\citenamefont
  {Brendel}, \citenamefont {Peano}, \citenamefont {Painter},\ and\
  \citenamefont {Marquardt}}]{brendel_snowflake_2018}%
  \BibitemOpen
  \bibfield  {author} {\bibinfo {author} {\bibfnamefont {C.}~\bibnamefont
  {Brendel}}, \bibinfo {author} {\bibfnamefont {V.}~\bibnamefont {Peano}},
  \bibinfo {author} {\bibfnamefont {O.}~\bibnamefont {Painter}}, \ and\
  \bibinfo {author} {\bibfnamefont {F.}~\bibnamefont {Marquardt}},\ }\href@noop
  {} {\bibfield  {journal} {\bibinfo  {journal} {Physical Review B}\ }\textbf
  {\bibinfo {volume} {97}},\ \bibinfo {pages} {020102} (\bibinfo {year}
  {2018})}\BibitemShut {NoStop}%
\bibitem [{\citenamefont {Yang}\ \emph {et~al.}(2016)\citenamefont {Yang},
  \citenamefont {Xu}, \citenamefont {Xu}, \citenamefont {Wang}, \citenamefont
  {Jiang}, \citenamefont {Hu},\ and\ \citenamefont
  {Hang}}]{yang_visualization_2016}%
  \BibitemOpen
  \bibfield  {author} {\bibinfo {author} {\bibfnamefont {Y.}~\bibnamefont
  {Yang}}, \bibinfo {author} {\bibfnamefont {Y.~F.}\ \bibnamefont {Xu}},
  \bibinfo {author} {\bibfnamefont {T.}~\bibnamefont {Xu}}, \bibinfo {author}
  {\bibfnamefont {H.-X.}\ \bibnamefont {Wang}}, \bibinfo {author}
  {\bibfnamefont {J.-H.}\ \bibnamefont {Jiang}}, \bibinfo {author}
  {\bibfnamefont {X.}~\bibnamefont {Hu}}, \ and\ \bibinfo {author}
  {\bibfnamefont {Z.~H.}\ \bibnamefont {Hang}},\ }\href
  {http://arxiv.org/abs/1610.07780} {\bibfield  {journal} {\bibinfo  {journal}
  {arXiv:1610.07780 [physics]}\ } (\bibinfo {year} {2016})}\BibitemShut
  {NoStop}%
\bibitem [{\citenamefont {Cha}\ \emph {et~al.}(2018)\citenamefont {Cha},
  \citenamefont {Kim},\ and\ \citenamefont {Daraio}}]{cha_experimental_2018}%
  \BibitemOpen
  \bibfield  {author} {\bibinfo {author} {\bibfnamefont {J.}~\bibnamefont
  {Cha}}, \bibinfo {author} {\bibfnamefont {K.~W.}\ \bibnamefont {Kim}}, \ and\
  \bibinfo {author} {\bibfnamefont {C.}~\bibnamefont {Daraio}},\ }\href
  {\doibase 10.1038/s41586-018-0764-0} {\bibfield  {journal} {\bibinfo
  {journal} {Nature}\ }\textbf {\bibinfo {volume} {564}},\ \bibinfo {pages}
  {229} (\bibinfo {year} {2018})}\BibitemShut {NoStop}%
\bibitem [{\citenamefont {Barik}\ \emph {et~al.}(2018)\citenamefont {Barik},
  \citenamefont {Karasahin}, \citenamefont {Flower}, \citenamefont {Cai},
  \citenamefont {Miyake}, \citenamefont {DeGottardi}, \citenamefont {Hafezi},\
  and\ \citenamefont {Waks}}]{barik_topological_2018}%
  \BibitemOpen
  \bibfield  {author} {\bibinfo {author} {\bibfnamefont {S.}~\bibnamefont
  {Barik}}, \bibinfo {author} {\bibfnamefont {A.}~\bibnamefont {Karasahin}},
  \bibinfo {author} {\bibfnamefont {C.}~\bibnamefont {Flower}}, \bibinfo
  {author} {\bibfnamefont {T.}~\bibnamefont {Cai}}, \bibinfo {author}
  {\bibfnamefont {H.}~\bibnamefont {Miyake}}, \bibinfo {author} {\bibfnamefont
  {W.}~\bibnamefont {DeGottardi}}, \bibinfo {author} {\bibfnamefont
  {M.}~\bibnamefont {Hafezi}}, \ and\ \bibinfo {author} {\bibfnamefont
  {E.}~\bibnamefont {Waks}},\ }\href {\doibase 10.1126/science.aaq0327}
  {\bibfield  {journal} {\bibinfo  {journal} {Science}\ }\textbf {\bibinfo
  {volume} {359}},\ \bibinfo {pages} {666} (\bibinfo {year}
  {2018})}\BibitemShut {NoStop}%
\bibitem [{\citenamefont {Parappurath}\ \emph {et~al.}(2020)\citenamefont
  {Parappurath}, \citenamefont {Alpeggiani}, \citenamefont {Kuipers},\ and\
  \citenamefont {Verhagen}}]{parappurath_direct_2020}%
  \BibitemOpen
  \bibfield  {author} {\bibinfo {author} {\bibfnamefont {N.}~\bibnamefont
  {Parappurath}}, \bibinfo {author} {\bibfnamefont {F.}~\bibnamefont
  {Alpeggiani}}, \bibinfo {author} {\bibfnamefont {L.}~\bibnamefont {Kuipers}},
  \ and\ \bibinfo {author} {\bibfnamefont {E.}~\bibnamefont {Verhagen}},\
  }\href {\doibase 10.1126/sciadv.aaw4137} {\bibfield  {journal} {\bibinfo
  {journal} {Science Advances}\ }\textbf {\bibinfo {volume} {6}},\ \bibinfo
  {pages} {eaaw4137} (\bibinfo {year} {2020})}\BibitemShut {NoStop}%
\bibitem [{\citenamefont {Shao}\ \emph {et~al.}(2020)\citenamefont {Shao},
  \citenamefont {Chen}, \citenamefont {Wang}, \citenamefont {Mao},
  \citenamefont {Yang}, \citenamefont {Wang}, \citenamefont {Wang},
  \citenamefont {Hu},\ and\ \citenamefont {Ma}}]{shao_high-performance_2020}%
  \BibitemOpen
  \bibfield  {author} {\bibinfo {author} {\bibfnamefont {Z.-K.}\ \bibnamefont
  {Shao}}, \bibinfo {author} {\bibfnamefont {H.-Z.}\ \bibnamefont {Chen}},
  \bibinfo {author} {\bibfnamefont {S.}~\bibnamefont {Wang}}, \bibinfo {author}
  {\bibfnamefont {X.-R.}\ \bibnamefont {Mao}}, \bibinfo {author} {\bibfnamefont
  {Z.-Q.}\ \bibnamefont {Yang}}, \bibinfo {author} {\bibfnamefont {S.-L.}\
  \bibnamefont {Wang}}, \bibinfo {author} {\bibfnamefont {X.-X.}\ \bibnamefont
  {Wang}}, \bibinfo {author} {\bibfnamefont {X.}~\bibnamefont {Hu}}, \ and\
  \bibinfo {author} {\bibfnamefont {R.-M.}\ \bibnamefont {Ma}},\ }\href
  {\doibase 10.1038/s41565-019-0584-x} {\bibfield  {journal} {\bibinfo
  {journal} {Nature Nanotechnology}\ }\textbf {\bibinfo {volume} {15}},\
  \bibinfo {pages} {67} (\bibinfo {year} {2020})}\BibitemShut {NoStop}%
\bibitem [{\citenamefont {Vogl}\ \emph {et~al.}(2017)\citenamefont {Vogl},
  \citenamefont {Pankratov},\ and\ \citenamefont
  {Shallcross}}]{vogl_semiclassics_2017}%
  \BibitemOpen
  \bibfield  {author} {\bibinfo {author} {\bibfnamefont {M.}~\bibnamefont
  {Vogl}}, \bibinfo {author} {\bibfnamefont {O.}~\bibnamefont {Pankratov}}, \
  and\ \bibinfo {author} {\bibfnamefont {S.}~\bibnamefont {Shallcross}},\
  }\href {\doibase 10.1103/PhysRevB.96.035442} {\bibfield  {journal} {\bibinfo
  {journal} {Physical Review B}\ }\textbf {\bibinfo {volume} {96}},\ \bibinfo
  {pages} {035442} (\bibinfo {year} {2017})}\BibitemShut {NoStop}%
\bibitem [{\citenamefont {Gosselin}\ \emph {et~al.}(2009)\citenamefont
  {Gosselin}, \citenamefont {B{\'e}rard}, \citenamefont {Mohrbach},\ and\
  \citenamefont {Ghosh}}]{gosselin_berry_2009}%
  \BibitemOpen
  \bibfield  {author} {\bibinfo {author} {\bibfnamefont {P.}~\bibnamefont
  {Gosselin}}, \bibinfo {author} {\bibfnamefont {A.}~\bibnamefont
  {B{\'e}rard}}, \bibinfo {author} {\bibfnamefont {H.}~\bibnamefont
  {Mohrbach}}, \ and\ \bibinfo {author} {\bibfnamefont {S.}~\bibnamefont
  {Ghosh}},\ }\href {\doibase 10.1140/epjc/s10052-008-0839-4} {\bibfield
  {journal} {\bibinfo  {journal} {The European Physical Journal C}\ }\textbf
  {\bibinfo {volume} {59}},\ \bibinfo {pages} {883} (\bibinfo {year}
  {2009})}\BibitemShut {NoStop}%
\bibitem [{\citenamefont {Fuchs}\ \emph {et~al.}(2010)\citenamefont {Fuchs},
  \citenamefont {Pi{\'e}chon}, \citenamefont {Goerbig},\ and\ \citenamefont
  {Montambaux}}]{fuchs_topological_2010}%
  \BibitemOpen
  \bibfield  {author} {\bibinfo {author} {\bibfnamefont {J.~N.}\ \bibnamefont
  {Fuchs}}, \bibinfo {author} {\bibfnamefont {F.}~\bibnamefont {Pi{\'e}chon}},
  \bibinfo {author} {\bibfnamefont {M.~O.}\ \bibnamefont {Goerbig}}, \ and\
  \bibinfo {author} {\bibfnamefont {G.}~\bibnamefont {Montambaux}},\ }\href
  {\doibase 10.1140/epjb/e2010-00259-2} {\bibfield  {journal} {\bibinfo
  {journal} {The European Physical Journal B}\ }\textbf {\bibinfo {volume}
  {77}},\ \bibinfo {pages} {351} (\bibinfo {year} {2010})}\BibitemShut
  {NoStop}%
\bibitem [{\citenamefont {Delplace}\ and\ \citenamefont
  {Montambaux}(2010)}]{delplace_wkb_2010}%
  \BibitemOpen
  \bibfield  {author} {\bibinfo {author} {\bibfnamefont {P.}~\bibnamefont
  {Delplace}}\ and\ \bibinfo {author} {\bibfnamefont {G.}~\bibnamefont
  {Montambaux}},\ }\href {\doibase 10.1103/PhysRevB.82.205412} {\bibfield
  {journal} {\bibinfo  {journal} {Physical Review B}\ }\textbf {\bibinfo
  {volume} {82}},\ \bibinfo {pages} {205412} (\bibinfo {year}
  {2010})}\BibitemShut {NoStop}%
\bibitem [{\citenamefont {Reijnders}\ \emph {et~al.}(2018)\citenamefont
  {Reijnders}, \citenamefont {Minenkov}, \citenamefont {Katsnelson},\ and\
  \citenamefont {Dobrokhotov}}]{reijnders_electronic_2018}%
  \BibitemOpen
  \bibfield  {author} {\bibinfo {author} {\bibfnamefont {K.}~\bibnamefont
  {Reijnders}}, \bibinfo {author} {\bibfnamefont {D.}~\bibnamefont {Minenkov}},
  \bibinfo {author} {\bibfnamefont {M.}~\bibnamefont {Katsnelson}}, \ and\
  \bibinfo {author} {\bibfnamefont {S.}~\bibnamefont {Dobrokhotov}},\ }\href
  {\doibase 10.1016/j.aop.2018.08.004} {\bibfield  {journal} {\bibinfo
  {journal} {Annals of Physics}\ }\textbf {\bibinfo {volume} {397}},\ \bibinfo
  {pages} {65} (\bibinfo {year} {2018})}\BibitemShut {NoStop}%
\bibitem [{\citenamefont {Doost}\ \emph {et~al.}(2021)\citenamefont {Doost},
  \citenamefont {Kasmaei},\ and\ \citenamefont
  {Beckwith}}]{doost_foldy-wouthuysen_2021}%
  \BibitemOpen
  \bibfield  {author} {\bibinfo {author} {\bibfnamefont {M.~B.}\ \bibnamefont
  {Doost}}, \bibinfo {author} {\bibfnamefont {H.~D.}\ \bibnamefont {Kasmaei}},
  \ and\ \bibinfo {author} {\bibfnamefont {A.~W.}\ \bibnamefont {Beckwith}},\
  }\href {\doibase 10.1016/j.physe.2021.114654} {\bibfield  {journal} {\bibinfo
   {journal} {Physica E: Low-dimensional Systems and Nanostructures}\ }\textbf
  {\bibinfo {volume} {130}},\ \bibinfo {pages} {114654} (\bibinfo {year}
  {2021})}\BibitemShut {NoStop}%
\bibitem [{\citenamefont {Mousavi}\ \emph {et~al.}(2015)\citenamefont
  {Mousavi}, \citenamefont {Khanikaev},\ and\ \citenamefont
  {Wang}}]{mousavi_topologically_2015}%
  \BibitemOpen
  \bibfield  {author} {\bibinfo {author} {\bibfnamefont {S.~H.}\ \bibnamefont
  {Mousavi}}, \bibinfo {author} {\bibfnamefont {A.~B.}\ \bibnamefont
  {Khanikaev}}, \ and\ \bibinfo {author} {\bibfnamefont {Z.}~\bibnamefont
  {Wang}},\ }\href {\doibase 10.1038/ncomms9682} {\bibfield  {journal}
  {\bibinfo  {journal} {Nature Communications}\ }\textbf {\bibinfo {volume}
  {6}},\ \bibinfo {pages} {8682} (\bibinfo {year} {2015})}\BibitemShut
  {NoStop}%
\bibitem [{\citenamefont {Miniaci}\ \emph {et~al.}(2018)\citenamefont
  {Miniaci}, \citenamefont {Pal}, \citenamefont {Morvan},\ and\ \citenamefont
  {Ruzzene}}]{miniaci_experimental_2018}%
  \BibitemOpen
  \bibfield  {author} {\bibinfo {author} {\bibfnamefont {M.}~\bibnamefont
  {Miniaci}}, \bibinfo {author} {\bibfnamefont {R.}~\bibnamefont {Pal}},
  \bibinfo {author} {\bibfnamefont {B.}~\bibnamefont {Morvan}}, \ and\ \bibinfo
  {author} {\bibfnamefont {M.}~\bibnamefont {Ruzzene}},\ }\href@noop {}
  {\bibfield  {journal} {\bibinfo  {journal} {Physical Review X}\ }\textbf
  {\bibinfo {volume} {8}},\ \bibinfo {pages} {031074} (\bibinfo {year}
  {2018})}\BibitemShut {NoStop}%
\bibitem [{\citenamefont {Jackiw}\ and\ \citenamefont
  {Rebbi}(1976)}]{jackiw_solitons_1976}%
  \BibitemOpen
  \bibfield  {author} {\bibinfo {author} {\bibfnamefont {R.}~\bibnamefont
  {Jackiw}}\ and\ \bibinfo {author} {\bibfnamefont {C.}~\bibnamefont {Rebbi}},\
  }\href {\doibase 10.1103/PhysRevD.13.3398} {\bibfield  {journal} {\bibinfo
  {journal} {Physical Review D}\ }\textbf {\bibinfo {volume} {13}},\ \bibinfo
  {pages} {3398} (\bibinfo {year} {1976})}\BibitemShut {NoStop}%
\bibitem [{\citenamefont {Zhang}\ \emph {et~al.}(2018)\citenamefont {Zhang},
  \citenamefont {Tian}, \citenamefont {Wang}, \citenamefont {Gao},
  \citenamefont {Cheng}, \citenamefont {Liu},\ and\ \citenamefont
  {Christensen}}]{zhang_directional_2018}%
  \BibitemOpen
  \bibfield  {author} {\bibinfo {author} {\bibfnamefont {Z.}~\bibnamefont
  {Zhang}}, \bibinfo {author} {\bibfnamefont {Y.}~\bibnamefont {Tian}},
  \bibinfo {author} {\bibfnamefont {Y.}~\bibnamefont {Wang}}, \bibinfo {author}
  {\bibfnamefont {S.}~\bibnamefont {Gao}}, \bibinfo {author} {\bibfnamefont
  {Y.}~\bibnamefont {Cheng}}, \bibinfo {author} {\bibfnamefont
  {X.}~\bibnamefont {Liu}}, \ and\ \bibinfo {author} {\bibfnamefont
  {J.}~\bibnamefont {Christensen}},\ }\href@noop {} {\bibfield  {journal}
  {\bibinfo  {journal} {Advanced Materials}\ }\textbf {\bibinfo {volume}
  {30}},\ \bibinfo {pages} {1803229} (\bibinfo {year} {2018})}\BibitemShut
  {NoStop}%
\bibitem [{\citenamefont {Charlier}\ \emph {et~al.}(2007)\citenamefont
  {Charlier}, \citenamefont {Blase},\ and\ \citenamefont
  {Roche}}]{charlier_electronic_2007}%
  \BibitemOpen
  \bibfield  {author} {\bibinfo {author} {\bibfnamefont {J.-C.}\ \bibnamefont
  {Charlier}}, \bibinfo {author} {\bibfnamefont {X.}~\bibnamefont {Blase}}, \
  and\ \bibinfo {author} {\bibfnamefont {S.}~\bibnamefont {Roche}},\ }\href
  {\doibase 10.1103/RevModPhys.79.677} {\bibfield  {journal} {\bibinfo
  {journal} {Reviews of Modern Physics}\ }\textbf {\bibinfo {volume} {79}},\
  \bibinfo {pages} {677} (\bibinfo {year} {2007})}\BibitemShut {NoStop}%
\bibitem [{\citenamefont {Akhmerov}\ and\ \citenamefont
  {Beenakker}(2008)}]{akhmerov_boundary_2008}%
  \BibitemOpen
  \bibfield  {author} {\bibinfo {author} {\bibfnamefont {A.~R.}\ \bibnamefont
  {Akhmerov}}\ and\ \bibinfo {author} {\bibfnamefont {C.~W.~J.}\ \bibnamefont
  {Beenakker}},\ }\href {\doibase 10.1103/PhysRevB.77.085423} {\bibfield
  {journal} {\bibinfo  {journal} {Physical Review B}\ }\textbf {\bibinfo
  {volume} {77}},\ \bibinfo {pages} {085423} (\bibinfo {year}
  {2008})}\BibitemShut {NoStop}%
\bibitem [{\citenamefont {Delplace}\ \emph {et~al.}(2011)\citenamefont
  {Delplace}, \citenamefont {Ullmo},\ and\ \citenamefont
  {Montambaux}}]{delplace_zak_2011}%
  \BibitemOpen
  \bibfield  {author} {\bibinfo {author} {\bibfnamefont {P.}~\bibnamefont
  {Delplace}}, \bibinfo {author} {\bibfnamefont {D.}~\bibnamefont {Ullmo}}, \
  and\ \bibinfo {author} {\bibfnamefont {G.}~\bibnamefont {Montambaux}},\
  }\href {\doibase 10.1103/PhysRevB.84.195452} {\bibfield  {journal} {\bibinfo
  {journal} {Physical Review B}\ }\textbf {\bibinfo {volume} {84}},\ \bibinfo
  {pages} {195452} (\bibinfo {year} {2011})}\BibitemShut {NoStop}%
\bibitem [{\citenamefont {Hofstadter}(1976)}]{hofstadter_energy_1976}%
  \BibitemOpen
  \bibfield  {author} {\bibinfo {author} {\bibfnamefont {D.~R.}\ \bibnamefont
  {Hofstadter}},\ }\href {\doibase 10.1103/PhysRevB.14.2239} {\bibfield
  {journal} {\bibinfo  {journal} {Physical Review B}\ }\textbf {\bibinfo
  {volume} {14}},\ \bibinfo {pages} {2239} (\bibinfo {year}
  {1976})}\BibitemShut {NoStop}%
\bibitem [{\citenamefont {Landau}\ and\ \citenamefont
  {Lifshitz}(1981)}]{landau_quantum_1981}%
  \BibitemOpen
  \bibfield  {author} {\bibinfo {author} {\bibfnamefont {L.~D.}\ \bibnamefont
  {Landau}}\ and\ \bibinfo {author} {\bibfnamefont {L.~M.}\ \bibnamefont
  {Lifshitz}},\ }\href {http://www.worldcat.org/isbn/0750635398} {\emph
  {\bibinfo {title} {Quantum {Mechanics} {Non}-{Relativistic} {Theory}, {Third}
  {Edition}: {Volume} 3}}},\ \bibinfo {edition} {3rd}\ ed.\ (\bibinfo
  {publisher} {Butterworth-Heinemann},\ \bibinfo {year} {1981})\BibitemShut
  {NoStop}%
\bibitem [{\citenamefont {Marthaler}\ and\ \citenamefont
  {Dykman}(2007)}]{marthaler_quantum_2007}%
  \BibitemOpen
  \bibfield  {author} {\bibinfo {author} {\bibfnamefont {M.}~\bibnamefont
  {Marthaler}}\ and\ \bibinfo {author} {\bibfnamefont {M.~I.}\ \bibnamefont
  {Dykman}},\ }\href {\doibase 10.1103/PhysRevA.76.010102} {\bibfield
  {journal} {\bibinfo  {journal} {Physical Review A}\ }\textbf {\bibinfo
  {volume} {76}},\ \bibinfo {pages} {010102} (\bibinfo {year}
  {2007})}\BibitemShut {NoStop}%
\bibitem [{\citenamefont {Zhang}\ \emph {et~al.}(2017)\citenamefont {Zhang},
  \citenamefont {Gosner}, \citenamefont {Girvin}, \citenamefont {Ankerhold},\
  and\ \citenamefont {Dykman}}]{zhang_time-translation-symmetry_2017}%
  \BibitemOpen
  \bibfield  {author} {\bibinfo {author} {\bibfnamefont {Y.}~\bibnamefont
  {Zhang}}, \bibinfo {author} {\bibfnamefont {J.}~\bibnamefont {Gosner}},
  \bibinfo {author} {\bibfnamefont {S.~M.}\ \bibnamefont {Girvin}}, \bibinfo
  {author} {\bibfnamefont {J.}~\bibnamefont {Ankerhold}}, \ and\ \bibinfo
  {author} {\bibfnamefont {M.~I.}\ \bibnamefont {Dykman}},\ }\href {\doibase
  10.1103/PhysRevA.96.052124} {\bibfield  {journal} {\bibinfo  {journal}
  {Physical Review A}\ }\textbf {\bibinfo {volume} {96}},\ \bibinfo {pages}
  {052124} (\bibinfo {year} {2017})}\BibitemShut {NoStop}%
\end{thebibliography}%
\clearpage
\onecolumngrid

\appendix

\section{Details of the semi-classical calculation of edge state spectrum neglecting tunneling}
\label{Appendix:details_classical_edge_state}

Here, we show how to calculate the topological edge state spectrum $E_{\bar{k}_s}$, neglecting tunneling. This is a generalization of the Jackiw-Rebbi solution in that it applies to the whole strip BZ  and not only to the vicinity of the high symmetry points.

\subsection{Details of the 1D ansatz}
\label{APP:Bloch_wave_ansatz}

Since the quasi-momentum is defined up to a reciprocal lattice vector, an appropriate solution $\tilde{\boldsymbol{\psi}}_{\bar{k}_s}(\mathbf{k})$ in quasi-momentum space should fulfill the periodic boundary conditions
\begin{equation}
\tilde{\boldsymbol{\psi}}_{\bar{k}_s}(\mathbf{k})=\tilde{\boldsymbol{\psi}}_{\bar{k}_s}(\mathbf{k}+\boldsymbol{b}_1)=\tilde{\boldsymbol{\psi}}_{\bar{k}_s}(\mathbf{k}+\boldsymbol{b}_2).
\end{equation}
where $\boldsymbol{b}_1$ and $\boldsymbol{b}_2$ are two primitive lattice vectors.
Strictly speaking, the simple ansatz Eq.~(\ref{eq:Blochwave_gen}) does not yield such a periodic solution. However, a periodic solution can always be build as a superposition of our solution and other solutions obtained  displacing it by a reciprocal lattice vector.  For an irrational $\alpha$, the formal expression for such   periodic solution is
\begin{equation}
 \tilde{\boldsymbol{\psi}}_{\bar{k}_s}(\mathbf{k})=\sum_{\mathbf{j}}\delta(k_s+(j_1\boldsymbol{b}_{\rm 1}+j_2\boldsymbol{b}_{\rm 2})\cdot\mathbf{e}_s-\bar{k}_s)\tilde{\boldsymbol{\phi}}_{\bar{k}_s}(k_r+(j_1\boldsymbol{b}_{\rm 1}+j_2\boldsymbol{b}_{\rm 2})\cdot\mathbf{e}_r).\label{eq:Blochwave_irrational}
\end{equation} 
where the multi-index $\mathbf{j}=(j_1,j_2)$ has integer components. We note that each term in the sum describes the wavefunction on an infinite line which is parallel to the line ($k_x\cos\varphi+k_y\sin\varphi=\bar{k}_s$) supporting the initial non-periodic solution.
For rational $\alpha$ the periodic solution corresponds to a Bloch wave and can be formally written as
\begin{equation}
 \tilde{\boldsymbol{\psi}}_{\bar{k}_s}(\mathbf{k})=\sum_m\delta(k_s+m\boldsymbol{b}_{\rm st}\cdot\mathbf{e}_s-\bar{k}_s)\tilde{\boldsymbol{\phi}}_{\bar{k}_s}(k_r+m\boldsymbol{b}_{\rm st}\cdot\mathbf{e}_r),\label{eq:Blochwave}
\end{equation} 
where  $\boldsymbol{b}_{\rm st}$  is a reciprocal lattice vector  fulfilling $\boldsymbol{a}_{\rm st}\cdot \boldsymbol{b}_{\rm st}=2\pi$, and $\boldsymbol{\phi}_{\bar{k}_s}(k_r+L_{\rm lp}(\varphi))=\boldsymbol{\phi}_{\bar{k}_s}(k_r)$. In this way, subsequent terms in the sum describe the wavefunction on parallel quasimomentum lines separated by the strip BZ width $2\pi/|\boldsymbol{a}_{\rm st}|$.

\subsection{Calculation of the edge band dispersion as a function of the classical quasi-momentum}

\label{app:edge_state_calculation}

The first step to obtain the semi-classical solution is to expand Eq.~(\ref{eq:H_WKB}) about the 'classical' quasimomentum $\bar{k}=(\bar{k}_s,\bar{k}_r)$. We remind that  $\bar{k}_r$ is chosen to be a local extremum of $|\mathbf{h}(\bar{k}_s,k_r)|$ for fixed $\bar{k}_s$. In other words, we require   that
\begin{equation}\label{eq:bar_k_definition}
\partial_{k_r} |\mathbf{h}(\bar{\mathbf{k}})|=0.    
\end{equation}
Taking into account that
\begin{equation}\label{eq:cond_straight}
\partial_{k_r} |\mathbf{h}|=  \frac{\mathbf{h}\cdot\partial_{k_r} \mathbf{h}}{|\mathbf{h}|},
\end{equation}
We see that $\mathbf{h}(\bar{\mathbf{k}})$ is orthogonal to $\partial_{k_r}\mathbf{h}(\bar{\mathbf{k}})$. If we also define the unit vectors
\begin{equation}
\mathbf{e}_{r,\bar{\mathbf{k}}}=
\left.
\frac{\partial_{k_r}\mathbf{h}}{|\partial_{k_r}\mathbf{h}|}
\right|_{\mathbf{k}=\bar{\mathbf{k}}}
,\quad \mathbf{e}_{s,\bar{\mathbf{k}}}=\mathbf{e}_{r,\bar{\mathbf{k}}}\wedge\mathbf{e}_{z},
\end{equation}
and the rotated Pauli  matrices $\hat{\sigma}_{r/s,\bar{\mathbf{k}}}=\hat{\boldsymbol{\sigma}}\cdot \mathbf{e}_{r/s,\bar{\mathbf{k}}}$, we can write the subleading order expansion of Hamiltonian Eq.~(\ref{eq:H_WKB}) about $\bar{k}$ as
\begin{equation}\label{eq:H_Dirac_WKB}
  \hat{H}\approx m(\hat{Q}_r)\hat{\sigma}_{z}+g(\bar{\mathbf{k}})|\mathbf{h}(\bar{\mathbf{k}})| \hat{\sigma}_{s,\bar{\mathbf{k}}}+(\hat{k}_r-\bar{k}_r)|\partial_{k_r}\mathbf{h}(\bar{\mathbf{k}})|\hat{\sigma}_{r,\bar{\mathbf{k}}}.
\end{equation}
where $g(\mathbf{k})={\rm sign}(\mathbf{e}_{s,\bar{\mathbf{k}}}\cdot\mathbf{h}(\mathbf{k}))$. We note that close to the high-symmetry point $\mathbf{K}$ we recover the Dirac equation substituting $\bar{k}_r\approx 0$, $\mathbf{h}(\mathbf{k})\approx v\bar{k}_s\mathbf{e}_s$, $\bar{k}_r\approx 0$, and $\partial_{k_r}\mathbf{h}(\bar{\mathbf{k}})=v\mathbf{e}_r$. Our more general expression Eq.~(\ref{eq:H_Dirac_WKB}) is similar to the Dirac equation for a straight domain wall in that the dependence on the radial quasi-momentum $k_r$ is linear and at the same time $[\hat{\sigma}_{s,\bar{\mathbf{k}}},\hat{\sigma}_{r,\bar{\mathbf{k}}}]=i\hat{\sigma}_z$ [but, here,  the vector $\mathbf{e}_{s,\bar{\mathbf{k}}}$ ($\mathbf{e}_{r,\bar{\mathbf{k}}}$) is not aligned with (perpendicular to) the domain wall]. The solution is most easily found in position space. Substituting $k_r-\bar{k}_r=-i\lambda d/dQ_r$, and looking for a solution whose pseudo-spin is aligned with the vector $\mathbf{e}_{s,\bar{\mathbf{k}}}$, we find the energy and envelope function,
\begin{equation}
E_{\bar{k}_s}=g(\bar{\mathbf{k}})|\mathbf{h}(\bar{\mathbf{k}})|,\quad\boldsymbol{\Psi}_{\bar{k}_s}= C\begin{pmatrix}e^{-i\varphi_{\bar{k}}/2}\\
e^{i\varphi_{\bar{k}}/2}
\end{pmatrix}
\exp\left[\int_0^{Q_r} \frac{dQ_r'}{\lambda} \frac{m(Q_r')}{|\partial_{k_r}\mathbf{h}(\bar{\mathbf{k}})|}\right] ,\label{eq:JackiewandRebbi}
\end{equation}
where $\varphi_{\bar{k}}$ is the angular coordinate of the vector $\mathbf{e}_{s,\bar{\mathbf{k}}}$. The corresponding edge state wavefunction $\quad\boldsymbol{\psi}_{\bar{k}_s}(\mathbf{Q})$ is obtained by multiplying the envelope by a planewave of quasimomentum $\bar{\mathbf{k}}$,
\begin{equation}\label{eq:straight_dom_WKB}
\boldsymbol{\psi}_{\bar{k}_s}(\mathbf{Q})=\exp[i\mathbf{\bar{k}}\cdot \mathbf{Q}/\lambda]\boldsymbol{\Psi}_{\bar{k}_s}(Q_r).
\end{equation} 
We can also calculate the wavefunction in quasimomentum space either by taking the Fourier transform and evaluating the integral with the steepest descent method or directly in quasi-momentum space by approximating $m(Q_r)$ as linear about the domain wall. Either way we obtain Eq.~(\ref{eq:Blochwave_gen}) with,
\begin{equation}
\tilde{\boldsymbol{\phi}}_{\bar{k}_s}(k_r)=\frac{1}{(8\pi\lambda)^{1/4}\sigma^{1/2}} \begin{pmatrix}e^{-i\varphi_{\bar{k}}/2}\\
e^{i\varphi_{\bar{k}}/2}
\end{pmatrix}  \exp\left[-\frac{(k_r-\bar{k}_r)^2}{4\lambda\sigma^2}\right],\label{eq:JackiewandRebbi_smooth}
\end{equation}
with  $\sigma^2\equiv m_{\rm bk}/(2|\partial_{k_r}\mathbf{h}(\bar{\mathbf{k}})|a)$. This expression is accurate in the region about the 'classical' quasimomentum $\bar{\mathbf{k}}$ but it is not valid for the tails. 

Using Eq.~(\ref{eq:JackiewandRebbi}) one can also compute the speed 
\begin{equation}
v_{\bar{k}_s}=\frac{dE_{\bar{k}_s}}{d\bar{k}_s} =g(\bar{\mathbf{k}})\partial_{k_s}|\mathbf{h}(\bar{\mathbf{k}})|.
\end{equation}
From this equation and the definition  Eq.~(\ref{eq:bar_k_definition}) of the classical quasimomentum  $\bar{\mathbf{k}}$,  we see that  the speed is zero  if and only if $\nabla|\mathbf{h}(\bar{\mathbf{k}})|=0$. This equality  is satisfied only  at the $\mathbf{M}$-points and the $\boldsymbol{\Gamma}$-point.

\subsection{Numerical calculation of the classical quasi-momentum $\bar{k}_r$}
\label{APP:num_calc_classical_quasimom}
We note that to evaluate the edge band spectrum and the underlying normal modes we need to calculate the classical quasi-momentum $\bar{k}_r(\bar{k}_s)$. In general, this can be done only numerically. In practice, we substitute the constraint that the longitudinal quasi-momentum $\bar{k}_s$ is conserved , $k_y=(\bar{k}_s-k_x\cos\varphi)/\sin\varphi$ into 
\begin{equation}\label{eq:h_explicit}
|\mathbf{h}(k_x,k_y)|=J\left\{1+4\cos\left(\frac{\sqrt{3}}{2}k_xa\right)\left[\cos\left(\frac{\sqrt{3}}{2}k_xa\right)+\cos\left(\frac{3}{2}k_ya\right)\right]\right\}^{1/2}
\end{equation}
and and look for a local minimum $\bar{k}_x$. [One can then calculate $\bar{k}_r$ from $\bar{k}_x$ and $\bar{k}_s$.]  We note that this function supports at least two local minima (one for each valley) and can even support an infinite number of minima if  the periods of the two sinusoidal functions in Eq.~(\ref{eq:h_explicit}) are incommensurate. This is the case if $\alpha=\sqrt{3}\cot\varphi$ is an irrational number. [In this case the domain wall configuration is not translationally invariant.] Nevertheless,  one can find a unique periodic solution $E_{\bar{k}_s}$ by following the same minimum as a function of $\bar{k}_s$, see discussion in the main text.

\subsection{Analytical solutions for 'armchair' and 'zig-zag' domain walls}
\label{APP:armchair-zigzag-anal}
Here, we calculate analytically the semi-classical band structure for 'armchair' and  'zig-zag' domain wall orientations. We preliminary note that since six-fold rotations leave the underlying honeycomb lattice invariant, there are six such configurations supporting the same edge band structure (when expressed in terms of the longitudinal quasi-momentum $\bar{k}_s$)

The armchair-orientations correspond to the angles $\varphi=\pi/2+n\pi/3$,
$n\in\mathbb{Z}$. Without loss of generality, we  focus  on the case $\varphi=\pi/2$. In this case, $k_s=k_y$ and $k_r=-k_x$ and
$|h(k_x,\bar{k}_y)|$ has exactly two minima for fixed $\bar{k}_y$ (one for each valley). By
deriving Eq.~(\ref{eq:h_explicit}) with respect of $k_x$, and imposing $\partial_{k_x}|h(k_x,k_y)|=0$ one finds
\begin{equation}
 2 \cos \left(\frac{\sqrt{3}}{2}\bar{k}_xa\right)  +\cos \left(\frac{3}{2}\bar{k}_ya\right) =0.
\end{equation}
Substituting this into Eqs.~(\ref{eq:h_explicit}) and (\ref{eq:En_WKB}) we  arrive at the edge band structure
\begin{equation}\label{eq:band_structure_armchair}
 E_{\bar{k}_y}=J\sin \left(\frac{3}{2}\bar{k}_ya\right).
\end{equation}


Next, we consider the zig-zag orientations $\varphi=n\pi/3$, $n\in\mathbb{Z}$. 
For concreteness  we focus on $\varphi=0$ such that $k_r=k_y$ and $k_s=k_x$. In this cases, $\bar{k}_y=2\pi/(3a)$ independent of  $\bar{k}_x$. By substituting into Eq.~(\ref{eq:En_WKB}), we find
\begin{equation}
E_{\bar{k}_x}=-J+2J\cos \left(\frac{\sqrt{3}}{2}k_xa\right).   
\end{equation}
We note that this solution  corresponds to a minimum of $|h(\bar{k}_x,k_y)|$  only for $-\pi/(\sqrt{3}a)<k_x<\pi/(\sqrt{3}a)$, corresponding to half of its period. At $\bar{k}_x=\pm \pi/(\sqrt{3}a)$ [corresponding to $\bar{\mathbf{k}}=\pi(\pm 1/\sqrt{3},2/3)/a$)], the quasi-momentum localization length   $\lambda^{1/2}\sigma$ diverges because $|\partial_{k_y}\mathbf{h}(\bar{k}_x,\bar{k}_y)|=0$, cf Eq.~(\ref{eq:JackiewandRebbi_smooth}). This indicates a breakdown of the WKB approximation. We note that the  zig-zag strip BZ width is $2\pi/(\sqrt{3}a)$. Thus, our semi-classical solution describes well the topological edge state band across the whole BZ except for the immediate vicinity of the $X$-point.

\subsection{Calculation of the period $T_r$ of the quasi-momentum loop $\mathbf{k}(\cdot)_{\bar{k}_s}$}
\label{APP:period}

Here, we calculate the length $T_r$ of the quasi-momentum  loops $\mathbf{k}(\cdot)_{\bar{k}_s}$ for translationally invariant strips. This allows also to derive  the length $|\boldsymbol{a}_{\rm st}|$ of the strip unit cell,  the width $2\pi/|\boldsymbol{a}_{\rm st}|$ of the strip BZ, and the number of edge bands $N$.

 We start finding a sufficient and necessary condition for the vector $\mathbf{h}(k_s,k_r)$, defined in Eq.~(\ref{eq:h_of_k}), to be a periodic function of $k_r$. We substitute in Eq.~(\ref{eq:h_of_k}), $k_s=k_x\cos\varphi+k_y\sin\varphi$, $k_r=k_y\cos\varphi-k_x\sin\varphi$. We note that  $\mathbf{h}(k_s,k_r)$ is a function of two distinct sinusoidal functions with periods
\begin{equation}
T_{r,1}=\frac{4\pi}{\sqrt{3}a\sin\varphi},\quad T_{r,2}= \frac{4\pi}{3a\cos\varphi}. 
\end{equation}
Thus, $\mathbf{h}(k_s,k_r)$ is a periodic function of $k_r$ if and only if the periods $T_{r,1}$ and $T_{r,2}$ of the two sinusoidal functions are commensurate,
\begin{equation}
   \alpha\equiv \frac{T_{r,1}}{T_{r,2}}=\sqrt{3}\cot\varphi=\frac{p}{q},
\end{equation}
with $p$ and $q$ being two relatively prime integers.
To calculate the overall period $L(\varphi)$ of the function $\mathbf{h}$ one has to distinguish two scenarios. In the first scenario both $q$ and $p$ are odd. In this case, 
the period is
\begin{equation}\label{eq:L_odd}
T_r=q\frac{T_{r,1}}{2}=p\frac{T_{r,2}}{2}=\frac{2\pi q}{\sqrt{3}a\sin\varphi}.
\end{equation}
This fulfills $h(k_s,k_r+L)=h(k_s,k_r+L)$ because  the argument of both sinusoidal functions increase by an odd integer-multiple of $\pi$ and the resulting factors of $-1$ are multiplied and, thus, drop out in Eq.~(\ref{eq:h_of_k}). In the second scenario $q$ or $p$ are even. In this case,  the sign cancellation does not take place because one of the two sinusoidal functions increases by an even multiple of $\pi$. Thus, in this case one finds
\begin{equation}\label{eq:L_even}
T_r=qT_{r,1}=pT_{r,2}=\frac{4\pi q}{\sqrt{3}a\sin\varphi}.
\end{equation}
Equations (\ref{eq:L_odd}) and (\ref{eq:L_even}) can be combined in a single formula
\begin{equation}
T_r=\frac{4\pi q}{[1+(pq\!\!\!\mod 2)]\sqrt{3}a\sin\varphi}.
\end{equation}
The length $|\boldsymbol{a}_{\rm st}|$ of the strip unit cell and  the width $2\pi/|\boldsymbol{a}_{\rm st}|$ of the strip BZ are directly related to $T_r$
\begin{equation}
  \frac{2\pi}{|\boldsymbol{a}_{\rm st}|}
  =\frac{A_{\rm BZ}}{T_r}=\frac{2\pi }{3aq}[1+(pq\!\!\!\mod 2)]\sin\varphi
\end{equation}
where $A_{\rm BZ}=8\pi^2/[a^23^{3/2}]$ is the area of the honeycomb lattice BZ. Dividing  the period of the semi-classical solution $T=4\pi/(3a)\sin\varphi$ by the width of the strip BZ, one finds the number of edge bands $N$,
\begin{equation}
N=\frac{T_rT}{A_{\rm BZ}}=2q/[1+(pq\!\!\!\mod 2)].
\end{equation}

\section{Details of the calculation of the WKB wavefunction}
\label{Appendix:details_WKB_wavefunction}

After calculating the classical quasi-momentum $\bar{k}_s$ and energy $E_{\bar{k}_s}$ in Appendix \ref{app:edge_state_calculation}, here,  we evaluate the tail of the semiclassical wavefunction far away from  $\bar{\mathbf{k}}$. This can be viewed as a preliminary step to calculate the tunneling rate.

We consider the quasimomentum-dependent pseudospin direction $\mathbf{e}_{s,(\bar{k}_s,k_r)}$ defined according to $\mathbf{e}_{s,(\bar{k}_s,k_r)}\cdot \mathbf{h}(\bar{k}_s,k_r)=g(\mathbf{\bar{k}})|h(\bar{\mathbf{k}})|$.
In other words,  the pseudo-spin is rotated to maintain  the projection of $\mathbf{h}(\bar{k}_s,k_r)$ in its direction  equal to $E_{\bar{k}_s}=g(\mathbf{\bar{k}})|h(\bar{\mathbf{k}})|$. 
It is then convenient to rewrite Hamiltonian Eq.~(\ref{eq:H_WKB}) in terms of the Pauli matrices $\hat{\sigma}_{s/r,(\bar{k}_s,k_r)}=\mathbf{e}_{s/r,(\bar{k}_s,k_r)}\cdot \hat{\boldsymbol{\sigma}}$ where $\mathbf{e}_{r,(\bar{k}_s,k_r)}=\mathbf{e}_z\wedge \mathbf{e}_{s,(\bar{k}_s,k_r)}$.
In this way, we generalize Eq.~(\ref{eq:H_Dirac_WKB})
far away from the classical quasi-momentum $\bar{k}$ as
\begin{equation}\label{eq:H_Dirac_WKB_2}
  \hat{H}\approx m(\hat{Q}_r)\hat{\sigma}_{z}+E_{\bar{k}_s} \hat{\sigma}_{s,(\bar{k}_s,k_r)}+\sqrt{|\mathbf{h}(\bar{k}_s,k_r)|^2-E_{\bar{k}_s}^2}\hat{\sigma}_{r,(\bar{k}_s,k_r)}.
\end{equation}
We then apply the WKB ansatz Eq.~(\ref{eq:WKB_ansatz_k}) to the time-independent Schr\"odinger equation choosing the leading order  azimuthal angle $\varphi_{0,\bar{k}_s}(k_r)$  to be the azimuthal angle of the vector $\mathbf{e}_{s,(\bar{k}_s,k_r)}$.
This leads to a simple scalar equation  for the leading order $S_{0,\bar{k}_s}$ of the action,
\begin{equation}\label{eq:Ham-Jacob_expl}
m(\partial_{k_r}S_{0,\bar{k}_s})-i\sqrt{|\mathbf{h}(\bar{k}_s,k_r)|^2-E_{\bar{k}_s}^2} =0.
\end{equation}
We note in passing that this can also be rewritten in the form 
\begin{equation}
   g(\mathbf{\bar{k}}) \sqrt{m(\partial_{k_r}S_{0,\bar{k}_s})^2+|\mathbf{h}(\bar{k}_s,k_r)|^2}=E_{\bar{k}_s}.
\end{equation}
This is the Hamilton-Jacobi equation for the effective classical Hamiltonian Eq.~(\ref{eq:effective_class_Ham}). Solving Eq.~(\ref{eq:Ham-Jacob_expl}), we arrive at the classical action Eq.~(\ref{eq:action_k_space}) with
\begin{equation}\label{eq:Q_of_k_tail_SM}
    Q_r=-ia\arctan\sqrt{
    \frac{|\mathbf{h}(\bar{k}_s,k_r)|^2-E_{\bar{k}_s}^2}{m^2_{\rm bk}}}.
\end{equation}
Here, one has to take the square-root branch that is positive for $k_r-\bar{k}_r>0$.

We note that  the pseudo-spin can rotate to keep the projection of  $\mathbf{h}(\bar{k}_s,k_r)$ constant only as long as $|\mathbf{h}(\bar{k}_s,k_r)|$ remains larger than $|\mathbf{h}(\bar{k}_s,\bar{k}_r)|$. This is possible over the whole quasi-momentum loop only if $\bar{k}_r$ is a global minimum of $|h(\bar{k}_s,k_r)|$ (for fixed $\bar{k}_s$). Even in situations when this is not the case, our solution might still apply to the wavefunction along the tunneling path. In particular, it will always apply to the more direct tunneling paths (leading to the largest tunneling rates) which traverse each valley only once; cf Fig.~3.

\subsection{The WKB wavefunction and band structure at an avoided crossing}
\label{eq:APP_avoided_crossings}

The approximate wavefunctions $\tilde{\boldsymbol{\phi}}_{\bar{k}_s}(k_r)$ calculated using the WKB ansatz Eq.~(\ref{eq:WKB_ansatz_k})  are peaked around the classical quasi-momentum $\bar{k}$ that lies within one of the two valleys. Due to tunneling, the Bloch edge waves $\tilde{\boldsymbol{\phi}}_{n,k}(k_r)$ for the $n$-th edge band are in general a superposition of two time-reversal-partner semi-classical solutions with classical quasi-momentum $\bar{k}_s$ fulfilling $k=\bar{k}_s\!\!\mod(2\pi/|\boldsymbol{a}_{\rm st}|)$. In the semi-classical limit, the admixture of the two semi-classical solution can be large and, thus, will modify the band structure only in a small quasi-momentum region about the strip high-symmetry points $\Gamma$ and $X$.

Exactly at a strip time-symmetric quasimomentum $\Gamma$ of $X$ (corresponding to $k=0$, or $\pi/|\boldsymbol{a}_{\rm st}|)$) the exact Bloch-waves with transverse wavefunction $\tilde{\boldsymbol{\phi}}_{n,k}(k_r)$  can be chosen to be eigenstates of the time-reversal symmetry ${\cal T}$ (because ${\cal T}^2=1$). In position space,  ${\cal T}$ is just the complex conjugation while in reciprocal space it also  change the sign of $\mathbf{k}$. Thus,  the transverse wavefunction $\tilde{\boldsymbol{\phi}}_{n,k}(k_r)$ of a time-reversal symmetric solutions fulfills the constraint
\begin{equation}
 \tilde{\boldsymbol{\phi}}_{n,k}(k_r^{(\rm T)}+k_r) =  \tilde{\boldsymbol{\phi}}^*_{n,k}(k_r^{(\rm T)}-k_r).\label{eq:time-rev-loop}
\end{equation}
where $k_r^{(\rm T)}$ is the transverse component of $\mathbf{k}^{(\rm T)}$ (one of the two  time-reversal symmetric quasimomenta  that lies on the path $\mathbf{k}(\cdot)_{\bar{k}_s}$). We can construct two orthogonal  time-reversal-symmetric  Bloch waves with transverse wavefunctions 
\begin{eqnarray}\label{eq:Bloch_waves_Gamma_X}
&&\tilde{\boldsymbol{\phi}}_{n+1,k}(k_r^{(\rm T)}+k_r)\approx\frac{1}{\sqrt{2}}\left(\tilde{\boldsymbol{\phi}}_{\bar{k}_s}(k_r^{(\rm T)}+k_r)+\tilde{\boldsymbol{\phi}}^{*}_{\bar{k}_s}(k_r^{(\rm T)}-k_r)\right),\nonumber\\
&&\tilde{\boldsymbol{\phi}}_{n,k+\delta k}(k_r^{(\rm T)}+k_r)\approx \frac{i}{\sqrt{2}}\left(\tilde{\boldsymbol{\phi}}_{\bar{k}_s}(k_r^{(\rm T)}+k_r)-\tilde{\boldsymbol{\phi}}^{*}_{\bar{k}_s}(k_r^{(\rm T)}-k_r)\right),
\end{eqnarray}
starting from the transverse wavefunction $\tilde{\boldsymbol{\phi}}_{\bar{k}_s}(k_r)$ of a semi-classical solution.
We note that both approximate solutions $\tilde{\boldsymbol{\psi}}_{n+1/2\pm 1/2,k}(k_r)$  can be viewed as equal superposition of the semi-classical solution $\tilde{\boldsymbol{\psi}}_{\bar{k}_s}(k_r)$ and its time-reversal-partner solution (the second term of each Bloch wave). The time-reversal symmetry does not fix the relative phase of the superposition, nevertheless, one can always cast the Bloch waves in the form Eq.~(\ref{eq:Bloch_waves_Gamma_X}) by appropriately choosing the complex phase   of the normalization constant $C'$ in the WKB ansatz  Eq.~(\ref{eq:WKB_ansatz_k}). We note further that the energy difference between the two Bloch-waves is by definition the tunneling rate $\Delta$.

Using perturbation theory for quasi-degenerate levels one can describe the doublet in the region of the avoided crossing with an effective $2\times2$ Hamiltonian. Using as a basis the two semi-classical time-reversal-partner solutions this effective Hamiltonian reads
\begin{equation}
H_{\delta k} =\begin{pmatrix}
E_{\bar{k}_s}+v_{\bar{k}_s}\delta k & \Delta/2 \\
\Delta/2 & E_{\bar{k}_s}-v_{\bar{k}_s}\delta k,
\end{pmatrix}   
\end{equation}
where $v_{\bar{k}_s}=dE_{\bar{k}_s}/d\bar{k}_s$ and $\delta k$ is the strip quasi-momentum counted off from the high-symmetry point  $k$.
We note that the phase of the off-diagonal matrix element is fixed by Eq.~(\ref{eq:Bloch_waves_Gamma_X}). By diagonalizing this Hamiltonian we obtain the Bloch waves and energy dispersion in the region of the avoided crossings
\begin{eqnarray}
&&\tilde{\boldsymbol{\phi}}_{n+1,k+\delta k}(k_r^{(\rm T)}+k_r)\approx\cos\left(\frac{\Theta}{2}\right)\tilde{\boldsymbol{\phi}}_{\bar{k}_s}(k_r^{(\rm T)}+k_r)+\sin\left(\frac{\Theta}{2}\right)\tilde{\boldsymbol{\phi}}^{*}_{\bar{k}_s}(k_r^{(\rm T)}-k_r),\nonumber\\
&&\tilde{\boldsymbol{\phi}}_{n,k+\delta k}(k_r^{(\rm T)}+k_r)\approx i\sin\left(\frac{\Theta}{2}\right)\tilde{\boldsymbol{\phi}}_{\bar{k}_s}(k_r^{(\rm T)}+k_r)-i\cos\left(\frac{\Theta}{2}\right)\tilde{\boldsymbol{\phi}}^{*}_{\bar{k}_s}(k_r^{(\rm T)}-k_r),\nonumber\\
&&\Theta=\arg(i\Delta/2-v_{\bar{k}_s}\delta\bar{k}_s),\quad E_{n+1/2\pm 1/2,k+\delta k}= E_{\bar{k}_s}\pm\sqrt{(\Delta/2)^2+(v_{\bar{k}_s}\delta k)^2}  .\nonumber\\\label{eq:JR_gen}
\end{eqnarray}

\section{Details of the calculation of the tunneling rate}

\subsection{Details of the calculation of the tunneling path and the tunneling energy}
\label{APP:details_tunn_path}

First, we prove that the paths $\mathbf{k}(\cdot)_{\bar{k}_s}$ that are time-reversal invariant pass at least through a time-reversal invariant high symmetry point $\mathbf{k}^{(\rm T)}$. Preliminary we note that the time-symmetric high symmetry points $\mathbf{M}_i$ are equal to half of a primitive lattice vector $\mathbf{b}_i$, $\mathbf{M}_i=\mathbf{b}_i/2$. Moreover,  the half of any reciprocal lattice vector $\mathbf{b}$ is either a lattice vector and, thus, equivalent to the $\boldsymbol{\Gamma}$-point or is equivalent to half a primitive lattice vector and, thus, to a $\mathbf{M}$-point. Thus, we need to prove that any time-symmetric path $\mathbf{k}(\cdot)_{\bar{k}_s}$ passes through $\mathbf{b}/2$ (the half of a reciprocal lattice vector). By definition the path is time-reversal invariant if for every $\mathbf{k}$ on the path $\mathbf{k}(\cdot)_{\bar{k}_s}$ also $-\mathbf{k}(\cdot)_{\bar{k}_s}$ lies on the same path. Equivalently,  if $\mathbf{k}$ lies on the line $k_x\cos\varphi+k_y\sin\varphi=\bar{k}_s$ (or in vector notation $\mathbf{k}\cdot\mathbf{e}_s=\bar{k}_s$) there is a $\mathbf{b}$ such that $\mathbf{b}-\mathbf{k}$ lies on the same line,  $(\mathbf{b}-\mathbf{k})\cdot\mathbf{e}_s=\bar{k}_s$. 
By summing the two equations and dividing by half we find that also  $\mathbf{b}/2$
lies on the same line, $(\mathbf{b}/2)\cdot\mathbf{e}_s=\bar{k}_s$ as we wanted to
prove. In the same way, we can prove that if the path passes by a time-symmetric high symmetry point it is time-symmetric. In addition, we note that if
$\mathbf{k}(\cdot)_{\bar{k}_s}$ passes through two distinct time-symmetric high-symmetry points then it is a periodic path. The path
$\mathbf{k}(\cdot)_{\bar{k}_s}$ can be periodic only if $\mathbf{h}(k_s,k_r)$ is a periodic function of $k_r$ for rational $\alpha$. Thus, for irrational $\alpha$ every
periodic path can be identified  with a time-reversal-symmetric high-symmetry point $\mathbf{k}^{(\rm T)}$.

Next,  we calculate the longitudinal quasi-momentum $\bar{k}^{(\rm tun)}_s$ for which the tunneling is resonant by requiring that the loop $\bar{k}^{(\rm tun)}_s=k_x\cos\varphi+k_y\sin\varphi$ pass through the relevant high-symmetry point $\mathbf{k}^{(\rm T)}$. For the dominant tunneling transition [corresponding to $\mathbf{k}^{(\rm T)}=\mathbf{M}_1=(0,2\pi/(3a))$], we find $\bar{k}^{(\rm tun)}_s=2\pi/(3a)\sin\varphi$. The $x$-component $\bar{k}^{(\rm tun)}_x(\varphi)$ of the 'classical' quasi-momentum $\bar{\mathbf{k}}_{\mathbf{M}_1,1}^{(\rm tun)}(\varphi)$  is, then, the local minimum of $|\mathbf{h}(k_x,2\pi/(3a)- k_x\coth\varphi)|$ that is closer to $k_x=0$, which we calculate numerically as discussed in Appendix \ref{Appendix:details_classical_edge_state}. From the components  $\bar{k}^{(\rm tun)}_s$ and $\bar{k}^{(\rm tun)}_x$, we  find the classical quasi-momentum  $\bar{\mathbf{k}}_{\mathbf{M}_1,1}^{(\rm tun)}(\varphi)=(\bar{k}^{(\rm tun)}_x,2\pi/3-\bar{k}^{(\rm tun)}_x\cot\varphi)$ and  
the tunneling energy $E_{\mathbf{M}_1,1}^{(\rm tun)}(\varphi)=g(\mathbf{k}^{(\rm tun)})|\mathbf{h}(\mathbf{k}^{(\rm tun)})|$. We note that $E_{\mathbf{M}_1,1}^{(\rm tun)}(\varphi)$ varies monotonically between $-J$ and $J$ in the $\varphi$-interval of length $\pi/3$ between two subsequent zig-zag domain walls, cf Fig.~3(j). Remarkably, the dependence in this interval is very nearly (but not exactly) linear, with an average slope of $6J /\pi\approx  1.91 J$ and a slope for $\varphi=\pi/2$ [corresponding to $E_{\mathbf{M}_1,1}^{(\rm tun)}(\pi/2)=0$] of $J \pi/\sqrt{3}\approx  1.81 J$.
  
\subsection{Calculation of the tunneling rate for the armchair strip}
\label{App:tunneling_armchair}
\begin{figure*}
\begin{center}
\includegraphics[width=1\columnwidth]{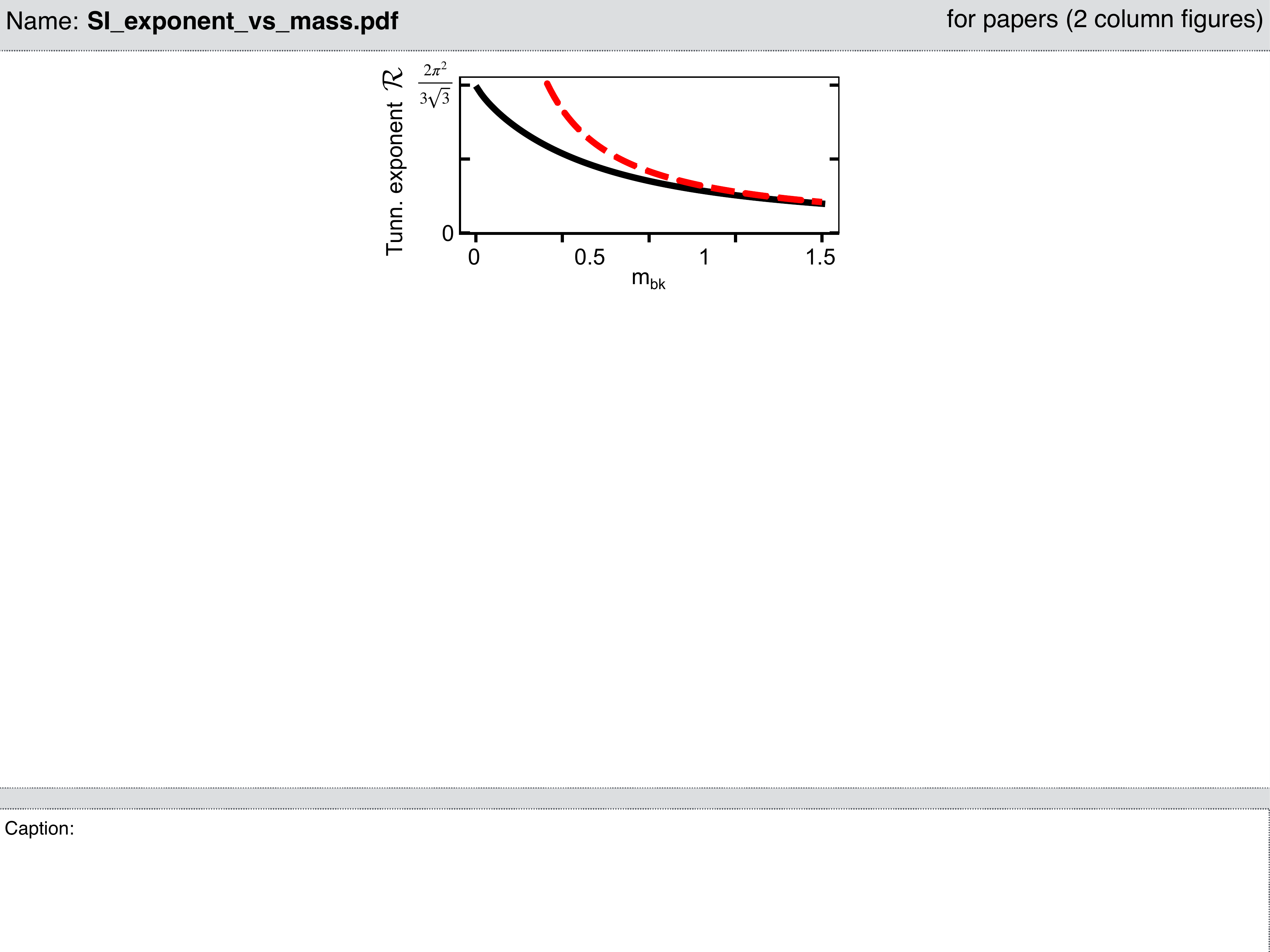}
\caption{
Tunneling exponent $\mathcal{R}$ as a function of the mass parameter $m_{bk}$ for an armchair domain wall. The full WKB result (black line), obtained evaluating numerically the integral in Eq.~(\ref{tunn_expo_eqn}),  is compared to the closed form large mass limit (red-dashed line), cf Eq.~(\ref{exponent_large_mass}). For small mass $m_{bk} \rightarrow 0$, the full result tends to $\mathcal{R}=2\pi^2/(3\sqrt{3})$ as predicted by Eq.~(\ref{exponent_small_mass}).
}
\end{center}
\end{figure*}
Here, we calculate the tunneling rate including the prefactor for an armchair domain wall configuration $\varphi=\pi/2+n\pi/3$.
For concreteness, we consider $\varphi=\pi/2$ (but the final result apply to any armchair configuration). In this case, the system is invariant under two-fold rotations,
\begin{equation}
\hat{\sigma}_xH(-\hat{Q}_x,-\hat{Q}_y,-\hat{k}_x,-\hat{k}_y)\hat{\sigma}_x=H(\hat{Q}_x,\hat{Q}_y,\hat{k}_x,\hat{k}_y).
\end{equation} In addition, the Hamiltonian has a non-local chiral symmetry, in
\begin{equation}
\hat{\sigma}_zH(-\hat{Q}_x,\hat{Q}_y,-\hat{k}_x,\hat{k}_y)\hat{\sigma}_z=-H(\hat{Q}_x,\hat{Q}_y,\hat{k}_x,\hat{k}_y).
\end{equation}   These additional symmetries make it possible to calculate the tunneling rate $\Delta_{\mathbf{M}_1,1}$ including the prefactor as shown below.

As one see from Fig.~3(j), for the armchair domain wall configuration there is a single edge band gap, which correspond to the dominant tunneling pathway $\gamma_{\mathbf{M}_1,1}$ (in the reminder of this section we drop out the indexes ${\mathbf{M}_1,1}$). As one can  read out from the analytical expression Eq.~(\ref{eq:band_structure_armchair})  of the band structure in the neglect of tunneling, the tunneling  energy  is  $E^{(\rm tun)}(\pi/2)=0$ with classical quasi-momentum $\mathbf{\bar{k}}^{(\rm tun)}=\mathbf{K}=2\pi(-3^{-1/2},1)/(3a)$. Thus, the resonant tunneling WKB wavefunction  $\tilde{\boldsymbol{\phi}}(k_x)$   is a solution of Eq.~(\ref{eq:H_WKB}) with $k_y=\bar{k}_y^{(\rm tun)}=2\pi/(3a)$ and $E=E^{(\rm tun)}=0$,
\begin{equation}
\left\{m(\hat{Q}_x)\hat{\sigma}_{z}-J\hat{\sigma}_{x}\left[1-2\cos\left(\frac{\sqrt{3}}{2}\hat{k}_xa\right)\right]\right\}\tilde{\boldsymbol{\phi}}(k_x)=0.
\end{equation}
We note that this equation is invariant under the unitary $\hat{\sigma}_y$. This symmetry is a consequence of the two-fold symmetry, the chiral symmetry and the fact that we are looking for a solution with zero energy. It has the important consequence that the pseudo-spin and orbital degrees of freedom factorize. This allows us to apply the  simpler ansatz,
\begin{equation}\label{eq:WKB_ansatz_armchair}
\tilde{\boldsymbol{\phi}}(k_x)=\frac{1}{\sqrt{2}}\begin{pmatrix}i^{-1/2}\\
i^{1/2}
\end{pmatrix}\tilde{\phi}(k_x),\quad \tilde{\phi}(k_x)=C'\exp[-\frac{i}{\lambda}S_\lambda(k_x)]
\end{equation}
In other words,  we plug in Eq.~(\ref{eq:WKB_ansatz_k}) the Bloch wave angle $\varphi_\lambda(k_x)=\pi/2$ independent of the radial coordinate $k_x$. This symmetry simplifies very much the   calculation of the WKB wavefunction. Up to subleading order we find
\begin{equation}\label{eq:WKB_armchair_tunn}
\tilde{\phi}(k)=
C'\left(-m^{(1)}\right)^{-1/2}
\exp[-i\int_{\bar{k}^{(\rm tun)}_x}^{k_x}Q_x(k'_x)dk'_x/\lambda].
\end{equation}
where $m^{(1)}$ indicates the derivative of $m$ and $\bar{k}^{(\rm tun)}_x=-2\pi/(3\sqrt{3}a)$. For $m(Q)$ as in Eq.~(\ref{eq:sigmoid}) we find 
\begin{eqnarray}
&&m'(Q_x)=-\frac{m_{\rm bk}}{a}{\rm sech}^2\left(\frac{Q_x}{a}\right)\\
&& Q_x=-ia\arctan\left[\frac{J}{m_{{\rm bk}}}\left(2\cos\left(\frac{\sqrt{3}}{2}k_xa\right)-1\right)\right],
\end{eqnarray}
cf  Eq.~(\ref{eq:Q_of_k_tail}) with $E_{\bar{k}_s}=E^{(\rm tun)}=0$ and $\mathbf{h}(\mathbf{k})$ as in  Eq.~(\ref{eq:h_of_k}) with $k_y=2\pi/3$. 
We note that $Q_x(\bar{k}^{(\rm tun)}_x)=0$ for the classical quasi-momentum   $\bar{\mathbf{k}}^{(\rm tun)}=\mathbf{K}$. By expanding about the $\mathbf{K}$-point, we  recover the Gaussian in Eq.~(\ref{eq:JackiewandRebbi_smooth}), here, with $k_r=k_x$, $\bar{k}_r=\bar{k}^{(\rm tun)}_x$, and   $|\partial_{k_r}\mathbf{h}(\bar{\mathbf{k}})|=v=3Ja/2$.   By comparing Eq.~(\ref{eq:JackiewandRebbi_smooth}) and Eq.~(\ref{eq:WKB_ansatz_armchair}) we also find
\begin{equation}
|C'|=\left(\frac{3Jm_{\rm bk}}{2\pi\lambda }\right)^{1/4}.
\end{equation}
We note that  the exact Bloch-waves $ \tilde{\boldsymbol{\phi}}_{S/A}(k_x)$  are eigenstates of the two-fold symmetry,
\begin{eqnarray}\label{eq:two-fold_symmetry}
 \sigma_x\tilde{\boldsymbol{\phi}}_{S/A}(-k_x)=\pm \tilde{\boldsymbol{\phi}}_{S/A}(k_x).
\end{eqnarray}
where $A$ and $S$  labels the symmetric and anti-symmetric Bloch waves, respectively. We denote the corresponding energies as $E_S$ and $E_A$, respectively. We note that because of the chiral symmetry 
\begin{equation}
 E_S=-E_A=\Delta/2.   
\end{equation}
Below, we show that the anti-symmetric state is the lowest energy state and, thus, $\Delta$ is positive consistent with its interpretation as  the tunneling rate (as in the main text). 
In addition, one can fix the global phases of $ \tilde{\boldsymbol{\phi}}_{S}(k_x)$ and $ \tilde{\boldsymbol{\phi}}_{A}(k_x)$  such that they are invariant under the time-reversal symmetry \begin{eqnarray}\label{eq:time-reversal_symmetry_armchair}
 \tilde{\boldsymbol{\phi}}^*_{S/A}(-k_x)=\tilde{\boldsymbol{\phi}}_{S/A}(k_x), 
\end{eqnarray}
and mapped one into the other via the chiral symmetry
\begin{eqnarray}\label{eq:chiral_symmetry}
\tilde{\boldsymbol{\phi}}_{S/A}(k_x)=  \hat{\sigma}_z\tilde{\boldsymbol{\phi}}_{A/S}(-k_x). 
\end{eqnarray}
Next, we want to find an approximate expression for $\tilde{\boldsymbol{\phi}}_{S/A}(k_x)$ in terms of the WKB wavefunction $\tilde{\boldsymbol{\phi}}(k_x)$. We can enforce the time-reversal symmetry Eq.~(\ref{eq:time-reversal_symmetry_armchair}) using Eq.~(\ref{eq:Bloch_waves_Gamma_X}), here, with  $k^{(\rm T)}_r=0$,
\begin{eqnarray}\label{eq:Bloch_waves_armchair}
&&\tilde{\boldsymbol{\phi}}_{S}(k_x)\approx\frac{1}{\sqrt{2}}\left(\tilde{\boldsymbol{\phi}}(k_x)+\tilde{\boldsymbol{\phi}}^{*}(-k_x)\right),\nonumber\\
&&\tilde{\boldsymbol{\phi}}_{A}(k_x)\approx\frac{i}{\sqrt{2}}\left(\tilde{\boldsymbol{\phi}}(k_x)-\tilde{\boldsymbol{\phi}}^{*}(-k_x)\right).
\end{eqnarray}
In order to fulfill also Eqs.~(\ref{eq:two-fold_symmetry}) and (\ref{eq:time-reversal_symmetry_armchair}) we need to fix the global phase of $C'$, $C'=|C'|$.

Next, we adopt to our problem a strategy invented by Landau  to calculate the tunneling rate for a double well potential \cite{landau_quantum_1981}. We apply the Schr\"odinger Equation to $\tilde{\boldsymbol{\phi}}_{S}(k)$, multiply on the left-hand side by $\tilde{\boldsymbol{\phi}}^{*}(k)$ and integrate over half of the strip Brillouin zone to obtain 
\begin{equation}\int_{-L/2}^{0}dk_x\tilde{\boldsymbol{\phi}}^{*}(k_x)E_{S}\tilde{\boldsymbol{\phi}}_{S}(k_x)=\int_{-L/2}^{0}dk\tilde{\boldsymbol{\phi}}^{*}(k_x)\left[m\left(i\lambda\frac{d}{dk_x}\right)\hat{\sigma}_{z}-J\hat{\sigma}_{x}\left(1-2\cos\left(\frac{\sqrt{3}}{2}k_xa\right)\right)\right]\tilde{\boldsymbol{\phi}}_{S}(k_x).\label{eq:int_sch_0}
\end{equation}
We note that $k_x=0$ ($-L/2=0$) corresponds to the $\mathbf{M}_1$-point ($\boldsymbol{\Gamma}$-point).
Taking into account Eq.~(\ref{eq:Bloch_waves_armchair}) and that the semi-classical solution is normalized we find
\begin{equation}
\int_{-L/2}^{0}dk\tilde{\boldsymbol{\phi}}^{*}(k_x)\tilde{\boldsymbol{\phi}}_{R}(k_x)=\frac{1}{\sqrt{2}}.
\end{equation}
Plugging this equation together with Eqs.~(\ref{eq:two-fold_symmetry}) and (\ref{eq:chiral_symmetry}) into Eq.~(\ref{eq:int_sch_0}) we find
\begin{equation}\label{eq:int_sch_1}
\frac{E_{S}}{\sqrt{2}}=\frac{\Delta}{2^{3/2}}=\int_{-L/2}^{0}dk_x\tilde{\boldsymbol{\phi}}^{*}(k_x)m\left(i\lambda\frac{d}{dk_x}\right)\tilde{\boldsymbol{\phi}}_{A}(-\hat{k}_x)-J\left(1-2\cos\left(\frac{\sqrt{3}}{2}k_xa\right)\right)\tilde{\boldsymbol{\phi}}^{*}(k_x)\tilde{\boldsymbol{\phi}}_{S}(-k_x).
\end{equation}

Likewise, when we apply the Schr\"odinger equation to $\tilde{\boldsymbol{\phi}}(k_x)$, multiply on the left-hand side by $\tilde{\boldsymbol{\phi}}_{S}^{*}(k_x)$ integrate and take the complex conjugate we obtain
\begin{equation}
0=\int_{-L/2}^{0}dk_x\tilde{\boldsymbol{\phi}}_{A}(-k_x)m\left(-i\lambda\frac{d}{dk_x}\right)\tilde{\boldsymbol{\phi}}^*(k_x)-J\left(1-2\cos\left(\frac{\sqrt{3}}{2}k_xa\right)\right)\tilde{\boldsymbol{\phi}}_{S}(-k_x)\tilde{\boldsymbol{\phi}}^{*}(k_x).  
\end{equation}
By substracting   this equation  to Eq.(\ref{eq:int_sch_1}), we find
\begin{equation}
\Delta=2^{3/2}\int_{-L/2}^{0}dk_x\left(\tilde{\boldsymbol{\phi}}^{*}(k_x)m\left(i\lambda\frac{d}{dk_x}\right)\tilde{\boldsymbol{\phi}}_{A}(-k_x)-\tilde{\boldsymbol{\phi}}_{A}(-k_x)m\left(-i\lambda\frac{d}{dk_x}\right)\tilde{\boldsymbol{\phi}}^*(k_x)\right).
\end{equation}
Next, we plug the Taylor expansion $m(Q_x)=\sum_{n=\rm odd} m^{(n)}(0)Q_x^n/n!$  inside the integral. The sum is over odd integers because $m(Q_x)$ is an odd function. Using   also $m(-Q)=-m(-Q)$ we find
\begin{equation}
\Delta=2^{3/2}\sum_{n=\rm odd}\frac{m^{(n)}(0)}{n!}(i\lambda)^n\int_{-L/2}^{0}dk_x\left(\tilde{\boldsymbol{\phi}}^*(k_x) \partial_{k_x}^{n}\tilde{\boldsymbol{\phi}}_{A}(-k_x)+\tilde{\boldsymbol{\phi}}_{A}(-k_x)\partial_{k_x}^{n}\tilde{\boldsymbol{\phi}}^*(k_x)\right).  
\end{equation}
For each term in the sum, we get rid of the integral by integrating $n$-times by part, 
\begin{equation}
\Delta=2^{3/2}\sum_{n=\rm odd}\frac{m^{(n)}(0)}{n!}(i\lambda)^n\sum_{l=1}^n\left.(-)^{l-1}\left(\partial_{k_x}^{l-1}\tilde{\boldsymbol{\phi}}_A(-k_x) \partial_{k_x}^{n-l}\tilde{\boldsymbol{\phi}}^*(k_x)\right)\right|_{-L/2}^{0}.  
\end{equation}
Finally by plugging Eqs.~(\ref{eq:WKB_armchair_tunn}) and (\ref{eq:Bloch_waves_armchair}), calculating the derivatives, and  keeping only the exponentially-larger boundary terms at $k_x=0$ we find
\begin{equation}
\label{tunn_expo_eqn}
\Delta=2\lambda\left(\sum_{n=\rm odd}\frac{m^{(n)}(0)}{(n-1)!}Q_x^{n-1}\right)\frac{|C'|^2}{m^{(1)}(Q_x)}e^{-{\cal R}/\lambda},\quad {\cal R}=2i\int_{\bar{k}^{(\rm tun)}_x}^{0}Q_x(k'_x)dk'_x.  
\end{equation}
We note that $m^{(n)}(0)Q_x^{n-1}/(n-1)!=m^{(1)}(Q_x)$ and, thus, we arrive to the simple expression
\begin{equation}
\Delta=2\lambda|C'|^2e^{-{\cal R}/\lambda}=\left(\frac{6Jm_{\rm bk}\lambda}{\pi }\right)^{1/2}e^{-{\cal R}/\lambda}.  
\end{equation}

For the limit of large mass $m_{\rm bk}\gg J$,  we can approximate the imaginary position $Q_x$ as  
\begin{equation}
 Q_x \approx-i\frac{aJ}{m_{{\rm bk}}}\left(2\cos\left(\frac{\sqrt{3}}{2}ka\right)-1\right)  
\end{equation}
and evaluate the classical action along the tunneling path analytically to find
\begin{equation}
\label{exponent_large_mass}
 {\cal R}\approx \frac{4J}{m_{\rm bk}}\left(1-\frac{\pi}{3\sqrt{3}} \right). 
\end{equation}
For the limit of small mass, $Q_x$ can be approximated as a step funtion with $Q_x=-ia\pi/2$ on the tunneling path. With this approximation we find 
\begin{equation}
\label{exponent_small_mass}
{\cal R}\approx \pi a|\bar{k}^{(\rm tun)}_x|=\frac{2\pi^2}{3\sqrt{3}}.
\end{equation}

\section{WKB edge-state solution for a  closed domain wall}
\label{app:WKB_closed}
\begin{figure*}
\begin{center}
\includegraphics[width=0.5\columnwidth]{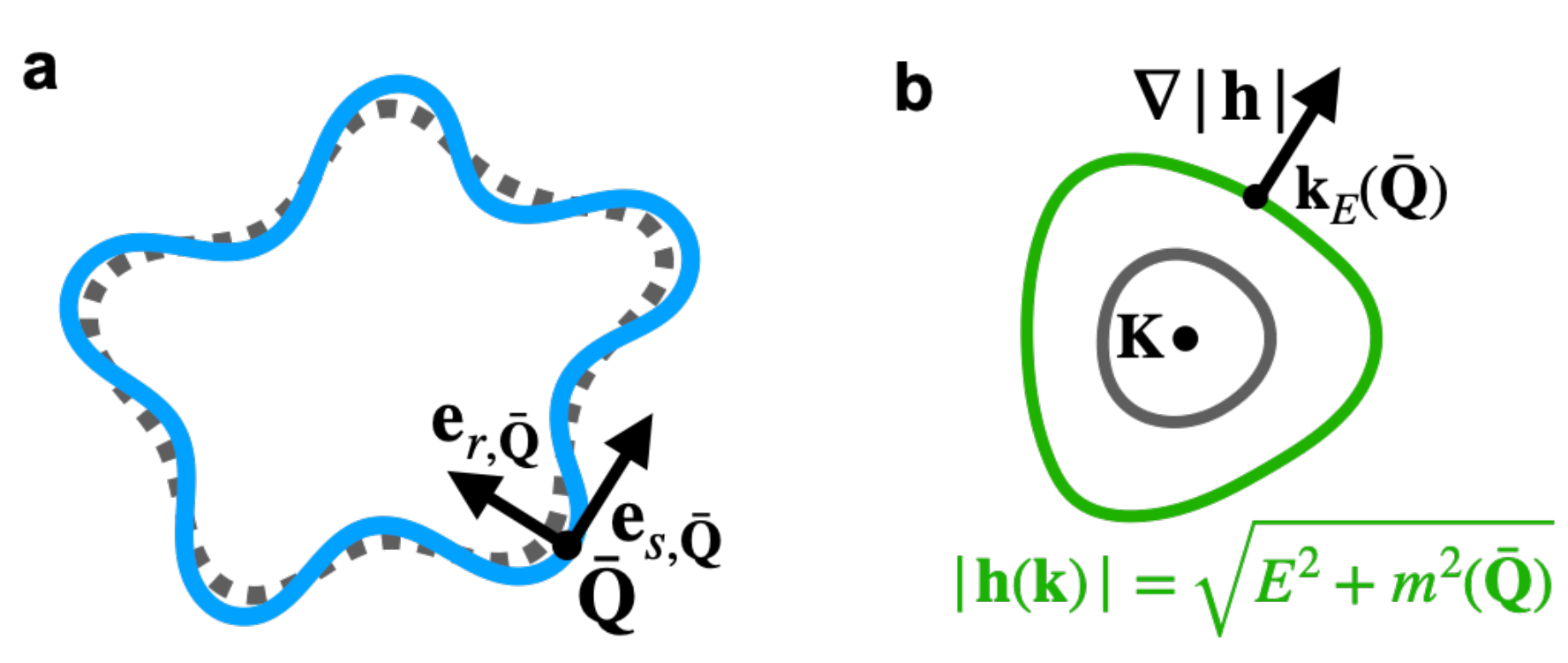}
\caption{
(a) Sketch of the classically accessible path.  In the presence of a finite curvature,  the classical trajectory (blue line) overshoots the domain wall (grey dashed line). (b) Contour plot of the function $|\mathbf{h}(\mathbf{k})|$ about the $\mathbf{K}$-point. Also shown is the classical 
  quasi-momentum $\mathbf{k}_E(\bar{\mathbf{Q}})$  for $E>0$ and the position $\bar{\mathbf{Q}}$ marked with a dot in (a). 
}
\end{center}
\end{figure*}
In this section we calculate the edge state spectrum  and WKB wavefunction for a closed smooth domain wall (neglecting tunneling). This results generalize the Jackiew and Rebbi solution including also the effects of a finite  curvature of the domain wall.  We show that the classical trajectory of an edge state wavepacket does not exactly follow the domain wall but rather tends to overshoot it. Also the wavepacket acquires a 'finite mass' that modifies it propagation speed.

Here, we  use the the WKB ansatz in position space
\begin{equation}\label{eq:WKB_ansatz_pos_0}
\boldsymbol{\psi}(\mathbf{Q})=C\begin{pmatrix}\cos(\theta_\lambda/2)e^{-i\varphi_\lambda(\mathbf{Q})/2}\\
\sin(\theta_\lambda/2)e^{i\varphi_\lambda(\mathbf{Q})/2}
\end{pmatrix}\exp\left[\frac{i}{\lambda}S_\lambda(\mathbf{Q})\right],
\end{equation}
with
\begin{equation}
\quad S_\lambda(\mathbf{Q})=\sum_{n=0}^{\infty}S_n(\mathbf{Q})\lambda^n,   
\end{equation}
and likewise for the Bloch sphere angles $\varphi_\lambda(\mathbf{Q})$ and $\theta_\lambda(\mathbf{Q})$. 

As usual for the WKB approach we seek to solve the time-independent Schr\"odinger equation (\ref{eq:H_WKB}) order by order in $\lambda$. We note that only the terms where the  derivative is applied to  the classical action  $S_0(\mathbf{Q})$ are independent of $\lambda$. Thus, up to leading order in $\lambda$ we arrive to the matrix Hamilton-Jacobi equation
\begin{equation}
\left[m(\mathbf{Q})\hat{\sigma}_{z}+\mathbf{h}(\boldsymbol{\nabla}S_0(\mathbf{Q}))\cdot\hat{\boldsymbol{\sigma}}\right]
\begin{pmatrix}\cos(\theta_0/2)e^{-i\varphi_0(\mathbf{Q})/2}\\
\sin(\theta_0/2)e^{i\varphi_0(\mathbf{Q})/2}
\end{pmatrix}
=E\begin{pmatrix}\cos(\theta_0/2)e^{-i\varphi_0(\mathbf{Q})/2}\\
\sin(\theta_0/2)e^{i\varphi_0(\mathbf{Q})/2}
\end{pmatrix}.
\label{eq:Hamilton_Jacobi_pos_space}  
\end{equation}
We then identify $\boldsymbol{\nabla}S_0(\mathbf{Q})$ with the classical quasi-momentum, $\mathbf{k}_E(\mathbf{Q})\equiv\boldsymbol{\nabla}S_0(\mathbf{Q})$, and formally write
\begin{equation}\label{eq:WKB_ansatz_2D}
\boldsymbol{\psi}(\mathbf{Q})\approx C\begin{pmatrix}\cos(\theta_0/2)e^{-i\varphi_0(\mathbf{Q})/2}\\
\sin(\theta_0/2)e^{i\varphi_0(\mathbf{Q})/2}
\end{pmatrix}\exp\left[\frac{i}{\lambda}\int_{{\cal C}_\mathbf{Q}}\mathbf{k}_E(\mathbf{Q}')\cdot d\mathbf{Q'}\right],  
\end{equation}
where ${\cal C}_\mathbf{Q}$ is a line connecting  a fixed reference point 
to $\mathbf{Q}$. 

Below, we show that one of the effects of a finite curvature $R$ is to slightly displace the edge state position away from the domain wall (as defined by the condition $m(\mathbf{Q})=0$). In anticipation of this we look for a solution of Eq.~(\ref{eq:Hamilton_Jacobi_pos_space}) with a real quasi-momentum $\mathbf{k}$ at position $\bar{\mathbf{Q}}$ away from the domain wall, such that  $m(\bar{\mathbf{Q}})\neq 0$.  The set of positions $\bar{\mathbf{Q}}$ with real quasi-momenta $\mathbf{k}$ will form a classically accessible closed path that is  to be determined in the course of our calculation. For large radius of curvature and/or small energy $|E|$ it will remain close to the domain wall. In general, we will only assume that the tangent to the classical path is orthogonal to the gradient of the mass function $m(\mathbf{Q})$.
From Eq.~(\ref{eq:Hamilton_Jacobi_pos_space}) and the condition that $\mathbf{k}_E$ should be real on the classical path, we find
\begin{equation}\label{eq:cond_dom_wall_1}
\sqrt{m^2(\bar{\mathbf{Q}})+ |\mathbf{h}(\mathbf{k}_E(\bar{\mathbf{Q}}))|^2}=|E|,
\end{equation}
with
\begin{equation}\label{eq:theta_of_Q}
\theta_0=\arg \left({\rm sign}(E)m(\bar{\mathbf{Q}})+i|\mathbf{h}(\mathbf{k}_E(\bar{\mathbf{Q}}))|\right),
\end{equation}
and $\varphi_0-\pi(1-{\rm sign}(E))/2$ being the angular coordinate of $\mathbf{h}$.
Inspired by the special solution for a straight domain wall (that can be viewed as the limit of zero curvature of a closed domain wall) we also restrict our ansatz, requiring  that the quasi-momentum $\mathbf{k}_E$ at a classical accessible position $\bar{\mathbf{Q}}$ obeys the additional constraint
\begin{equation}\label{eq:cond_dom_wall_2}
 \mathbf{e}_{r,\bar{\mathbf{Q}}}\wedge\left.\frac{\nabla_\mathbf{k}{|\mathbf{h}|}}{|\nabla_\mathbf{k}{|\mathbf{h}|}|}\right|_{\mathbf{k}=\mathbf{k}_E(\bar{\mathbf{Q}})}={\rm sign}(E)\mathbf{e}_z,\quad \mathbf{e}_{r,\mathbf{Q}}=-\frac{\nabla_\mathbf{Q}m}{| \nabla_\mathbf{Q}m|}. 
\end{equation}

This is a natural generalization of Eq.~(\ref{eq:cond_straight}) with the radial direction being determined by the direction of  the gradient of the mass function. 
We note that for $\mathbf{h}(\mathbf{k})$ as given by Eq.~(\ref{eq:h_of_k}), a quasimomentum $\mathbf{k}$ on the contour $|\mathbf{h}(\mathbf{k})|=\sqrt{|E|^2-m^2(\bar{\mathbf{Q}})}<J$ is uniquely identified by the direction of the gradient $\nabla_\mathbf{k}{|\mathbf{h}|}$, cf Fig.~7.
Thus, Eqs.~(\ref{eq:cond_dom_wall_1},\ref{eq:cond_dom_wall_2}) uniquely identify $\mathbf{k}_E$ for a fixed $\bar{\mathbf{Q}}$. In order to find a complete solution we need to find  the classically accessible path and to calculate the quasi-momentum $\mathbf{k}_E$  in the vicinity of  this path. The two problems are related because the wavefunctions has to fall off going away from the classically accessible path.

We parametrize the classical path with the arclength $Q_s$ counted off from a reference point on the path. We  the unit vector tangent to the classical path $\mathbf{e}_{s,Q_s}=\mathbf{e}_{r,\bar{\mathbf{Q}}(Q_s)}\wedge\mathbf{e}_z$. We also define the  angle $\varphi$ to be the azimuthal angle for the vector $\mathbf{e}_{s,Q_s}$. This allows to define the (rescaled) radius of curvature of the classical path $R_{\rm rs}(Q_s)=1/(\frac{d\varphi}{dQ_s})$. Likewise we denote as $\tilde{R}(Q_s)$ the radius of curvature  for the corresponding path $\mathbf{k}_{E}(Q_s)$ in reciprocal space   (which lies on the contour $|h(\mathbf{k})|=\sqrt{|E|^2-m^2(Q_s)}<J$). Since, we look for a classical path that is orthogonal to the gradient of $m(\mathbf{Q})$,
from Eq.~(\ref{eq:cond_dom_wall_1}), it follows that
\begin{equation}
    \nabla_\mathbf{k}|\mathbf{h}|\cdot\frac{d\mathbf{k}_E}{dQ_s}=0.
\end{equation}
From the above equation and Eq.~(\ref{eq:cond_dom_wall_2}) it directly follows
\begin{equation}
\frac{dk_{E,s}}{dQ_s}=0,\quad  k_{E,s}=\mathbf{k}_E\cdot \mathbf{e}_{s,Q_s}
\end{equation}
In addition, for Eq.~(\ref{eq:cond_dom_wall_2}) to be valid along the whole classical path  $\mathbf{Q}(Q_s)$  the change of azimuthal angle $d\varphi$  should be the same for  both $\mathbf{Q}(Q_s)$ and $\mathbf{k}_E(Q_s)$ and, thus,
\begin{equation}\label{eq:dkr/dqs}
\frac{dk_{E,r}}{dQ_s}=\frac{\tilde{R}}{R_{\rm rs}}.   
\end{equation}
Next we require that Eq.~(\ref{eq:Hamilton_Jacobi_pos_space}) also holds  away from the contour. We introduce the coordinate $Q_r$ orthogonal to the classically accessible path, $Q_r=(\mathbf{Q}-\bar{\mathbf{Q}})\cdot\mathbf{e}_{r,Q_s}$ where $\bar{\mathbf{Q}}$ is the position of the classical path that is closest to $\mathbf{Q}$. By expanding Eq.~(\ref{eq:cond_dom_wall_2}) about $\bar{\mathbf{Q}}(Q_s)$ we find
\begin{eqnarray}
&&\left[(m(\bar{\mathbf{Q}})-|\boldsymbol{\nabla}_\mathbf{Q}m|dQ_r)\hat{\sigma}_{z}+\hat{\boldsymbol{\sigma}}\cdot\left[\mathbf{h}+dQ_r\left(\frac{dk_{E,r}}{dQ_r}\partial_{k_r}\mathbf{h}+\frac{dk_{E,s}}{dQ_r}\partial_{k_s}\mathbf{h}\right)\right]_{\mathbf{k}_E(Q_s)}\right]
\begin{pmatrix}\cos(\theta_0/2)e^{-i\varphi_0/2}\\
\sin(\theta_0/2)e^{i\varphi_0/2}
\end{pmatrix}\nonumber\\
&&=E
\begin{pmatrix}\cos(\theta_0/2)e^{-i\varphi_0/2}\\
\sin(\theta_0/2)e^{i\varphi_0/2}
\end{pmatrix}.
\label{eq:Hamilton_Jacobi_pos_space_2}  
\end{eqnarray}
It is convenient  to rewrite the psudospin vector in Eq.~(\ref{eq:Hamilton_Jacobi_pos_space_2})
in terms of the eigenstates 
\begin{equation}
\mathbf{V}_{\pm}=\frac{1}{\sqrt{2}} \begin{pmatrix}e^{-i\varphi_0/2}\\
\pm e^{i\varphi_0/2}
\end{pmatrix}.   
\end{equation}
of $\hat{\boldsymbol{\sigma}}\cdot\mathbf{h}(\mathbf{k}_E(Q_s))$,
\begin{eqnarray}
\begin{pmatrix}\cos(\theta_0/2)e^{-i\varphi_0/2}\\
\sin(\theta_0/2)e^{i\varphi_0/2}
\end{pmatrix}=\cos(\tilde{\theta}/2)\mathbf{V}_+-\sin(\tilde{\theta}/2)\mathbf{V}_-,\nonumber\\
\hat{\boldsymbol{\sigma}}\cdot\mathbf{h}(\mathbf{k}_E(Q_s))\mathbf{V}_{\pm}=\pm {\rm sign}(E)\sqrt{E^2-m^2(Q_s)}\mathbf{V}_{\pm}.
\end{eqnarray}
Here,  $\tilde{\theta}=\theta_0-\pi/2$ is the polar angle counted off from the equator of the Bloch sphere. Grouping all the terms in Eq.~(\ref{eq:Hamilton_Jacobi_pos_space_2}) that are proportional to  $dQ_r$ into two separated groups containing  the terms proportional to either the vector $\mathbf{V}_{+}$ or $\mathbf{V}_{-}$ (the remaining terms drop out),  and requiring that the terms in each group  add up to zero  we obtain two  scalar complex algebraic equations,
\begin{eqnarray}
\sin\left(\frac{\tilde{\theta}}{2}\right)|\boldsymbol{\nabla}_\mathbf{Q}m|+\cos\left(\frac{\tilde{\theta}}{2}\right){\rm sign}(E)\frac{dk_{E,s}}{dQ_r}\partial_{k_s}|\mathbf{h}|-i\sin\left(\frac{\tilde{\theta}}{2}\right){\rm sign}(E)\frac{(\mathbf{e}_z\wedge\mathbf{h})}{|\mathbf{h}|}\cdot\left(\frac{dk_{E,s}}{dQ_r}\partial_{k_s}\mathbf{h}+\frac{dk_{E,r}}{dQ_r}\partial_{k_r}\mathbf{h}\right)=0\nonumber\\
-\cos\left(\frac{\tilde{\theta}}{2}\right)|\boldsymbol{\nabla}_\mathbf{Q}m|+\sin\left(\frac{\tilde{\theta}}{2}\right){\rm sign}(E)\frac{dk_{E,s}}{dQ_r}\partial_{k_s}|\mathbf{h}|-i\cos\left(\frac{\tilde{\theta}}{2}\right){\rm sign}(E)\frac{(\mathbf{e}_z\wedge\mathbf{h})}{|\mathbf{h}|}\cdot\left(\frac{dk_{E,s}}{dQ_r}\partial_{k_s}\mathbf{h}+\frac{dk_{E,r}}{dQ_r}\partial_{k_r}\mathbf{h}\right)=0.\nonumber\\\label{eq:Hamilton_Jacobi_4}
\end{eqnarray}
Taking into account Eqs.~(\ref{eq:WKB_ansatz_pos_0},\ref{eq:WKB_ansatz_2D} ,\ref{eq:dkr/dqs}) we find
\begin{equation}
\frac{dk_{E,s}}{dQ_r}=\frac{d^2S_0}{dQ_rdQ_s}=\frac{dk_{E,r}}{dQ_s}=\frac{\tilde{R}}{R_{\rm rs}}.
\end{equation}
Substituting into Eqs.~(\ref{eq:Hamilton_Jacobi_4}) we find
\begin{equation}
{\rm Re}\left[\frac{dk_{E,r}}{dQ_r}\right]=\frac{(\mathbf{e}_z\wedge\mathbf{h})\cdot\partial_{k_s}\mathbf{h}}{(\mathbf{e}_z\wedge\mathbf{h})\cdot\partial_{k_r}\mathbf{h}}\frac{\tilde{R}}{R_{\rm rs}}
\end{equation}
and are left with two real equations of two independent variables
\begin{eqnarray}
\sin\left(\frac{\tilde{\theta}}{2}\right)|\boldsymbol{\nabla}_\mathbf{Q}m|+\cos\left(\frac{\tilde{\theta}}{2}\right){\rm sign}(E)\frac{\tilde{R}}{R_{\rm rs}}\partial_{k_s}|\mathbf{h}|+\sin\left(\frac{\tilde{\theta}}{2}\right){\rm sign}(E)\frac{(\mathbf{e}_z\wedge\mathbf{h})\cdot\partial_{k_r}\mathbf{h}}{|\mathbf{h}|}{\rm Im}\left[\frac{dk_{E,r}}{dQ_r}\right]=0\\
-\cos\left(\frac{\tilde{\theta}}{2}\right)|\boldsymbol{\nabla}_\mathbf{Q}m|+\sin\left(\frac{\tilde{\theta}}{2}\right){\rm sign}(E)\frac{\tilde{R}}{R_{\rm rs}}\partial_{k_s}|\mathbf{h}|+\cos\left(\frac{\tilde{\theta}}{2}\right){\rm sign}(E)\frac{(\mathbf{e}_z\wedge\mathbf{h})\cdot\partial_{k_r}\mathbf{h}}{|\mathbf{h}|}{\rm Im}\left[\frac{dk_{E,r}}{dQ_r}\right]=0
.\label{eq:Hamilton_Jacobi_6}
\end{eqnarray}
For large radius of curvature $R$ this equation can be solved approximating the radius of curvature $R$ and the tangential vector $\mathbf{e}_{r,Q_s}$ of the classical path with the one of the domain wall to calculate the quasi-momentum on the classical path $\mathbf{k}_E$ and allowing to calculate $\tilde{\omega}$ and $dk_{E,r}/dQ_r$.  Outside the perurbative regime the equation can be solved iteratively  using $\tilde{\theta}(Q_s)$  as calculated with the perturbative procedure to calculate the displacement $\bar{Q}_r$ of the classical path from the domain wall and use  $R$ and the tangential vector $\mathbf{e}_{r,Q_s}$ taking into account this displacement to start a new iterative step. 

If we consider a small energy such that the Hamiltonian can be well approximated with the Dirac equation the perturbative solution can be found in closed form. In this case, $\mathbf{h}(Q_s)=v\mathbf{k}_E(Q_s)$, $\mathbf{k}_E(Q_s)=E/v\mathbf{e}_{s,Q_s}$, $\partial_{k_r}\mathbf{h}=v\mathbf{e}_{r,Q_s}$ $\tilde{R}\approx |E|/v=2|E|/(3Ja)$, $\partial_{k_s}|\mathbf{h}|\approx v$, $(\mathbf{e}_z\wedge\mathbf{h})\cdot\partial_{k_r}\mathbf{h}/|\mathbf{h}|\approx v$, $|\boldsymbol{\nabla}_\mathbf{Q}m|\approx m_{\rm bk}/a$, $\sin(\tilde{\theta}/2)\approx\tilde{\theta}/2$, $\cos(\tilde{\theta}/2)\approx 1$, $\tilde{\theta}\approx m_{\rm bk}\bar{Q}_r/(Ea)$ and obtain
\begin{eqnarray}\label{eq:Hamilton_Jacobi_5}
\frac{m_{\rm bk}^2\bar{Q}_r}{2Ea^2}+\frac{E}{R_{\rm rs}}+\frac{m_{\rm bk}\bar{Q}_r}{2Ea}v{\rm Im}\left[\frac{dk_{E,r}}{dQ_r}\right]=0\\
-\frac{m_{\rm bk}}{a}+\frac{m_{\rm bk}\bar{Q}_r}{2R_{\rm rs}a}+v{\rm Im}\left[\frac{dk_{E,r}}{dQ_r}\right]=0
.
\end{eqnarray}
This can be solved to obtain
\begin{equation}
 {\rm Im}\left[\frac{dk_{E,r}}{dQ_r}\right]=\frac{m_{\rm bk}}{av}\left(1+\frac{E^2a^2}{2R_{\rm rs}^2m_{\rm bk}^2}\right),\quad   
 \bar{Q}_r=-\frac{E^2a^2}{m^2_{\rm bk}R_{\rm rs}}.
\end{equation}
We note that in the limit of large radius we recover the result for a straight domain wall both for the quasi-momentum in the vicinity  of the classical path (which now coincide with the domain wall). The presence of a finite curvature tends to broaden the wavepacket and displaces the classical trajectory to overshoot the domain wall, see sketch in Fig.~7.

By enforcing periodic boundary condition we find the Bohr-Sommerfeld quantization condition
\begin{equation}
\oint \mathbf{k}_E(\mathbf{Q}')\cdot d\mathbf{Q'}=\lambda(2\pi n-I_1).
\end{equation}
Here, $I_1$ is  $\lambda$-independent and should be calculated including also the sub-leading contributions to the action. This allows to calculate the energy spacing between subsequent quasi-degenerate doublet
\begin{equation}
E_{n+1}-E_n=2\pi\lambda\left[\oint \left(\frac{dE}{dk_s}\right)^{-1}dQ_s\right]^{-1}.
\end{equation}
For small energy  and large radii of curvatures one finds
\begin{equation}
E_{n+1}-E_n=2\pi \lambda v/L_{\rm rs},
\end{equation}
where $L_{\rm rs}$ is the (rescaled) perimeter of the domain wall.
\section{Setting up the tight binding simulation}
\label{Appendix:tight_binding_setup}

\begin{figure*}
\begin{center}
\includegraphics[width=1\columnwidth]{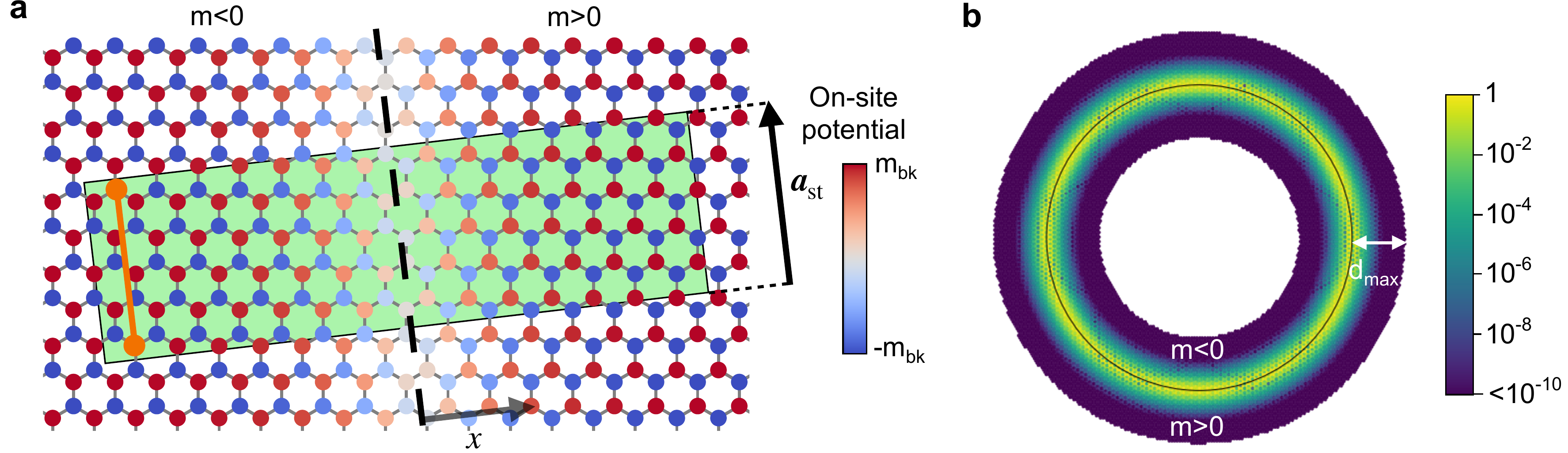}
\caption{
\textbf{a}, Setup of the tight binding strip simulation. Green box is the strip unit cell. Domain wall is indicated with thick black dashed line. The on-site potential depends on the sublattices A and B and the distance $x$ from the domain wall. An example of a pair of sites that are coupled via periodic boundary conditions is indicated in orange. 
\textbf{b}, Typical standing wave mode intensity $\vert \psi_n \vert ^2$ (maximum value normalized to 1) of the circular domain wall with reduced imprint. The maximum length $d_{\rm max}$ is chosen such that the eigenmode intensity decays by a factor greater than $10^{-10}$.
}
\label{SI_simulation_setup}
\end{center}
\end{figure*}

In this section, we describe the details of the tight binding numerical simulations of the translationally invariant strip (Figs.~\ref{fig2},\ref{fig3} of the main text) and the circular closed domain wall (Fig.~\ref{fig4} of the main text). 

All the tight binding numerical simulations in the main text requires defining a Hamiltonian matrix of the considered geometry on the honeycomb lattice, and obtaining its eigenvalues and eigenstates. A tight binding model of $N$ sites on a honeycomb lattice can be described by the Hamiltonian matrix of dimension $N$. The on-site potential at the $\rm n^{th}$ site is given by the diagonal matrix element $H_{\rm n,n}$. The coupling between the site $\rm n$ and the site $\rm m$ is given by the matrix element $H_{\rm n,m}=H_{\rm m,n}^*$. Below, we show in detail the algorithm to define the Hamiltonian matrix for the two geometries considered in the main text: the translationally invariant strip and the circular domain wall.

\subsection{Tight binding simulation of strip}
\label{Appendix:tight_binding_setup_strip}
Here, we outline the steps to define the Hamiltonian $H_{k}$ for the conserved quasi-momentum $k \in [-\pi/\vert {\bf a}_{\rm st} \vert, \pi/\vert {\bf a}_{\rm st} \vert)$ (${\bf a}_{\rm st}$ is the strip unit vector) in the strip Brillouin zone. First, we assign the nearest-neighbor couplings, and the on-site potentials at all the sites of the honeycomb lattice within a fixed region. The on-site potential at a site depends on the sublattices A and B and the distance $x$ (See Fig.~\ref{SI_simulation_setup}a) of the site from the domain wall $H_{\rm n,n}=\pm m_{bk} \tanh(\lambda x/a)$, where positive (negative) sign corresponds to the sublattice A (B). Next, we define the strip unit cell, that will be a rectangular box (indicated  in green in Fig~\ref{SI_simulation_setup}a) with one edge parallel to ${\bf a}_{\rm st}$. The domain wall passes through the middle of the box, the longitudinal edge length is $\vert {\bf a}_{\rm st} \vert$ and the transverse edge length is set to be sufficiently large such that the edge state decays considerably within this length. We consider, for the evaluation, only those sites that are located inside the strip unit cell. Thus, if N sites are located in the strip unit cell, then the dimension of the Hamiltonian matrix is $N$. The next step is to identify all the pairs of sites within the strip unit cell that are coupled via periodic boundary conditions. An example of one such pair is shown in Fig.~\ref{SI_simulation_setup}a. For the site $\rm n$ and the site $\rm m$ that are coupled with periodic boundary condition, such that $({\bf r}_n - {\bf r}_m).{\bf a}_{\rm st}>0$, the coupling matrix element is given by $H_{\rm n,m}=Je^{-ik \vert {\bf a}_{\rm st} \vert }=H_{\rm m,n}^*$. The Hamiltonian $H_{k}$ is diagonalized to obtain all the eigenvalues and eigenstates corresponding to the quasi-momentum $k$.
\subsection{Tight binding simulation of circular domain wall}
\label{Appendix:tight_binding_setup_circle_domain_wall}
Analogous to the case of the strip that contains a straight domain wall, we can define the Hamiltonian of the geometry that contains a circular domain wall of radius $R$. For the setup shown in Fig.~\ref{fig4}a of the main text, the dimension $N$ of the Hamiltonian matrix scales as $N \propto R^2 \propto \lambda^{-2}$. This scaling behavior prevents us to efficiently diagonalize $H$ for larger radii that are required to obtain the results of Fig.~\ref{fig4}d ($N=5 \times 10^4$ for $\lambda^{-1}=1.1$) and Fig.~\ref{fig4}j ($N=7 \times 10^5$). We came up with two strategies to solve this problem:- 

(i) Reducing the imprint: We consider, for the evaluation, only those sites whose distance from the domain wall is less than a maximum distance $d_{\rm max}\propto \lambda^{-1}$. The maximum distance $d_{\rm max}$ is set to be sufficiently large such that the edge state decays considerably within this length (See the illustration in Fig.~\ref{SI_simulation_setup}b). With this solution, the dimension of the Hamiltonian still scales in a similar manner $N \propto Rd_{\rm max} \propto \lambda^{-2}$. However, the proportionality factor is reduced.

(ii) Using sparse matrix diagonalization: The Hamiltonian matrix has very few non-zero elements. Hence, it is not only memory efficient to store it as a sparse matrix, but also time efficient to diagonalize it. We use scipy.sparse package in Python for the numerical simulations.


\section{Edge states for different domain wall orientations}
\label{Appendix:fourier_transform_different_orientations}

\begin{figure*}
\begin{center}
\includegraphics[width=1\columnwidth]{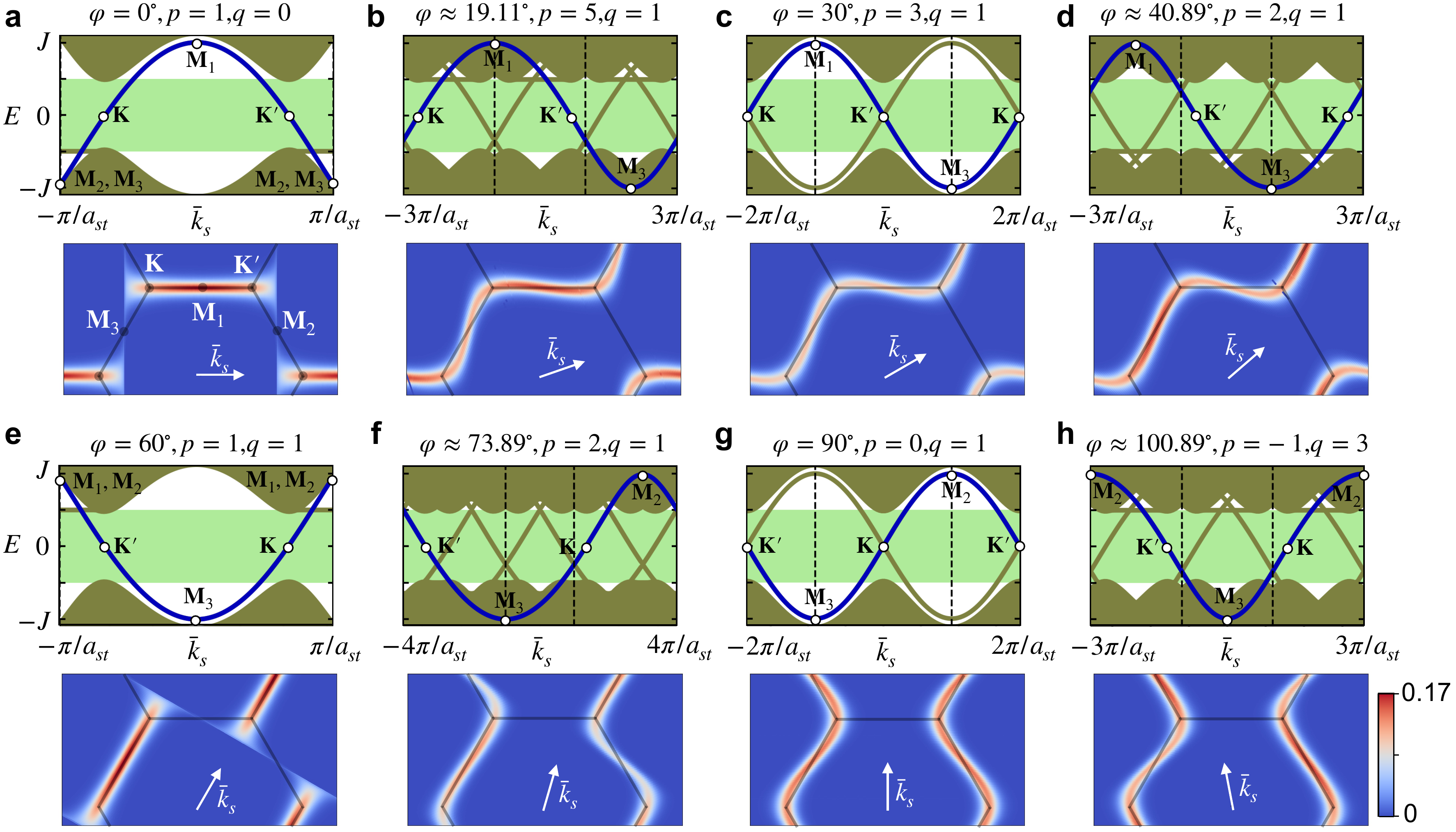}
\caption{Strip bandstructures and edge-band wavefunctions $\vert \tilde{\boldsymbol{\psi}}(\mathbf{k}) \vert ^2$ for few hand picked and increasing values of $\varphi$ from $0^\circ$ to $120^\circ$. The translationally invariant strip is defined for the condition $\sqrt{3}\cot \varphi = p/q$. The wavefunctions are plotted for the edge bands highlighted in blue. Note the closed loop traversed by the edge band in the bandstructure (indicated in blue) as well as in the wavefunctions (red regions). Labels of the high-symmetric points ($\bf K,K',M_1,M_2,M_3$) beside the edge band maps the corresponding location of the band in the wavefunctions.
}
\label{SI_strip_bandstructures}
\end{center}
\end{figure*}

In Fig.~\ref{fig2} of the main text, we show the band structures and the wavefunctions $\vert \tilde{\boldsymbol{\phi}}_{\bar{k}_s}(k_r) \vert ^2$ for the two domain wall orientations $\varphi = 90^\circ$ $(p=0, q=1)$ and $\varphi \approx 96.7^\circ$ $(p=-1, q=5)$. In this section, we investigate the same quantities for other domain wall orientations $\varphi$, cf Fig.~\ref{SI_strip_bandstructures}. Note that due to the $120^\circ$ rotation symmetry of the graphene tight-binding Hamiltonian with non-zero mass, the strip band structures for the orientation $\varphi$ and $\varphi + 120^\circ $ are identical. 

The edge state traverses a periodic loop in both the strip bandstructure and the wavefunction (See Fig~\ref{SI_strip_bandstructures}). The period of the loop $T(\varphi)=4\pi/(3a)\sin \varphi$  is a continuous function of $\varphi$. At $E=0$, this loop is at the valley $\bf K$ ($\bf K'$) for positive (negative) velocity of the edge state. At $E=\pm J$, the loop is at either of the three time-symmetric points $\bf M_1,M_2,M_3$. For $\varphi \in (0^\circ,60^\circ)$, the closed loop changes continuously and connects the two valleys via the $M_1$ and $M_3$ points. At $\varphi = 60^\circ$, corresponding to the zigzag domain wall orientation, the localization length of the wavefunction diverges at $E=+ J$, indicating the breakdown of the WKB approximation (See Appendix \ref{APP:armchair-zigzag-anal}). 
For $\varphi \in (60^\circ,120^\circ)$, the closed path of the wavefunction varies continuously and connects the two valleys via the $\bf M_2$ and $\bf M_3$ points.

\section{Position of scatterers at non-zero energy on the circular domain wall}
\label{Appendix:scat_points_non_0_E}

\begin{figure*}
\begin{center}
\includegraphics[width=1\columnwidth]{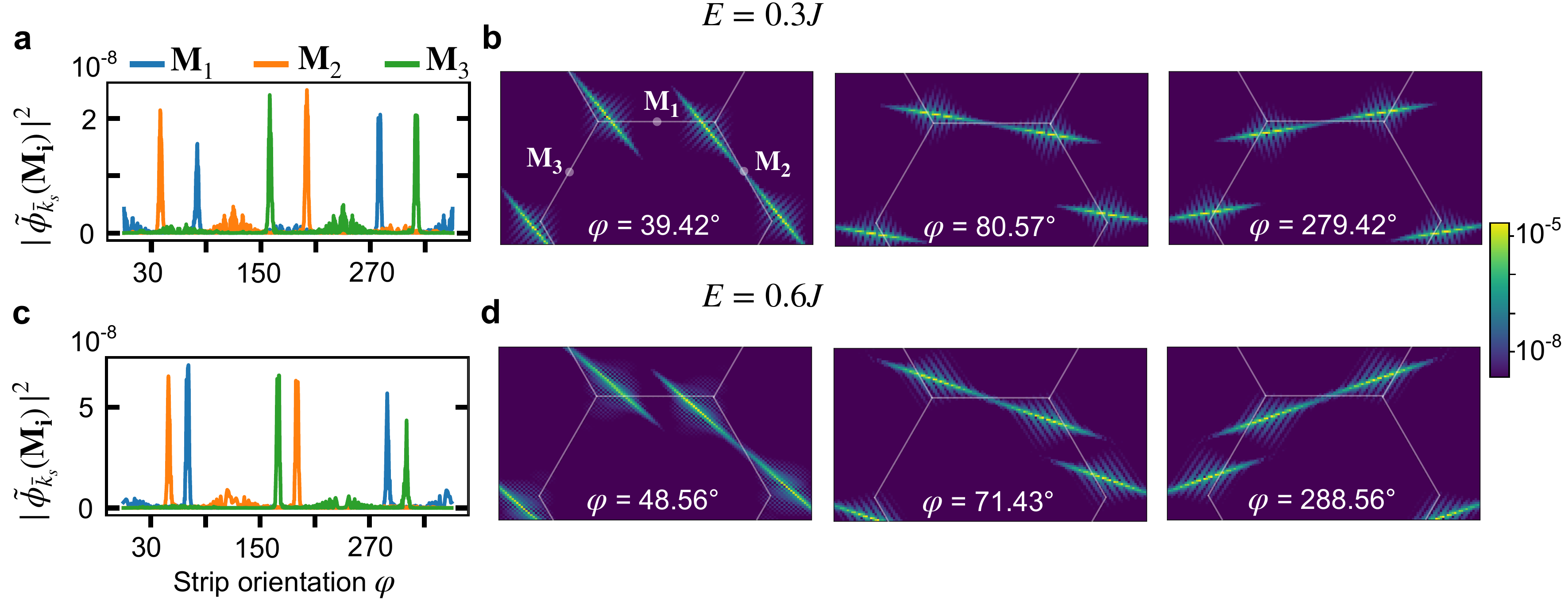}
\caption{
\textbf{a} (\textbf{c}), Fourier transform of the eigenmode at $E=0.3J$ $(0.6J)$ at the three \textbf{M} points as a function of position along the domain wall. The probability density shoots up at the scatterer locations. 
\textbf{b} (\textbf{d}), Local Fourier transforms of wave function at some of the scatterer locations along the domain wall, for $E=0.3J$ $(0.6J)$. For different scatterers, the tunneling path between the two valleys passes through the different $\bf M$ points.
}
\label{SI_circle_cavity_non_zero_contours}
\end{center}
\end{figure*}

In Fig~\ref{fig4}f of the main text, we show the location of the scatterers on the circular domain wall interface for three different energies. We also show that {\it local} fourier transforms of the zero energy eigenmode at the $\bf M_1, M_2, M_3$ points shoots up at the scatterer locations. This can be understood from the fact that the tunneling path in the reciprocal space between the two valleys is through the time-symmetric $\bf M$ points. In Fig.~\ref{SI_circle_cavity_non_zero_contours}, we demonstrate this fact for the non-zero energy eigenmodes.



\end{document}